%% file: nulling.tex
\documentclass[twocolumn, twocolappendix]{aastex631}
\usepackage{float}
\usepackage{amsmath}
\usepackage{hyperref}

\newcommand{\nf}{\ensuremath{\mathrm{NF}}}
\newcommand{\nfr}{\ensuremath{\mathrm{NF}_{\mathrm{r}}}}
\newcommand{\plmi}{$\pm$}

\shorttitle{Pulsar Nulling with Mixture Models}
\shortauthors{Anumarlapudi et al.}
\graphicspath{{./}{figures/}}

\begin{document}

\title{A Pilot Study of Nulling in 22 Pulsars Using  Mixture Modeling}

\correspondingauthor{Akash Anumarlapudi}
\email{aakash@uwm.edu}

\author[0000-0002-8935-9882]{Akash Anumarlapudi}
\affiliation{Center for Gravitation, Cosmology, and Astrophysics,
  Department of Physics, University of Wisconsin-Milwaukee, PO Box
  413, Milwaukee, WI, 53201, USA}

\author[0000-0002-1075-3837]{Joseph~K.~Swiggum}
\affiliation{Center for Gravitation, Cosmology, and Astrophysics,
  Department of Physics, University of Wisconsin-Milwaukee, PO Box
  413, Milwaukee, WI, 53201, USA}
\affiliation{Dept. of Physics, 730 High St., Lafayette College, Easton, PA 18042, USA}

\author[0000-0001-6295-2881]{David~L.~Kaplan}
\affiliation{Center for Gravitation, Cosmology, and Astrophysics,
  Department of Physics, University of Wisconsin-Milwaukee, PO Box
  413, Milwaukee, WI, 53201, USA}

\author{Travis D.~J.\ Fichtenbauer}
\affiliation{Center for Gravitation, Cosmology, and Astrophysics,
  Department of Physics, University of Wisconsin-Milwaukee, PO Box
  413, Milwaukee, WI, 53201, USA}






\begin{abstract}
The phenomenon of pulsar nulling, observed as the temporary inactivity of a pulsar, remains poorly understood both observationally and theoretically. Most observational studies that quantify nulling employ a variant of Ritchings (1976)'s algorithm which can suffer significant biases for pulsars where the emission is weak. Using a more robust mixture model method, we study pulsar nulling in a sample of 22 recently discovered pulsars, for which we publish the nulling fractions for the first time.  These data clearly demonstrate  biases of the former approach and show how an otherwise non-nulling pulsar can be classified as having significant nulls. We show that the population-wide studies that find a positive correlation of nulling with pulsar period/characteristic age can similarly be biased because of the bias in estimating the nulling fraction. We use our probabilistic approach to find the evidence for periodicity in the nulls in a subset of three pulsars in our sample. In addition, we also provide improved timing parameters for 17 of the 22 pulsars that had no prior follow-up.

\end{abstract}



\section{Introduction}\label{sec:intro}
Pulsar nulling, initially observed by \citet{backer}, is the absence of observed emission from a pulsar for one or more pulse periods. Observationally, the phenomenon of pulsar nulling remains poorly understood. It is clear that nulling is a broadband phenomenon, observed from 102\,MHz \citep{Davies1984} to 8.35\,GHz \citep{sneha2012}. However, it is not 
firmly established whether nulling is simultaneous over this frequency range using a large sample of nulling pulsars. Prior studies found contradictory conclusions.  For example, observing over two frequency ranges, 50-140~MHz and 275-430~MHz, \citet{Taylor1975} found that nulls are  simultaneous in two different pulsars (PSR B0031$-$07,  PSR B0809+74), while \citet{Davies1984} found the evidence for excessive nulls in single pulses at 102~MHz compared to 406~MHz in PSR B0809+74. 
A more recent study by \citet{Vishal2014} found that the nulls are highly coherent in three pulsars at four different frequencies --- 313, 607, 1380, and 4850~MHz. In addition, it is also not clear whether pulsars null randomly. \citet{Redman2009} and \citet{Vishal2012} found that 
nulls might not occur randomly but might be clustered, where nulls and bursts tend to occur in groups, but the latter found that the null durations can be random. However, for many of these results the dependency of the nulling inferences on signal-to-noise ratio makes it hard to robustly interpret their findings.

Although the formation of a pair cascade and the radiation from these accelerated pairs in the pulsar magnetosphere is often invoked to explain the observed emission from a pulsar \citep{ruderman}, a full theory of pulsar magnetospheres and its emission to explain the diverse morphology in pulse profiles and phenomenology is yet to be developed. As such, the theory of pulsar nulling remains largely speculative, though it is often attributed to one of two classes: i) inherent to the magnetosphere itself such as loss of coherence condition required for radio emission, e.g., \citet[]{Filip&Radha}, or the depletion of pairs in the magnetosphere themselves, e.g., \citet{kramer2006} or ii) geometrical factors external to the magnetosphere such as the line of sight traversing through the `empty' region between rotating emission carousels, e.g., \citet{herfindal07, herfindal09}.  Further progress may require additional observational data to understand how the properties of nulling relate to the properties of the pulsars themselves.

Nulling as a phenomenon may be related to other more extreme forms of intensity modulation, where the pulses can disappear for hours to months in the cases of rotating radio transients (RRATs; \citealt{rrats}) or intermittent pulsars \citep{kramer2006,Lyne2009}.  However, the connection between these populations is not clear.  Furthermore, pulsar nulling is often discussed in tandem with two other forms of single pulse variations: mode changing -- a phenomenon in which an otherwise stable pulse profile switches between multiple shapes (or modes) \citep{modechanging} and sub-pulse drifting -- a phenomenon in which the single pulse phase shows a uniform periodic drift \citep{subpulsedrift}. Regardless, in all of these cases the appearance of these phenomena can be limited by instrumental sensitivity: without enough sensitivity to probe single pulses at high significance,  one cannot be certain whether the pulsar emission is truly missing during the nulls or the pulsar switches to an alternate mode with lower intensity. Together all three are often thought of as different representatives of a larger underlying phenomenon of sub-pulse intensity variations \citep{handbook}.

Nulling is usually quantified by the fraction of pulses where there is no discernible emission, called the Nulling Fraction (\nf). \nf\ can vary from 0 -- in the case of standard emission picture that shows no nulls -- to 1, in the extreme case where the pulsar emission is visible only between long nulls (intermittent pulsars and RRATs). \nf\, has been measured in roughly 8\% of pulsars, but this has more to do with the lack of single pulse studies as opposed to nulling being restricted to a small subset of pulsars. This smaller data set of nulling pulsars  is entirely restricted to normal (not recycled) pulsars, owing to the high sensitivity demands that would be needed to observe single pulses of millisecond pulsars (MSPs), although some recent studies \citep{Kaustubh2014} have been conducted in a  sample of bright MSPs which did not find a signature of nulling with high confidence. In addition, there can be a bias against discovering normal pulsars which tend to have a high \nf, or are intermittent. Hence the fraction (8\%), can only be considered as a conservative lower limit. 

Such a small data set restricts our ability to infer population-wide properties, which might give clues to the origin of the phenomenon, and hence studies done thus far have not reached a consensus. An initial study done by \citet{Ritchings76} claimed a correlation between \nf\ and pulsar period (with longer period pulsars experiencing higher \nf) and also a stronger correlation with the characteristic age. \citet{Wang2007} also suggested a correlation with  spin-down age, albeit  qualitatively, with older pulsars experiencing higher \nf, before eventually crossing the death line. \citet{Konar2019} found that there may be two different populations of pulsars separated by a \nf\ of $\sim$ 40\% but did not find correlations with any intrinsic pulsar properties, while \citet{Sofia2021} claimed that there is no strong evidence for the existence of two sub-populations. All of these studies may be significantly biased since the samples used are restricted to the pulsars that explicitly showed nulling.

In general, most studies \citep{Wang2007,Vishal2012,Vishal2013,Vishal2014,herfindal09} estimate \nf\ using the methodology (or a variant) proposed by \citet{Ritchings76}. But as \cite{Kaplan2018} demonstrated, this method can suffer strong biases in the case of weaker pulsars which can lead to overestimating the \nf\ and classifying an otherwise standard weak pulsar as a nulling pulsar. This can also lead to systematic biases in population inferences. In addition, \citet{Kaplan2018} proposed an alternate method in which they use Gaussian Mixtures to model the single pulse intensities and estimate the \nf\,, and demonstrate the reliability of this method in accurately measuring the \nf\, in weaker pulsars. In this study, we expand on the Gaussian Mixture Model (GMM) of \cite{Kaplan2018}\footnote{As noted in \citet{Kaplan2018}, a similar method may have been used in \citet{Arjunwadkar2014}.} to generalize their method and apply it to a larger sample of 22 pulsars\footnote{All of our code is available at \url{https://github.com/AkashA98/pulsar_nulling}}. 

Pulsars selected for this study were discovered as a part of the Green Bank North Celestial Cap (GBNCC) pulsar survey \citep{gbncc} in 2-min drift scans at 350~MHz with a 100\,MHz bandwidth and with data sampled  every 81.92\,$\mu$s. 
At 350~MHz the beam size is 36\arcmin\ (Full Width at Half Maximum; FWHM) and hence the astrometric precision prior to a coherent timing solution is limited by the beam size depending on the Signal-to-noise ratio (SNR) of the discovery candidate. These were later followed up at the Green Bank Telescope (GBT) and Arecibo Observatory (AO) to improve their timing solutions and establish their nulling characteristics.

The structure of this paper is as follows: In Section \ref{sec:data_anly}, we detail our data acquisition and reduction methods, and provide  updated timing solutions for the pulsars in this study. We then describe the mixture model and provide our basic results in Section \ref{sec:methods}. Finally, we present the implications of the results in Section \ref{sec:discussion} and conclude in Section~\ref{sec:conc}.



\section{Data Analysis} \label{sec:data_anly}

\subsection{Observations and Data Reduction} \label{sec:data_red}
A sample of 22 recently discovered pulsars was selected for this pilot study if they showed any signs of intermittency in their discovery plots\footnote{See the GBNCC discovery page: \url{http://astro.phys.wvu.edu/GBNCC}.}. Data for 15 out of 22 pulsars were collected using the 100-m Robert C.\ Byrd Green Bank Telescope (GBT) (hereafter referred to as the GBT sample), operating at 820\,MHz with a bandwidth of 200\,MHz, in 2\,hr contiguous scans, with the primary aim of determining the pulsars' nulling characteristics (project code 18A$-$436; PI: J.~Swiggum). Data for another nine pulsars were collected at the 300-m William E.~Gordon Arecibo Observatory (AO) operating at 430\,MHz over a bandwidth of 24\,MHz, with the goals to both establish coherent timing solutions and determine nulling characteristics (project code P3436; PI: J.~Swiggum) (hereafter referred to as the AO sample). Two pulsars in our sample, PSR J0414+31, and PSR J1829+25, were observed at both observatories. 

Six of the 15 pulsars in the GBT sample already had coherent timing solutions \citep{Lynch2018} and the data for these were collected in coherent search mode using the Green Bank Ultimate Pulsar Processing Instrument (GUPPI; \citealt{guppi}) with 128 frequency channels sampled at 10.24\,$\mu$s and  retaining full polarization information. The remaining nine pulsars had no prior follow-up campaigns and so we first improved their positions using gridding observations and then observed them in incoherent search mode with 2048 frequency channels sampled at 40.96\,$\mu$s. Data for the AO sample were collected in coherent search mode using the Puerto Rico Ultimate Pulsar Processing Instrument\footnote{\href{ http://www.naic.edu/puppi-observing/}{http://www.naic.edu/puppi-observing/}} (PUPPI), with 64 channels sampled at 40.96\,$\mu$s, over a span of $\sim$ six months to establish coherent timing solutions in addition to studying the nulling properties. A summary of observations for each pulsar is provided in Tables \ref{tab:gbt_data} and \ref{tab:ao_data}.
\input{GBT_obs_schedule}
\input{AO_obs_schedule}

Starting with the raw search mode data, we used \texttt{dspsr} \citep{dspsr} to fold the data. We then used \texttt{pazi}, the interactive zapping routine in \texttt{psrchive} \citep{psrchive} to remove radio frequency interference (RFI)-affected frequency channels and single pulses. For GBT data, we also made use of  RFI scans taken at the observatory\footnote{\href{https://greenbankobservatory.org/rfi-gui-user-guide/}{https://greenbankobservatory.org/rfi-gui-user-guide/}}, when available, to identify the frequency bands that are affected by RFI, which are otherwise  not obvious visually. In some cases, we found that one of the polarization channels was persistently affected by RFI, and in such cases we excluded data from that polarization channel at the cost of SNR. Fortunately, this did not have a significant impact on the determination of the nulling fractions. 
Some of the AO data had periodic ``drop-outs''  in the data  with sub-millisecond periodicity at zero dispersion measure (DM), caused by data rate overflow during the observations.  We cleaned these ``drop-outs" by replacing the data with \texttt{NaN} values and being careful to exclude those when folding/averaging.
After cleaning  the RFI, both for timing and estimating nulling, we averaged polarizations to measure the total intensity. 


\subsection{Timing}\label{sec:timing}
For the 16 pulsars in our sample that had no prior follow-up, we first tried to improve the timing parameters. We used \texttt{paas} from \texttt{psrchive} \citep{psrchive} to make a standard template and then used \texttt{pat} to extract the Times of Arrival (TOAs) from the data. For the GBT data, our goal was to improve the spin frequency ($F_0$) and DM measurements since we had only 2\,hour scan at a single epoch for each source. For the AO data, the data spanned $\sim$3--6 months depending on the pulsar and hence we can generate a phase-connected solution. However, the relatively narrow bandwidth of the observations (24\,MHz) restricted our ability to fit for DM using sub-banded TOAs and hence we used the DM of the discovery candidate found on the GBNCC discovery page. 


The timing solutions for all the pulsars in this study are given in Table~\ref{tab:tim_param}. For pulsars observed at GBT we improved the positions through gridding, and $F_0$ and DM estimates through timing. For pulsars observed at AO, we improved the gridded positions, $F_0$ and the frequency derivative $F_1 = \dot F_0$ through coherent timing. For the two overlapping pulsars observed at both GBT and AO, a timing solution was obtained by combining the TOAs from both  observatories. In the case of pulsars observed at AO for only $\sim$3 months (J0355+28, J0414+31, J1822+02), and pulsars where a combination of low SNR and nulling resulted in  few TOAs with SNR $>8$ (J1928+28), it is difficult to estimate both position and $F_1$ precisely (they are highly covariant). In such cases, we rely on the

\begin{figure*}
    \plotone{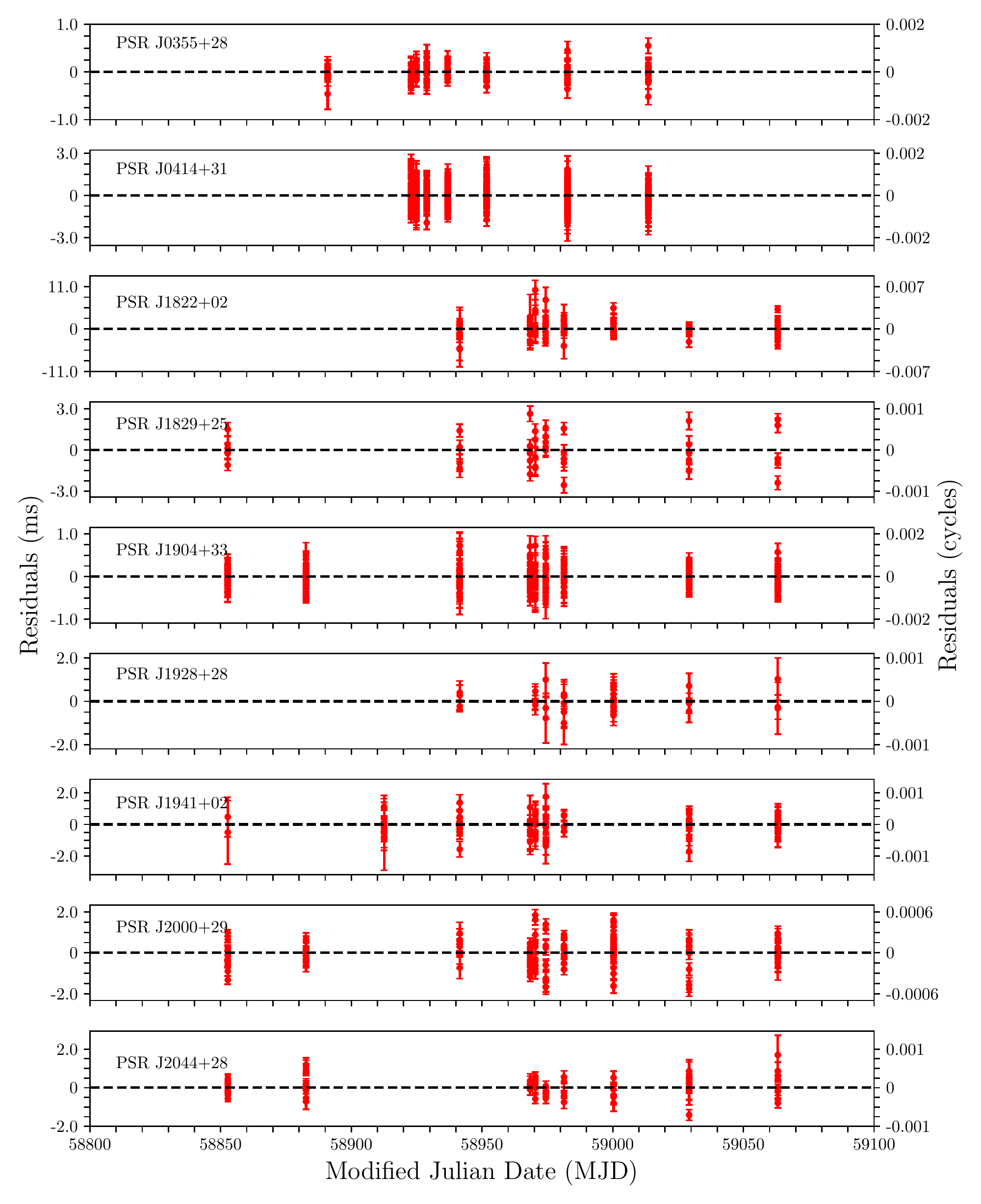}
    \caption{Timing residuals for the pulsars observed in the timing/nulling campaign at the AO. The red dots are the residuals (in milliseconds) from the timing model with the error bars representing the 1-$\sigma$ error on the TOAs. The timing model solutions are presented in Table \ref{tab:tim_param}.}
    \label{fig:resids}
\end{figure*}

F-statistic, given by 
\[
F = \frac{(\chi_0^2 - \chi^2)/(p - p_0)}{\chi^2/p}
\]
where $\chi_0^2$ and $\chi^2$ are the chi-squared values of the timing residuals, and $p_0$ and the $p$ are the degrees of freedom before and after the addition of $F_1$ (or any additional parameter(s), in general). This F-statistic follows an F-distribution \citep{lomax2007statistical} and hence we include $F_1$ in the fit if the improvement in the goodness of fit ($\chi^2$) due to $F_1$ is $<$1\% by chance.
The resulting timing residuals are shown in Figure \ref{fig:resids}.

\input{timing_params}

\subsection{ON/OFF histograms}
\begin{figure}[!h]
    \gridline{\fig{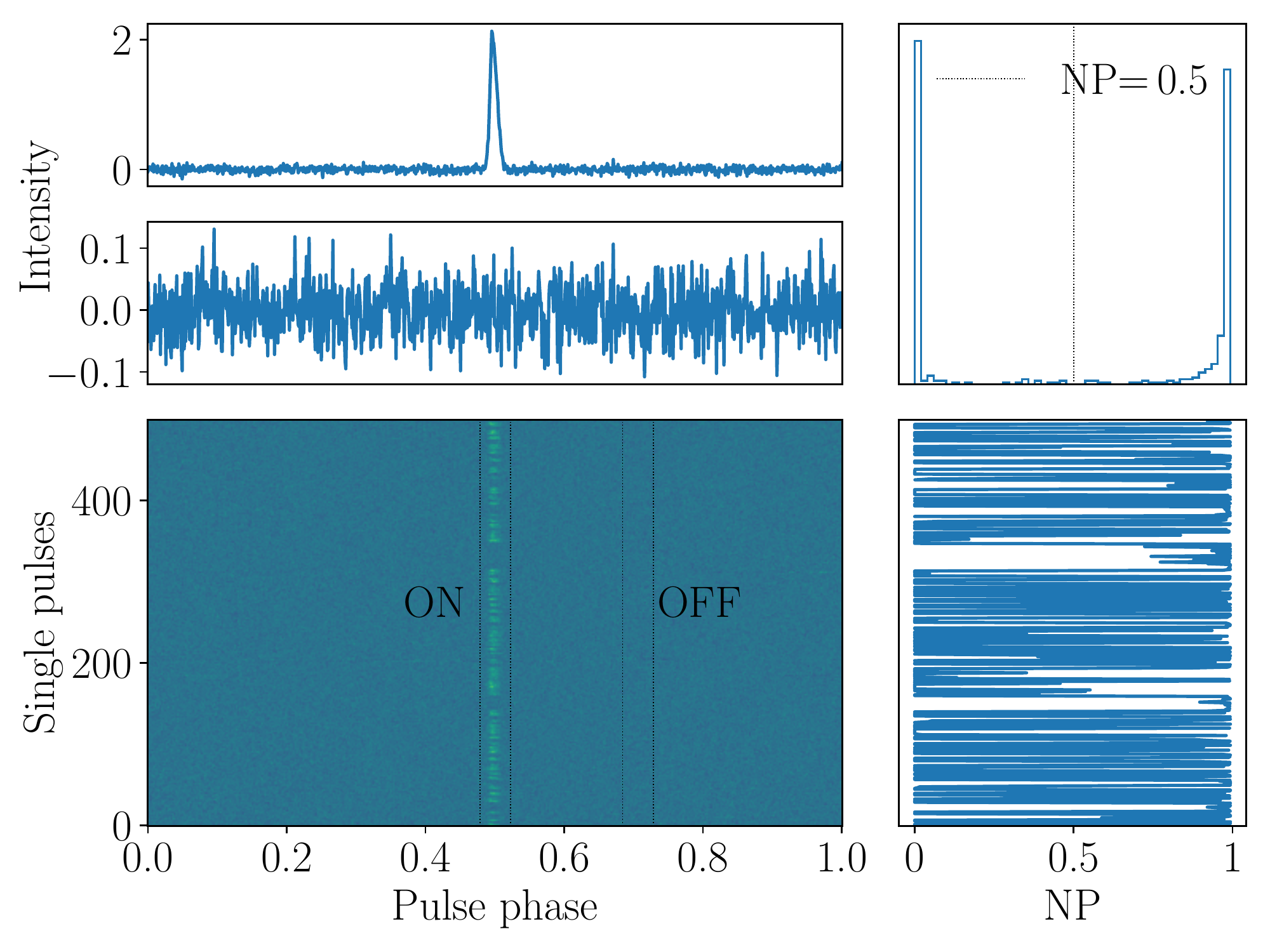}{0.45\textwidth}{(a) Single pulse stack of PSR J0325+6744}}
    \gridline{\fig{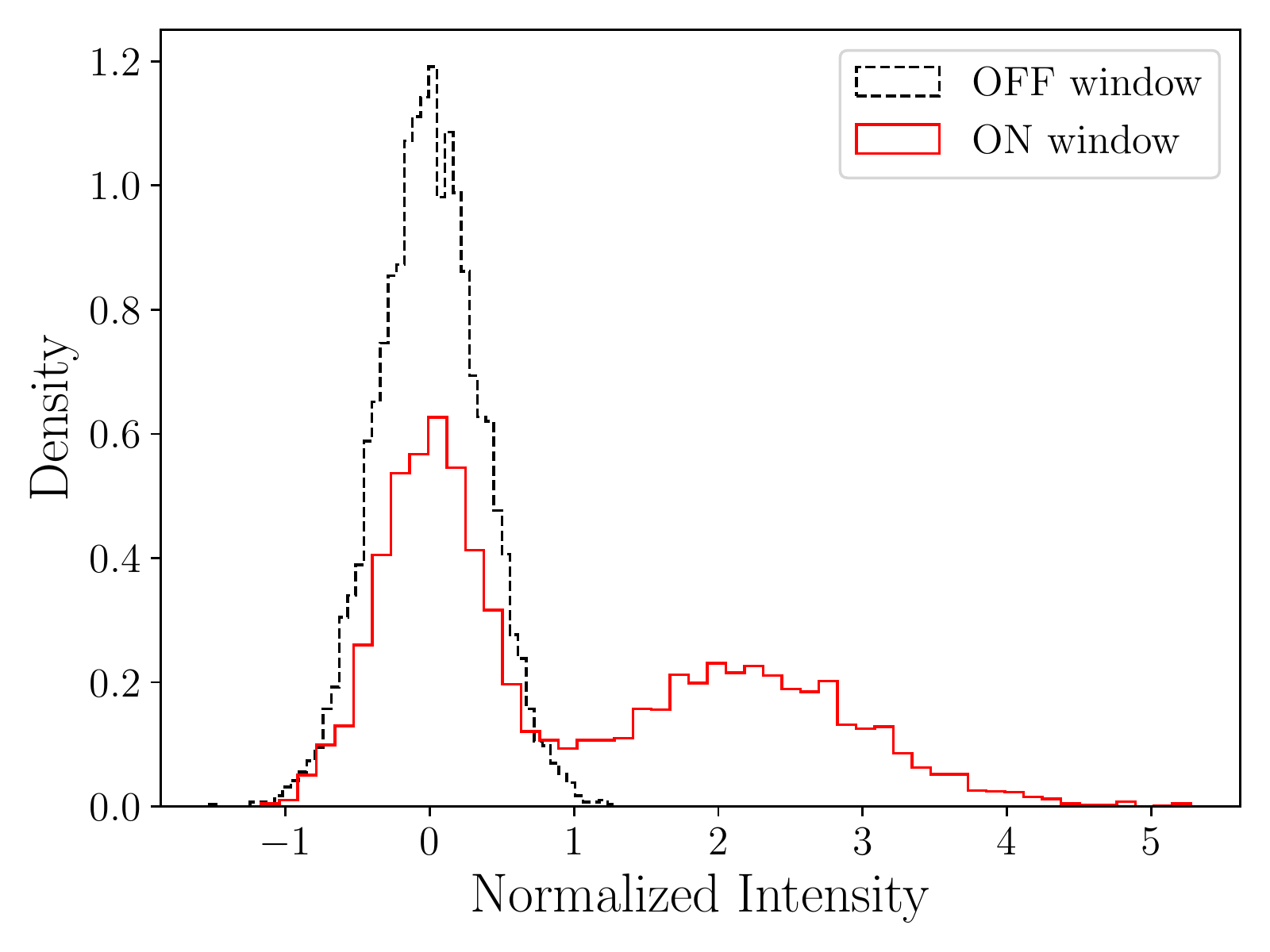}{0.45\textwidth}{(b) Pulse intensity histogram for PSR J0325+6744}}
    \caption{(a)The bottom left panel shows the single pulse stack with the ON and OFF windows marked with black dashed lines. Null probabilities (NP) for every single pulse are calculated using the method described in \S\ref{sec:nf_prob} and are shown in the bottom right plot. The distribution of NP is shown in the top right panel where we can clearly see the evidence for two classes of pulses. The summed profile of all the single pulses with null probability $<$\,0.5 is shown in the top left panel, while the summed profile for pulses with null probability  $>$\,0.5 is shown in the middle panel. (b) The pulse intensities in the OFF and ON windows are shown in blue and orange histograms. The presence of excessive counts in the ON histogram (the null component) at the background noise level separated from a second component at higher intensities (the emission component) is evidence for the nulling behavior.}
    \label{fig:pul_stack}
\end{figure}

Once we had improved the timing solution, we used \texttt{dspsr} in single pulse mode to generate single pulses for all scans and used \texttt{psradd}, from \texttt{psrchive}, to phase align pulses from different scans after cleaning the data for RFI. We then averaged the data along the polarization and frequency axes to obtain the  pulse intensity of the single pulses as a function of the rotational phase and generated  single pulse stacks such as that shown in Figure \ref{fig:pul_stack}.


The most important aspect in estimating the nulling fraction is determining the ``ON"-pulse and ``OFF"-pulse phase windows.
The single pulse intensities in the ``OFF"-pulse window should be entirely due to radiometer noise, while the intensities in the ``ON"-pulse window should be the sum of the radiometer noise component (same as the ``OFF"-pulse window) and the pulsar emission component.
We first generated the average pulse profile to visually select on and off windows of the same widths.  We then fit a sixth-order polynomial as a function of pulse phase to each single pulse \citep[similar to][]{Rosen2013, Lynch2013, Kaplan2018} after masking the ON/OFF windows to remove any trends and construct a flat baseline. We recorded the ON/OFF intensities as the sum of the baseline-subtracted intensities across the windows. Finally,  we constructed histograms of the ON/OFF intensities which we used to determine the nulling properties.
Figure~\ref{fig:pul_stack} shows the single pulse intensity distribution in the ON/OFF window. The OFF histogram can be accurately described by a single component (Gaussian noise), but the ON histogram can have multiple components --- ``null'' and ``emission'' components. The presence of nulling manifests in the ON histogram as an excess of samples at levels consistent with the OFF component, which we refer to as the null component. The residual distribution, after removing the null component, is supposed to be a realization of pulsar's emission distribution (hereafter referred to as `emission' component). The emission component can be a single distribution or a combination of multiple distributions. The ON distribution can be thought of as the sum of the null and the emission components.


\section{Methods \& Results}\label{sec:methods}
\subsection{Determining Nulling Frations}\label{sec:nulling}
As demonstrated by \citet{Kaplan2018}, Ritchings' method can give biased estimates for \nf\, (hereafter referred as \nfr) in pulsars where the emission component is close to the noise level.  Therefore, following \citet{Kaplan2018} we adopt a method which models the ON/OFF histograms using a mixture model (MM). This means that the intensities $x$ can be considered as random draws from the probability density function (PDF)
\begin{equation}
    p(x| \Bar{\theta}) =  \sum_{n=1}^{m} c_n \ \mathcal{F}_{n} (x| \{\theta_n\}),
    \label{eqn:pdf}
\end{equation}
where the $\mathcal{F}_{n}$ functions are the individual probability density functions parameterized by the set $\{\theta_n\}$, $c_n$ are the weights. In the case where all the $\mathcal{F}_{n}$ functions are the same and are normal distributions 
\[
\mathcal{F}_{n}(x; \mu_n, \sigma_n) = \mathcal{N}(x; \mu_n, \sigma_n) = \frac{1}{\sqrt{2 \pi} \sigma_n} e^{-\frac{1}{2}\left(\frac{x - \mu_n}{\sigma_n} \right)^2},
\]
where \{$\mu_n$\} and \{$\sigma_n$\} are the means and standard deviations of component $n$, this reduces to a Gaussian mixture model (GMM), but more general models are considered. There is an additional constraint that the weights $c_n$ add to one: 
\[
\sum_{n=1}^m c_n = 1,
\] 
which comes from the normalization of the PDF, which leaves the total number of free parameters to be determined as $\sum_{n=1}^{m} \mathrm{dim} (\{\theta_n\})$ model parameters, and $m-1$ latent parameters. 


In general, the OFF histogram can be well-described by a Gaussian as expected of radiometer noise (assuming that RFI has been sufficiently removed), and this is what we observe in our data. The emission component usually can be  described by a single Gaussian as well. However, there are cases when it deviates from a single Gaussian component. More than one component is a possibility considered in \citet{Kaplan2018}, which can be tested against the single-component model through a model comparison test.  However, we also consider non-Gaussian models here.  Specifically, multi-path propagation of the pulses through the interstellar medium (ISM) \citep{exptails,bhat2003,handbook}, can result in the emission distribution having long tails towards higher intensities. This effect can be reasonably well described  by the intensity  distribution 
\begin{align}
\mathcal{F}(x; \mu, \sigma, \tau) = \frac{1}{2\tau} \mathrm{exp}&\left(\frac{\sigma^2}{2\tau^2} \right) \mathrm{exp}\left(-\frac{x-\mu}{\tau} \right) \nonumber\\
& \mathrm{erfc}\left(- \frac{x - (\mu + \sigma^2/\tau)}{\sqrt{2}\sigma} \right)
\label{eqn:egauss}
\end{align}
which is a convolution of a Gaussian $\mathcal{N}(x; \mu, \sigma)$ and a one-sided exponential $\frac{1}{\tau} \mathrm{exp}(-x/\tau) \mathrm{U(x)}$, where $U(x)$ is the Heaviside or step function, ${\rm erfc}(x)$ is the complementary error function, and $\tau$ is the decay time of the exponential \citep{scatteringtail}. Hence we try to model the emission component using multi-component Gaussians and Gaussians with exponential tails and rank them using their Bayesian Information Criterion (BIC) values to choose the best-fit model.

We employ the \texttt{scikit-learn} Gaussian mixture model \citep{scikit}  to derive an initial fit for the ON and OFF histograms. This is based on the expectation–maximization (EM) algorithm, in which parameters are estimated by maximizing the likelihood function $\mathcal{L}$(data $|\, {\rm \Bar{\theta}}$) \citep[see][for details]{ivezic_book}.
This produces a very good fit for the OFF  histogram. However, in the case of weaker pulsars where the emission can be confused with the background, \citet{Kaplan2018} showed that this method can still fail in producing a reliable fit for the null and emission components of the ON histogram simultaneously, although this bias can be small compared to the Ritchings' algorithm. As such, a refined fit for the null and emission components can be obtained by performing a Markov-Chain Monte Carlo (MCMC) analysis.

For MCMC analysis, the likelihood function is given by 
\begin{equation}
    \mathcal{L}(\Bar{x}| \Bar{\theta}) = \prod_{i} p(x_i| \Bar{\theta})
\end{equation}
following $p(x_i|\Bar{\theta})$ from Equation~\ref{eqn:pdf}.

The priors chosen are:
\begin{itemize}
    \item Initial Gaussian fit from the EM algorithm for the off-pulse mean and standard deviation
    \item Uniform between the bounds dictated by the on-pulse intensities for the parameters governing the pulsar emission component
    \item Dirichlet distribution for the $m$ coefficients $c_m$ \citep{wilks2008mathematical}
\end{itemize}

We use  the \texttt{emcee} \citep{emcee} ensemble sampler to  sample  the posterior. We initialize 32 walkers within a $\pm 5\sigma$ range of the initial fit values of the parameters.
To account for the finite correlation length of the chains and produce independent samples, we first let the walkers ``burn-in" to erase their starting conditions, and we then let the walkers explore the parameter space until we have at least 100 independent samples for each walker.

\begin{figure*}
    \gridline{\leftfig{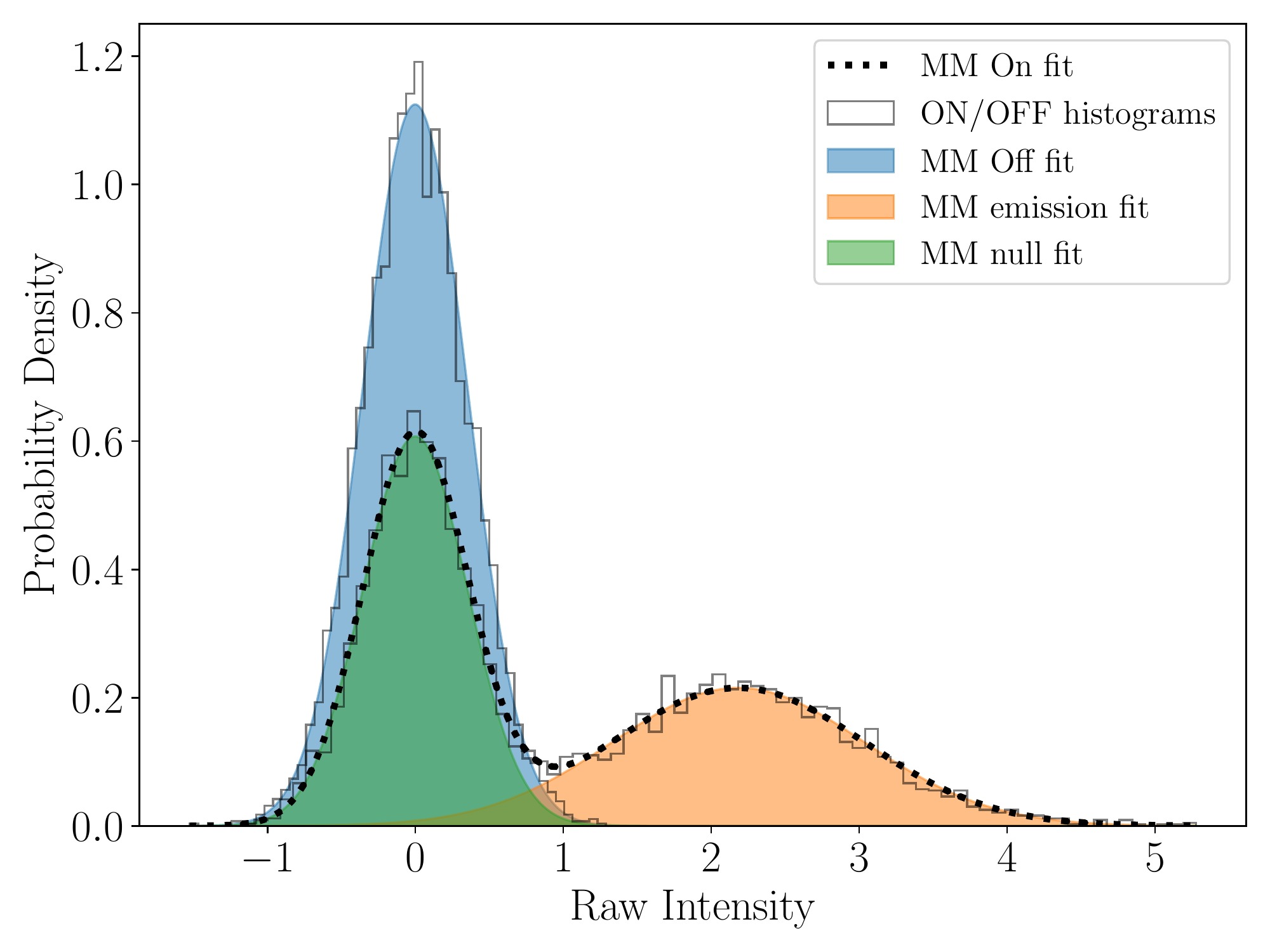}{0.48\textwidth}{(a1) 
                PSR~J0325+6744 -- $\nf=53.92 \pm 0.81$\% vs \nfr\,= 55.01\%}
                \rightfig{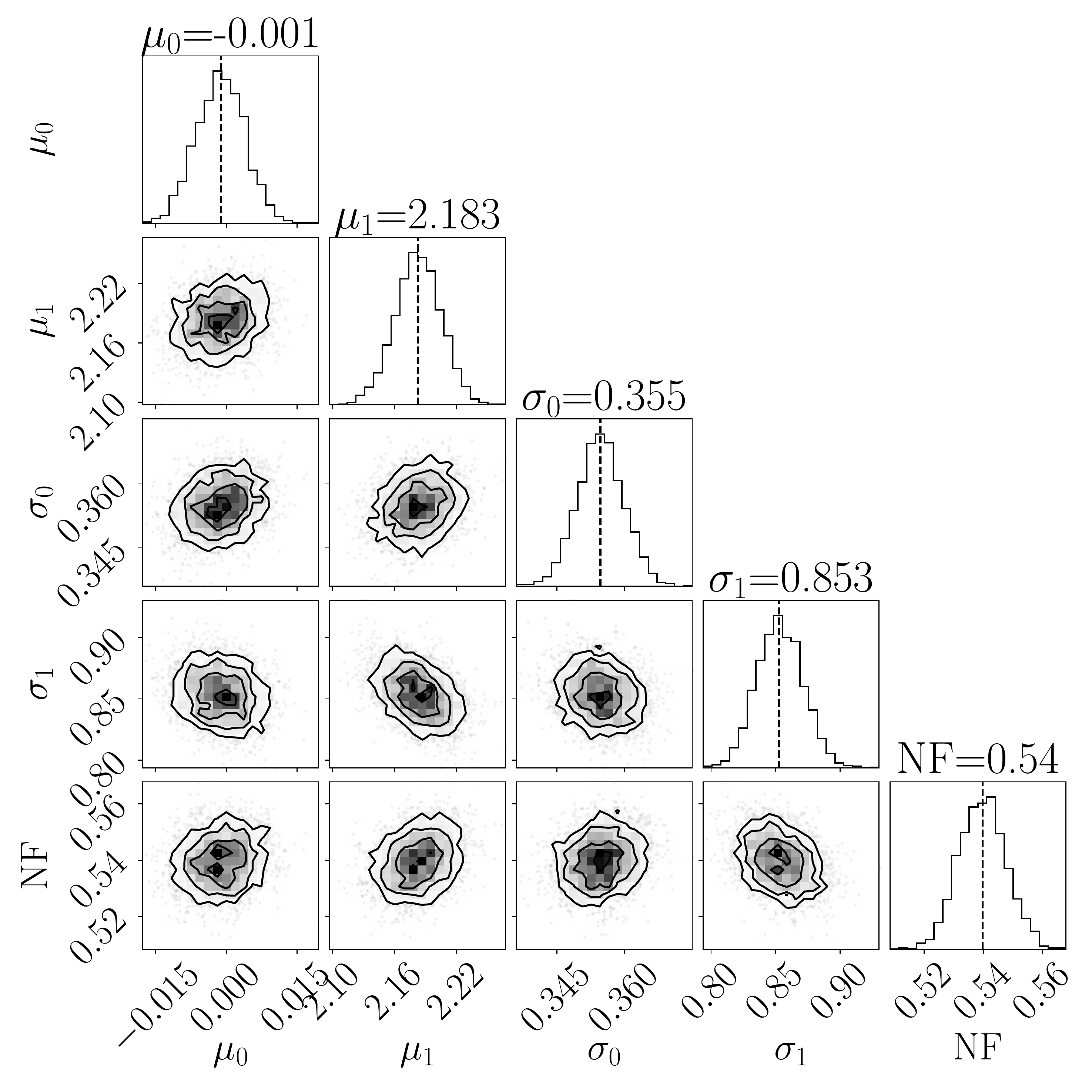}{0.5\textwidth}{(a2)             Model parameter posteriors for PSR~J0325+6744}
              }
              
    \gridline{\leftfig{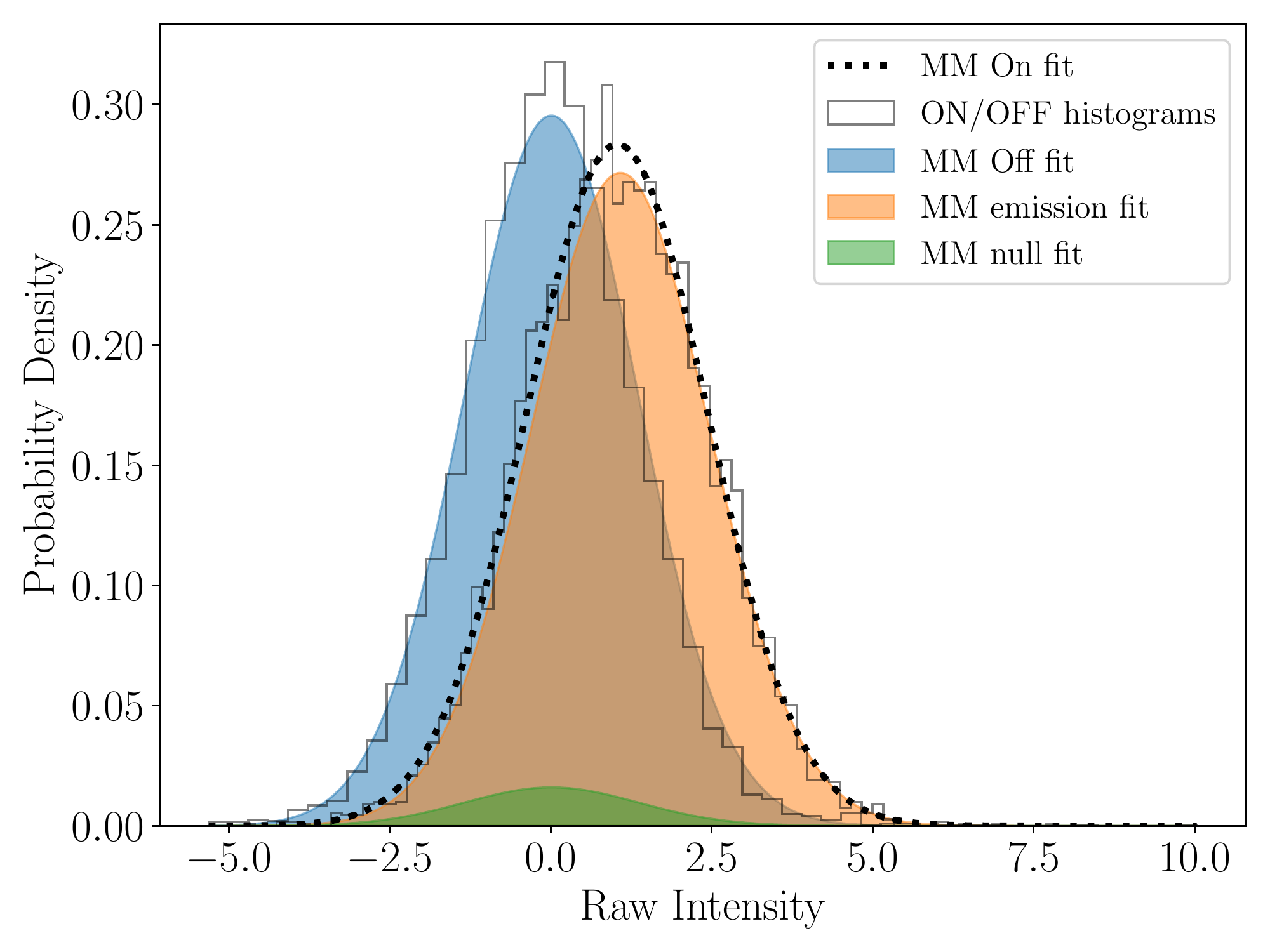}{0.48\textwidth}{(b1)
                PSR~J1529$-$26 -- $\nf=5.4 \pm 4.4$\% vs \nfr\,=48.5\%}
              \rightfig{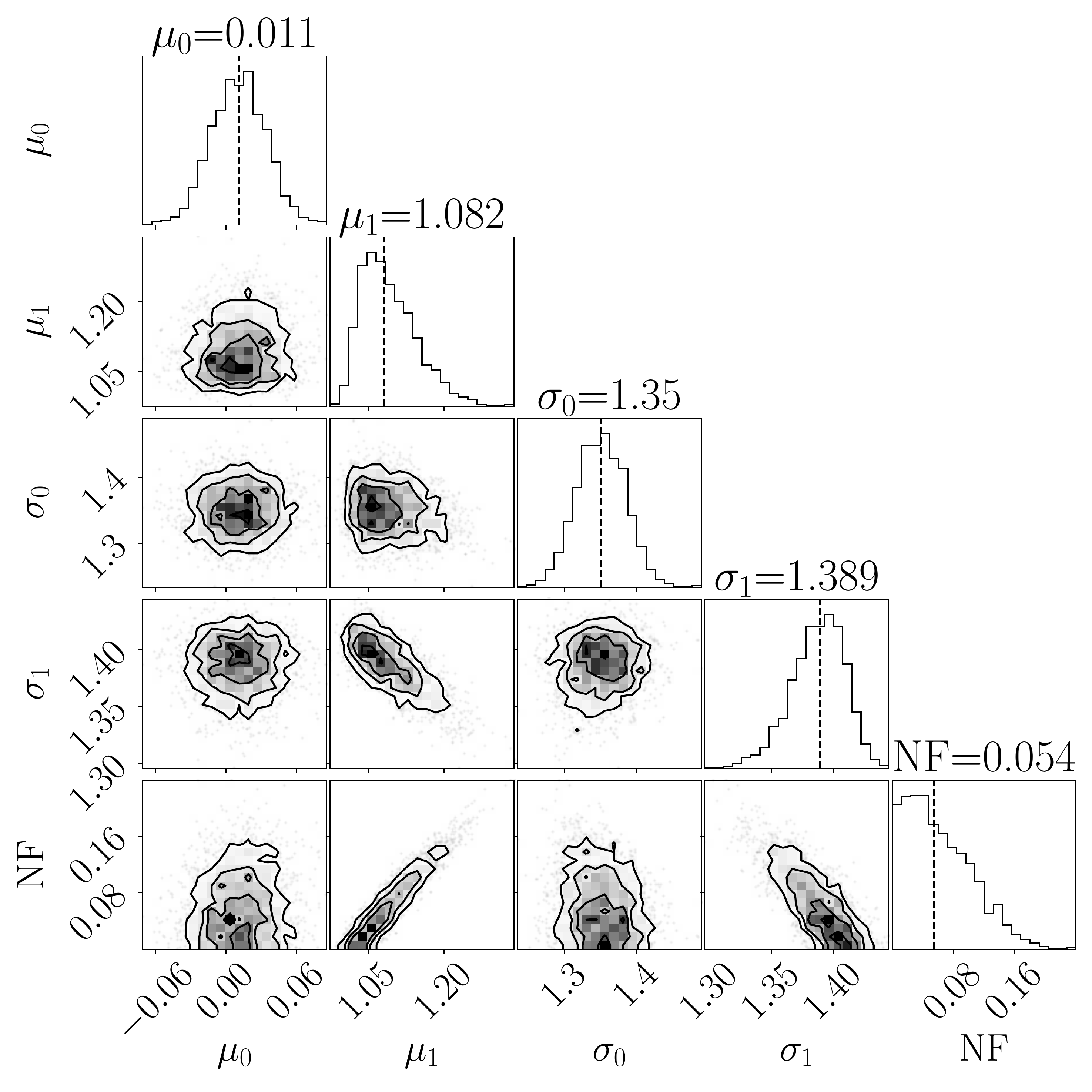}{0.5\textwidth}{(b2) Model parameter posteriors for PSR~J1529$-$26}}
    \caption{\textit{Left }(a1, b1) Two-component Gaussian model fits for the ON and OFF histograms. Individual ON/OFF histograms are shown in solid black lines. The blue, green and orange-filled regions shows the OFF, the null ($\nf \times {\rm OFF}$) and the emission (${\rm ON} - \nf \times {\rm OFF}$) components respectively, where this estimate of \nf\ is obtained using the mixture model. 
    The black dotted line shows the overall fit for the ON pulse distribution. \textit{Right }(a2, b2) Corner plots for 2-component Gaussian fit to the ON/OFF histograms parameterized by the means \{$\mu_1$, $\mu_2$\}, standard deviations \{$\sigma_1$, $\sigma_2$\} and the nulling fraction NF. The dashed vertical lines are the quoted median point estimates of the parameters}
    \label{fig:hist_fit}
\end{figure*}

Figure~(\ref{fig:hist_fit}, left column) shows the pulse intensity histograms for PSR J0325+6744: a pulsar in which the emission component is easily discernible from the noise; and PSR J1529$-$26: a pulsar where these two start to blend into each other. Looking at the null component in the ON histogram for the two pulsars, the evidence for nulling is clear in J0325+6744 while J1529$-$26 behaves like a non-nulling pulsar whose emission is weak. The blue, green and orange-filled regions show the fit for the OFF, null, and emission components respectively, and the black dotted line shows the overall fit for the ON component. The posteriors for the model parameters are presented in Figure~(\ref{fig:hist_fit}, right column) with the point estimates (median\footnote{In the case of non-nulling pulsars where the distribution of \nf\, is one-sided, the median will be over-estimated compared to the true value. Even so, the uncertainty on \nf\, is larger than the difference between the median and mode and hence \nf\, is still consistent with 0.}) of the \nf\ from MM given in Table~\ref{tab:nf_res}.

\begin{deluxetable}{ccccccc}[h]
\tablehead{\colhead{Pulsar} & \colhead{Model} &\colhead{NF}  & \colhead{\nfr} & \colhead{Null period} & \multicolumn{2}{c}{Lengths}\\
\colhead{} & \colhead{} & \colhead{} & \colhead{}  & \colhead{} & \colhead{Null} & \colhead{Em.}\\
\cline{6-7}\\
\colhead{} & \colhead{} & \colhead{(\%)} & \colhead{(\%)} & \multicolumn{3}{c}{(pulse periods)}}
\tablecaption{Nulling properties of the GBNCC pulsars \label{tab:nf_res}}
\startdata
\sidehead{GBT sample}
J0054+6946 & G3 & 27.5\plmi5.1 & 36.8 & \nodata & 2 & 3 \\
J0111+6624 & G2 & 10.2\plmi1.7 & 17.9 & \nodata & 2 & 7 \\
J0325+6744 & G2 & 53.9\plmi0.8 & 55.1 & \nodata & 3 & 4 \\
J0414+31\phm{00} & G2 & 27.5\plmi1.9 & 40.7 & 
28.4\tablenotemark{c} & 2 & 4 \\
J0614+83\phm{00} & G2 & 06.7\plmi3.1 & 52.3 & \nodata & 1-2\tablenotemark{a} & \nodata \\
J0738+6904 & Eg2 & 66.6\plmi1.5 & 64.9 & 
42.7\tablenotemark{c} & 9 & 4 \\
J1529$-$26\phm{00} & G2 & 05.4\plmi4.3 & 48.5 & \nodata & 1-2\tablenotemark{a} & \nodata \\
J1536$-$30\phm{00} & G2 & 43.1\plmi2.2 & 57.5 & \nodata & 4 &  \\
J1629+33\phm{00} & G2 & 83.8\plmi1.9 & 83.9 & \nodata & 12 & 1-2\tablenotemark{a} \\
J1821+4147 & G2 & 00.0\plmi0.6 & 20.9 & \nodata & 1-2\tablenotemark{a} & \nodata \\
J1829+25\phm{00} & G2 & 00.0\plmi0.6 & 07.8 & \nodata & 0\tablenotemark{b} & \nodata \\
J1901$-$04\phm{00} & G2 & 13.9\plmi4.1 & 50.4 &  & 1\tablenotemark{a} & \nodata \\
J2040$-$21\phm{00} & G2 & 25.4\plmi1.8 & 42.4 & 
23.3\tablenotemark{c} & 2 & 5 \\
J2131$-$31\phm{00} & G2 & 49.8\plmi8.6 & 54.2 & \nodata & 3 & 3 \\
J2310+6706 & Eg2 & 54.1\plmi2.7 & 52.7 &  & 3 & 3 \\
\tableline
\sidehead{AO sample}
J0355+28\phm{00} & G2 & 01.6\plmi1.1 & 30.3 & \nodata & 1-2\tablenotemark{a} & \nodata \\
J0414+31\phm{00} & G2 & 33.0\plmi0.7 & 37.1 & 
28.4\tablenotemark{c} & 2 & 4 \\
J1822+02\phm{00} & G2 & 00.1\plmi0.7 & 09.3 & \nodata & 1\tablenotemark{a} & \nodata \\
J1829+25\phm{00} & G2 & 00.0\plmi0.6 & 05.5 & \nodata & 0\tablenotemark{b} & \nodata \\
J1904+33\phm{00} & G2 & 00.0\plmi0.1 & 09.4 & \nodata & 1\tablenotemark{a} & \nodata \\
J1928+28\phm{00} & G2 & 47.6\plmi2.4 & 71.9 & \nodata & 3 & 3 \\
J1941+02\phm{00} & G2 & 00.2\plmi1.7 & 31.1 & \nodata & 1-3\tablenotemark{a} & \nodata \\
J2000+29\phm{00} & G2 & 19.3\plmi1.1 & 23.4 & \nodata & 1-2\tablenotemark{a} & 3 \\
J2044+28\phm{00} & G2 & 15.2\plmi0.9 & 17.4 & \nodata & 1-2\tablenotemark{a} & 6 \\
\enddata
\tablecomments{Naming convention for the model represents the model used to describe the emission histogram (G=Gaussian, Eg=Exponentially modified Gaussian) followed by the number of components in the ON histogram.}
\tablenotetext{a}{We find that in extreme cases (non-nulling/highly-nulling), one of the distributions is confined to very few bins and so we quote this range rather than fitting for it.}
\tablenotetext{b}{We find that there are no single pulses with NP$>$0.5.}
\tablenotetext{c}{ We observe quasi-periodicity in these cases.}
\end{deluxetable}

For PSR J0325+6744, where the null and emission components are well separated (bright pulsars), our method yields a $\nf=53.92 \pm 0.81$\% while Ritchings' method \citep[see][for implementation]{Ritchings76, Wang2007, Kaplan2018} gives a comparable estimate of 55.01\%. However in the case of a weaker pulsar, PSR J1529$-$26, where the emission component is closer to the background noise, our method gives a best-fit value of $\nf=5.55 \pm 4.4$\% compared to 48.1\% given by the Ritchings' method. The latter is significantly overestimated and can easily lead to (mis)classifying the source as a nulling pulsar, further illuminating the bias of Ritchings' method in weaker pulsars.

Full results for all the 23 pulsars, including the single pulse stacks, posteriors from the MCMC run and the resultant ON/OFF histogram model fits are shown in Appendix \ref{app:nulling}.

\subsection{Nulling Correlations}\label{sec:nf_prob}
After determining the nulling properties we wish to know whether the locations and durations of nulls are completely random, or if there is any correlation between different nulling and emission episodes
in a pulsar.  Specifically, given a single pulse that shows emission (or that nulls), how likely are we to see emission for the next pulse, and are there any patterns of longer duration?  

We test this using the probability of a null (the nulling ``responsibility") evaluated for each individual pulse, given by

\begin{equation}
    {\rm NP}_I = \frac{c_1 \mathcal{F}_{1} (I| \{\theta_1\}) }{\sum_{n=1}^{m} c_n \ \mathcal{F}_{n} (I| \{\theta_n\})}.
    \label{eqn:nfprob}
\end{equation}
We divided the data into stacks of 256 pulses \citep[similar to][]{Ritchings76, herfindal09} to calculate more robust estimates and to be less sensitive to long-term variations like scintillation and system temperature changes, and use equation \ref{eqn:nfprob} to calculate the probability of a given single pulse being a null. We then looked for periodic signature by taking the Fourier transform (FT) within each stack and co-adding the power from all  stacks incoherently. Figure \ref{fig:nfperiod} shows the resultant spectrum for PSR J0414+31, in which a certain pattern of combination of emission and nulls seems to be periodic over $\sim$28 pulse periods. We estimate the significance of peaks in the stacked power spectra assuming that the null distribution from $n$ stacks follows a $\chi^2$ distribution with $2n$ degrees of freedom (this assumes white noise). We see significant periodic or quasi-periodic (a significant broad peak in the power spectrum) signatures in a few other pulsars, and tabulate their periods in Table \ref{tab:nf_res}. In the case of precise period measurements, we estimate the uncertainty as described in \citet{scott2002}.

\input{period_plots}

\begin{figure}
    \centering
    \plotone{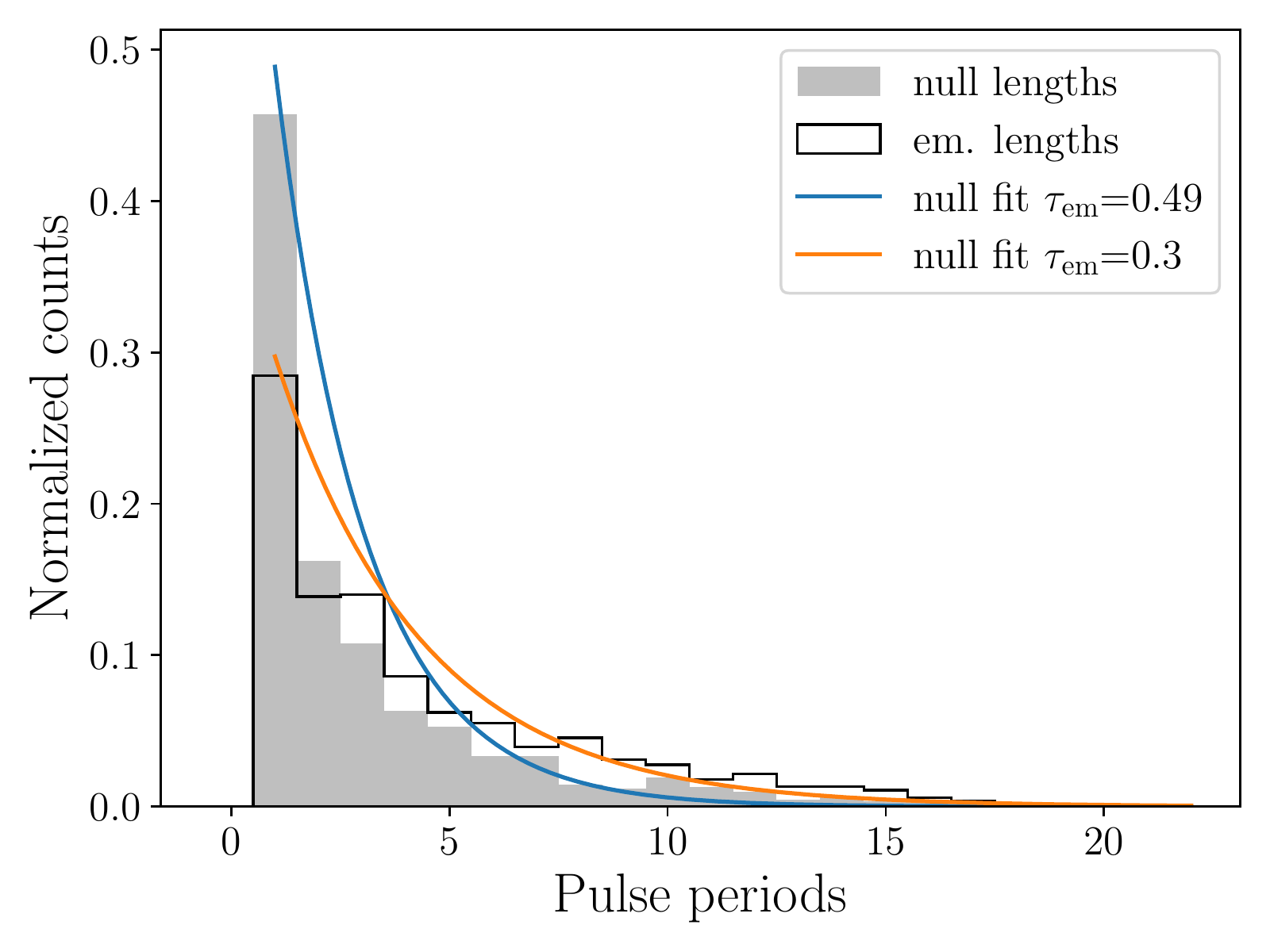}
    \caption{Distribution of emission lengths and null lengths for J0414+31. The gray-filled and the black-open histograms show the distribution of null and emission episodes respectively. The orange curve shows an exponential fit for the emission length distribution with decay constant $\tau_{\rm em}$=0.3, whereas the blue curve shoes the same for the null length distribution with $\tau_{\rm null}$=0.49.}
    \label{fig:len_hist}
\end{figure}

However, this only points to the periodic nature of a certain pattern of emission and nulls. To find how emissions and nulls are `bunched', we look for the distribution of continuous emissions and nulls, where we use $\rm NP_I$=0.5 to be the boundary between an emission and a null. Figure~\ref{fig:len_hist} shows the emission and null length distributions for the single pulses of PSR~J0414+31. We find that these distributions can be well described by an exponential distribution ($p(x) = \tau^{-1} \exp(-x/\lambda)$), where $x$ is the null or emission length and the mean duration of the episode is $\lambda$. We find that for PSR J0414+31, the emission episodes have a characteristic period of four periods, whereas the nulls are two periods long, which is consistent with the observed nulling fraction of $\sim$\,33\% (see Table~\ref{tab:nf_res}). We repeat this for all the pulsars and the results are tabulated in Table~\ref{tab:nf_res}.

\subsection{Sub-pulse Drifting}\label{sec:drift}
\begin{figure*}[!htb]
    \centering
    \includegraphics[scale=0.5]{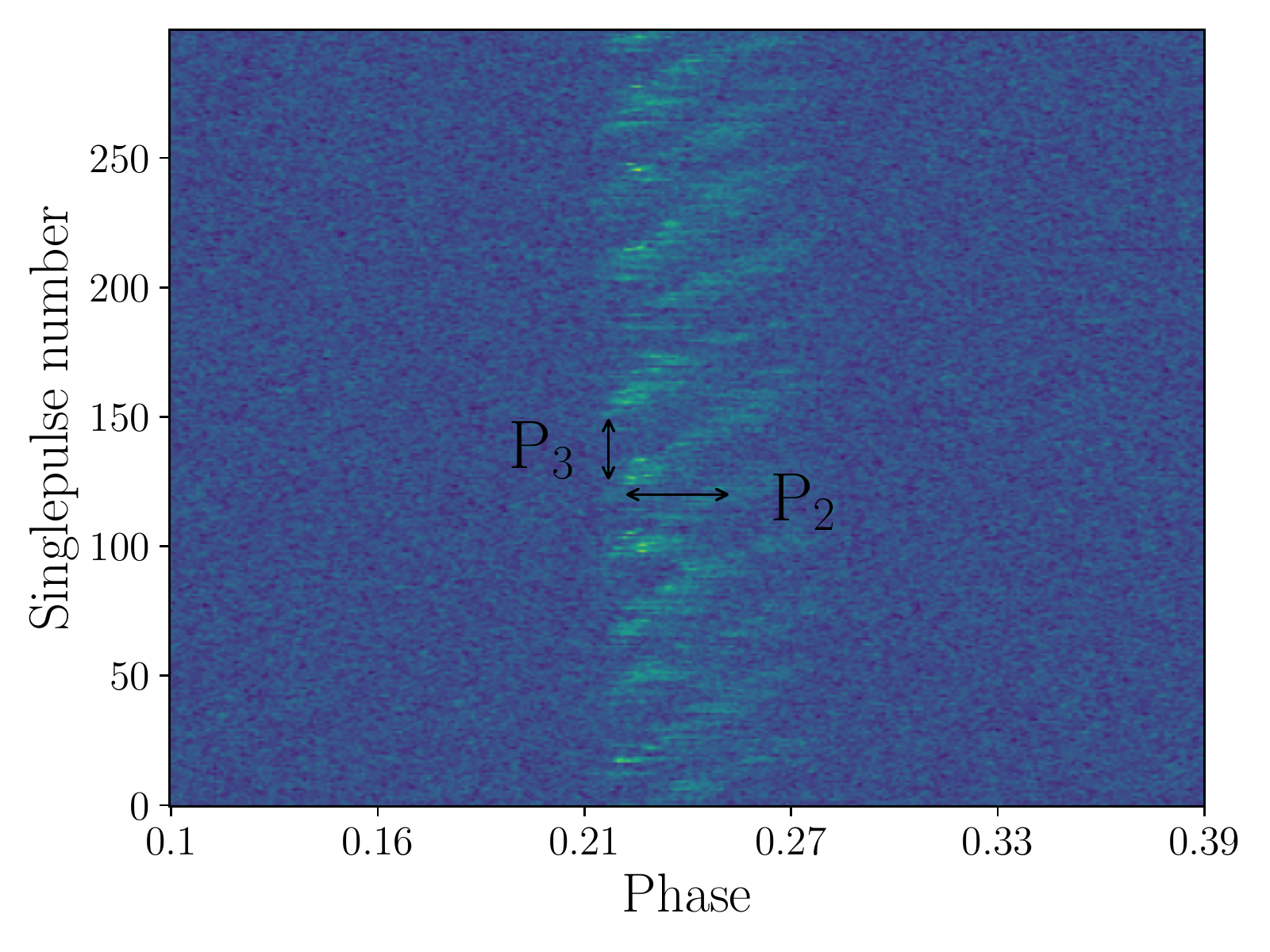}
    \includegraphics[scale=0.5]{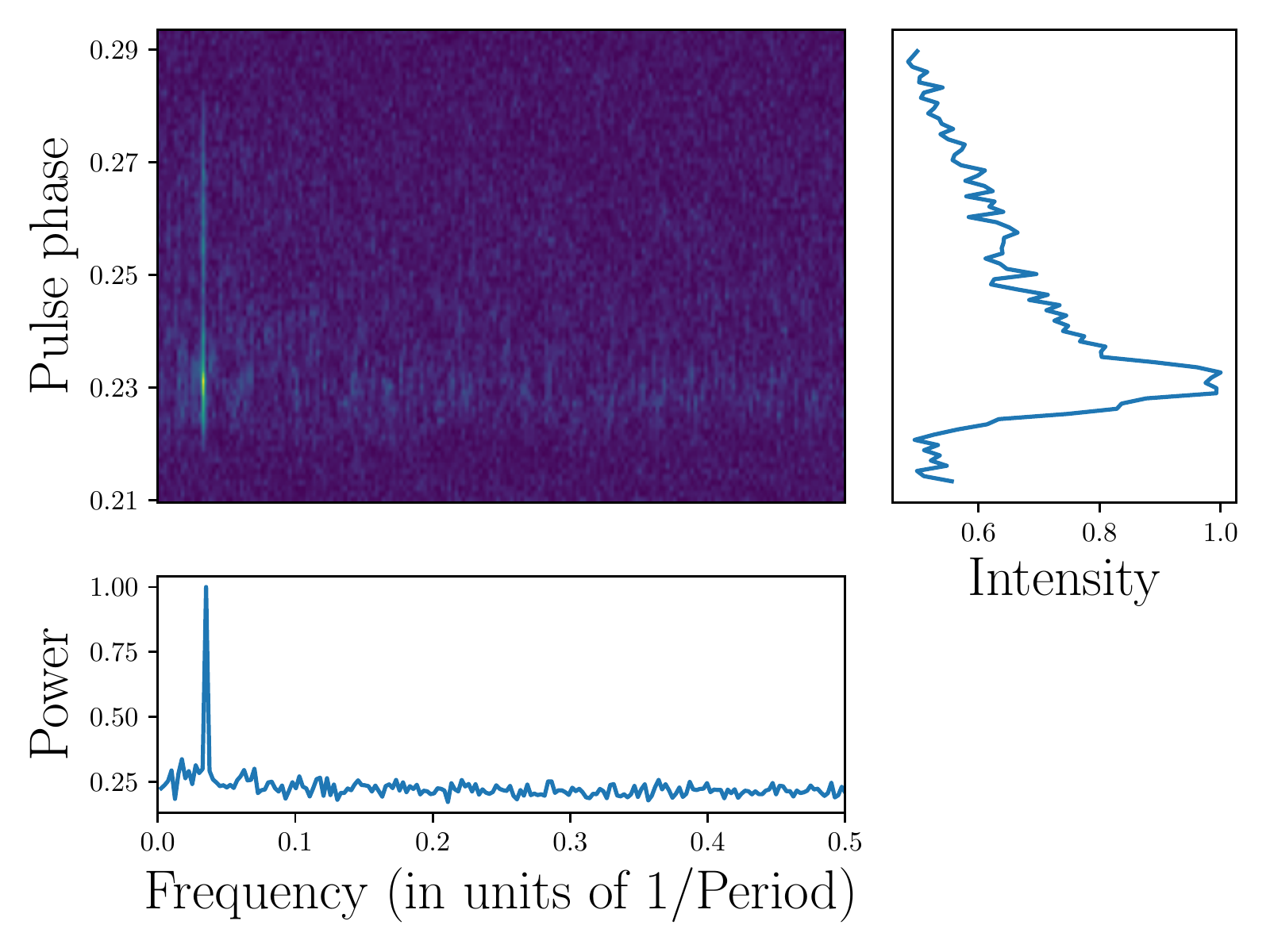}
    \caption{\textit{Left:} A stack of 300 single pulses of PSR J1822+02 clearly showing the sub-pulse drifting phenomenon. The drifting periods $P_2$ and $P_3$ are shown. \textit{Right: }LFRS of the single pulse stack of J1822+02. The 2D spectrogram shows the Fourier transform of data along the axis of single pulses. The evidence of a single drifting frequency across the phase bins is evident from the spectrogram. The bottom panel shows the 2D spectrogram scrunched along the phase axis and the right-hand plot shows the same scrunched along the frequency axis.}
    \label{fig:drift}
\end{figure*}

\begin{figure*}
    \centering
    \gridline{\fig{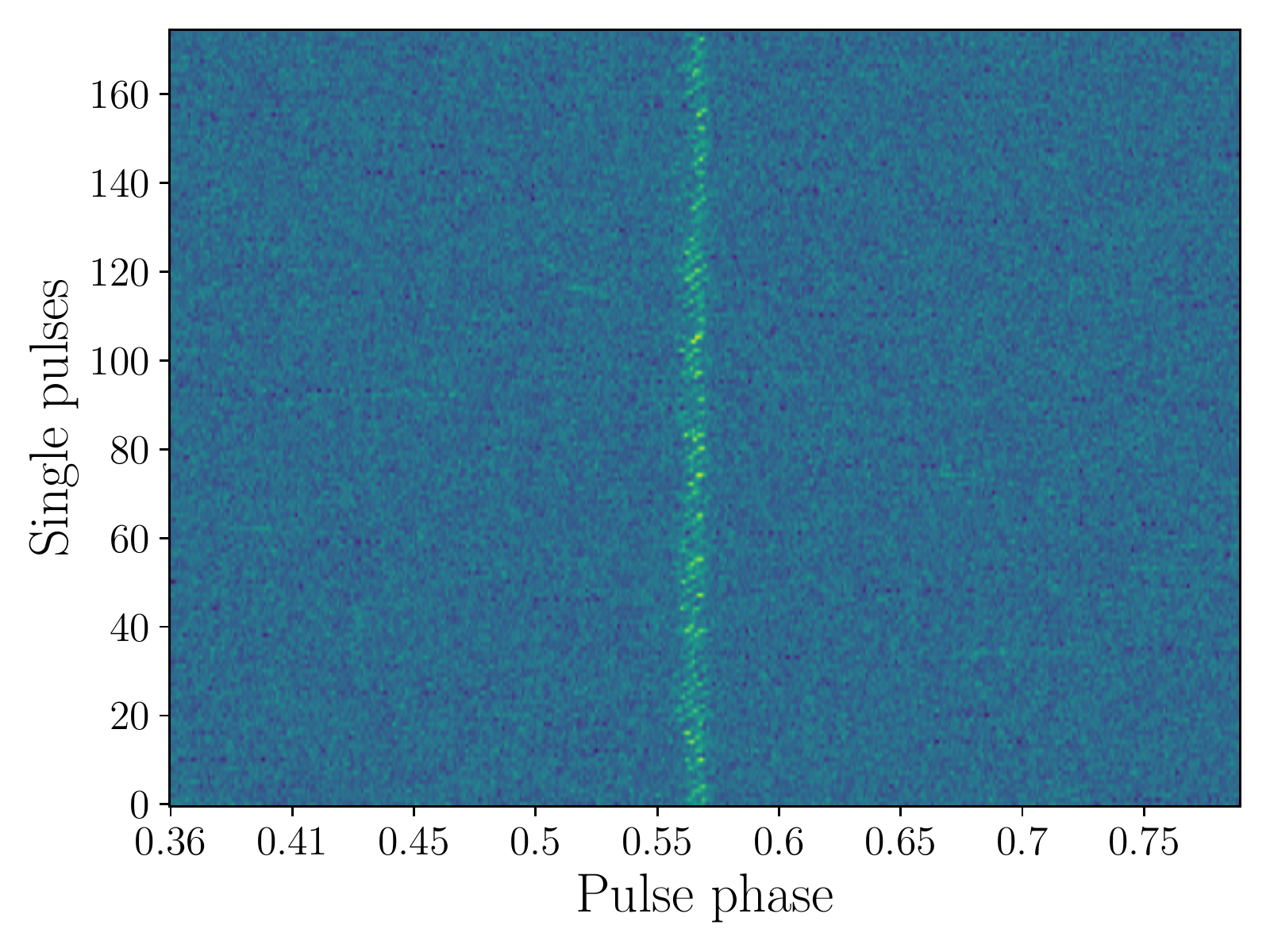}{0.5\textwidth}{(a) AO data single pulse stack}
    \fig{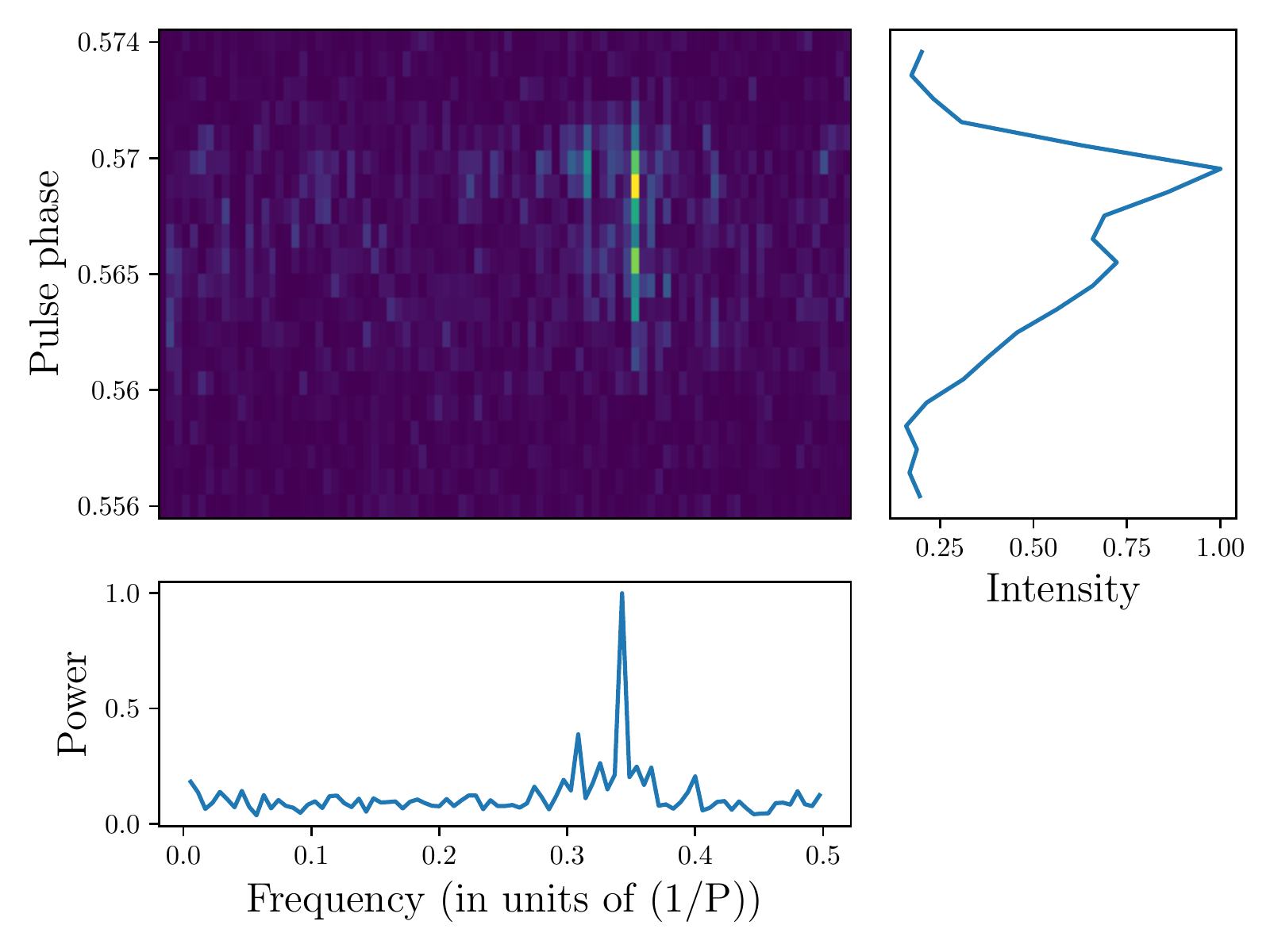}{0.5\textwidth}{(b) LRFS (AO data)}}
    \gridline{\fig{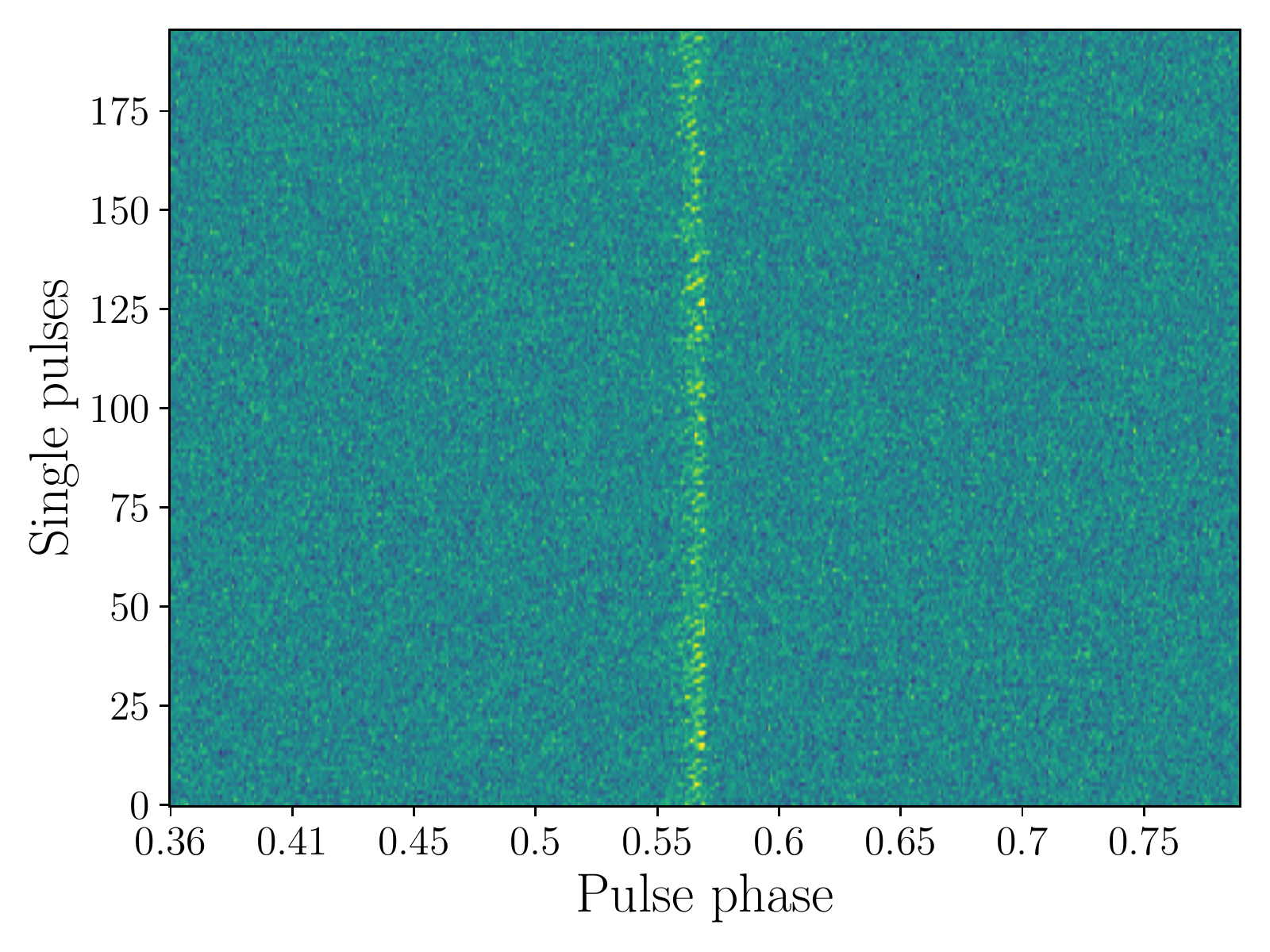}{0.5\textwidth}{(a) GBT data single pulse stack}
    \fig{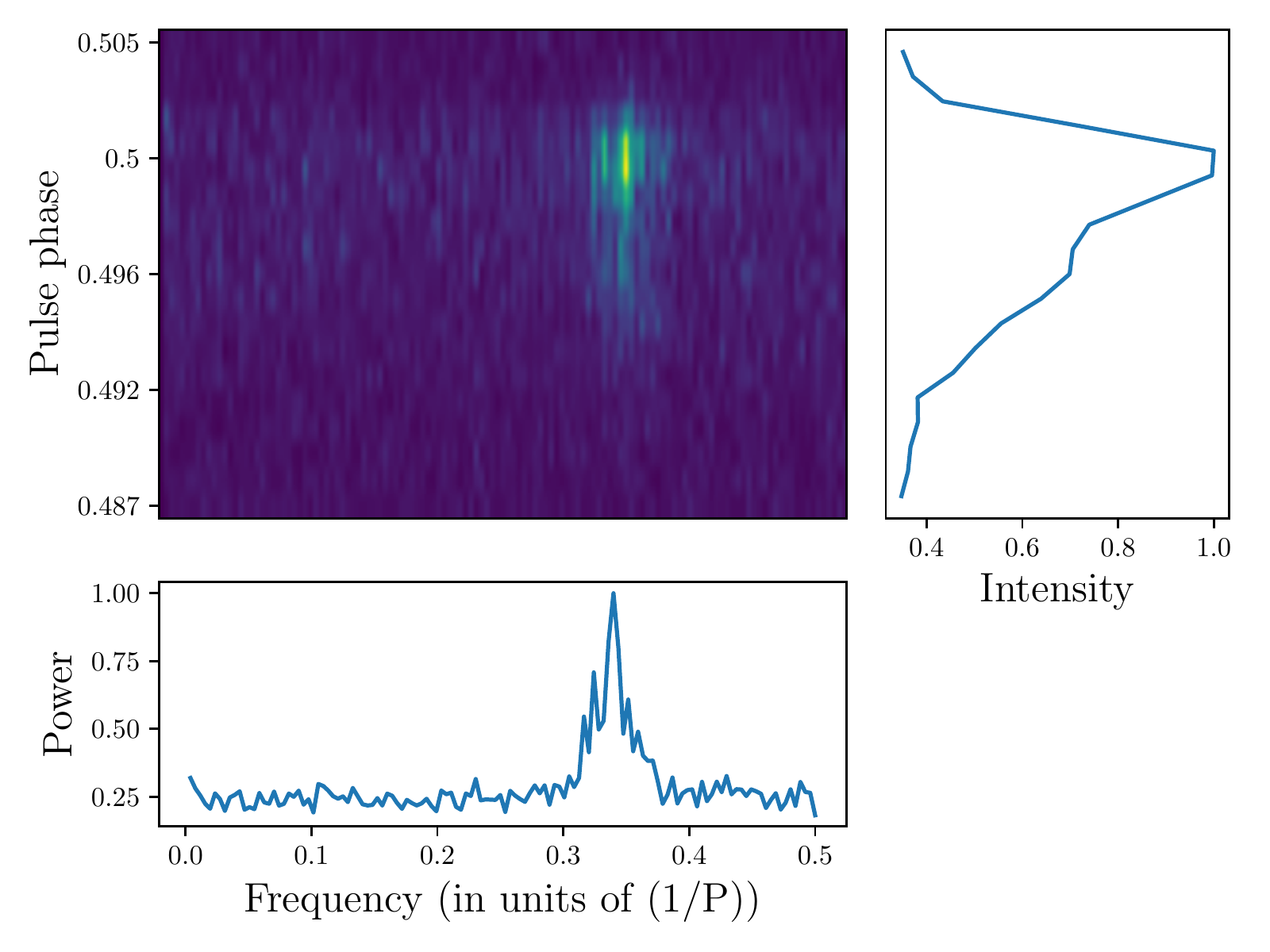}{0.5\textwidth}{(b) LRFS (GBT data)}}
    \caption{Sub-pulse drifting in PSR J1829+25: The left panels shows the stack of single pulses, in the data taken at AO and GBT, which shows the signature of drifting phenomenon. The right panels shows the LRFS (see \S\ref{sec:drift}) of the single pulse stacks. Data from AO (top right) shows a strong feature with a periodicity $\sim 3$ pulse periods. Data from GBT (bottom right) shows a quasi-periodic (broad) peak consistent with the period from AO data.}
    \label{fig:J1829drift}
\end{figure*}

Beyond nulling, we also look for any correlations between nulling and sub-pulse drifting. Drifting is usually characterized by two periods: the drifting period $P_3$, defined as the period for which the pulse is seen at the same longitude (phase), and $P_2$, the spacing between two sub-pluses within the same single pulse (see Figure \ref{fig:drift}). To estimate both, we prepared the data by selecting only the on-pulse window of data ($n_p$ phase bins) for all the single pulses ($n_s$ single pulses). We then calculated Longitude Resolved Fluctuation Spectra (LRFS, \citealt{backer1970drift}), where we take a 1-D Fourier transform of the ($n_s \times n_p$) data along the $n_s$ axis. Figure \ref{fig:drift} shows one of the two pulsars in our sample, J1822+02, that shows clear signs of drifting. A period $P_3$ of $\sim$ 28 pulse periods and $P_2$ of $\sim$ 35/1024 pulse periods can be clearly seen. We also find the evidence for drifting in PSR J1829+25 (see figure~\ref{fig:J1829drift}), with a $P_3$ of $\sim$ three pulse periods and a $P_2$ of 1/128 pulse periods, with similar inferences in the data from both AO and GBT.

\section{Discussion}\label{sec:discussion}

\subsection{Biases in Nulling Models}
\begin{figure}
    \centering
    \includegraphics[scale=0.5]{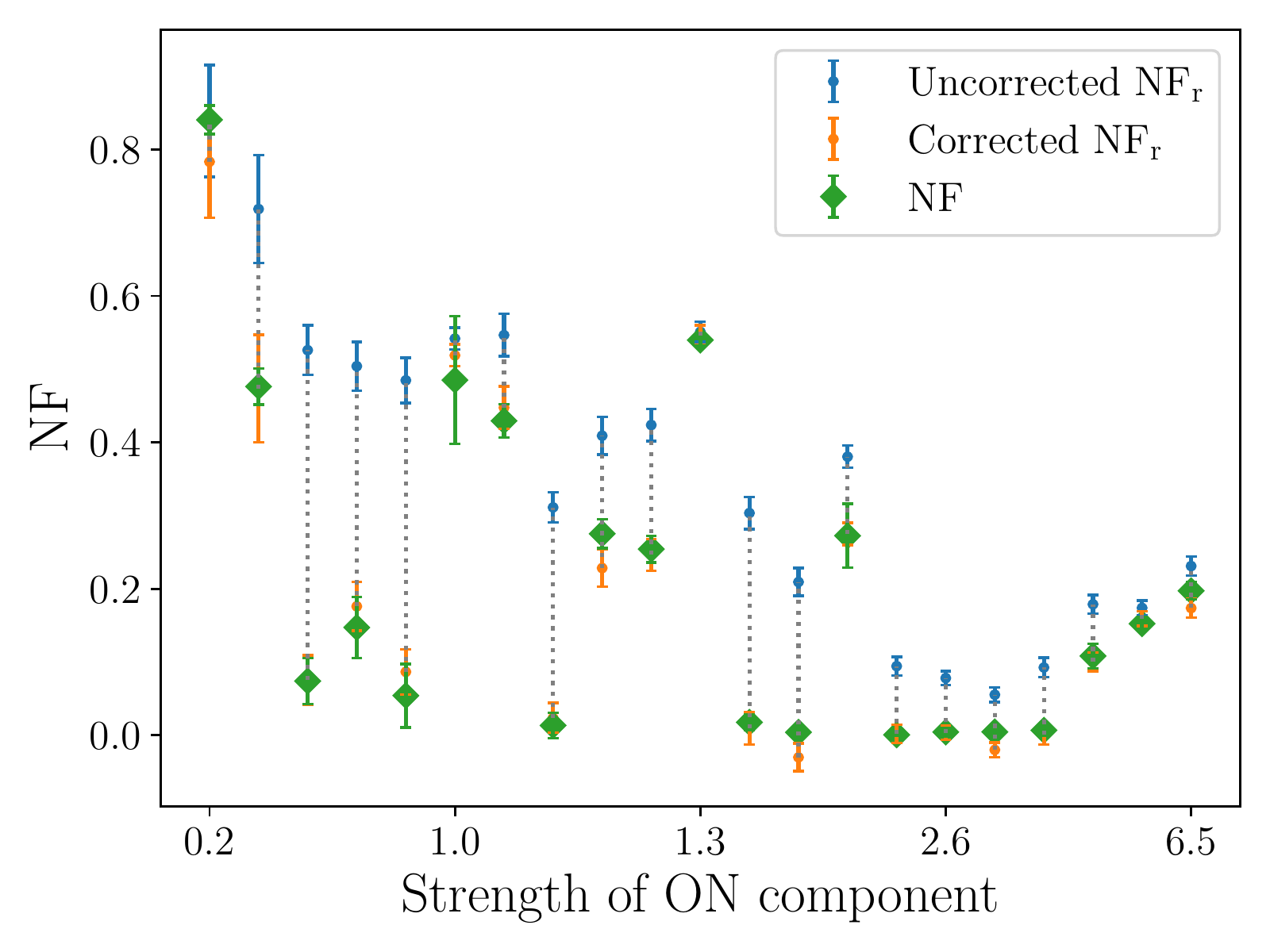}
    \caption{Comparison of \nf\, estimates from Ritchings' algorithm and mixture model as a function of pulsar emission component (significance; in  units of $\sigma_{\rm OFF}$). The blue error bars show the estimates from Ritchings' algorithm while the orange error bars are from mixture model. The green error bars are derived by estimating the systematic bias from the Ritchings' method and clearly depict the bias in the cases where the emission component is weak compared to the background.}
    \label{fig:nf_mm_vs_rit}
\end{figure}

\citet{Kaplan2018} demonstrated the bias of Ritchings' method for weaker pulsars through simulated data, where the mixture model was able to recover the true injected nulling fraction.  
They also showed that for Gaussian mixtures, an analytical correction can  correct the biased estimate of Ritchings' method to find the true value. We extend the same technique using our sample of 22 pulsars. Figure \ref{fig:nf_mm_vs_rit} shows the comparison of the \nf\, estimates derived using both methods. The blue points show the \nf\, estimate derived using Ritchings' algorithm (\nfr), the orange points show \nfr\, estimate corrected for the bias \citep[as in][]{Kaplan2018}, and the green points show the \nf\, derived using mixture modeling. In the case of highly nulling pulsars, the contamination of the null component from the emission component can be small, and both methods perform comparably. However, in the case of pulsars with small \nf\, a systematic bias can be seen as the pulsar emission component becomes blended with the background noise, and the fact that the green and orange points agree quite well demonstrates our confidence in estimating the bias in the Ritchings method and the utility of mixture models.

\subsection{Is the Nulling Fraction Correlated with Pulsar Properties?}
\begin{figure}[!htb]
    \includegraphics[scale=0.45]{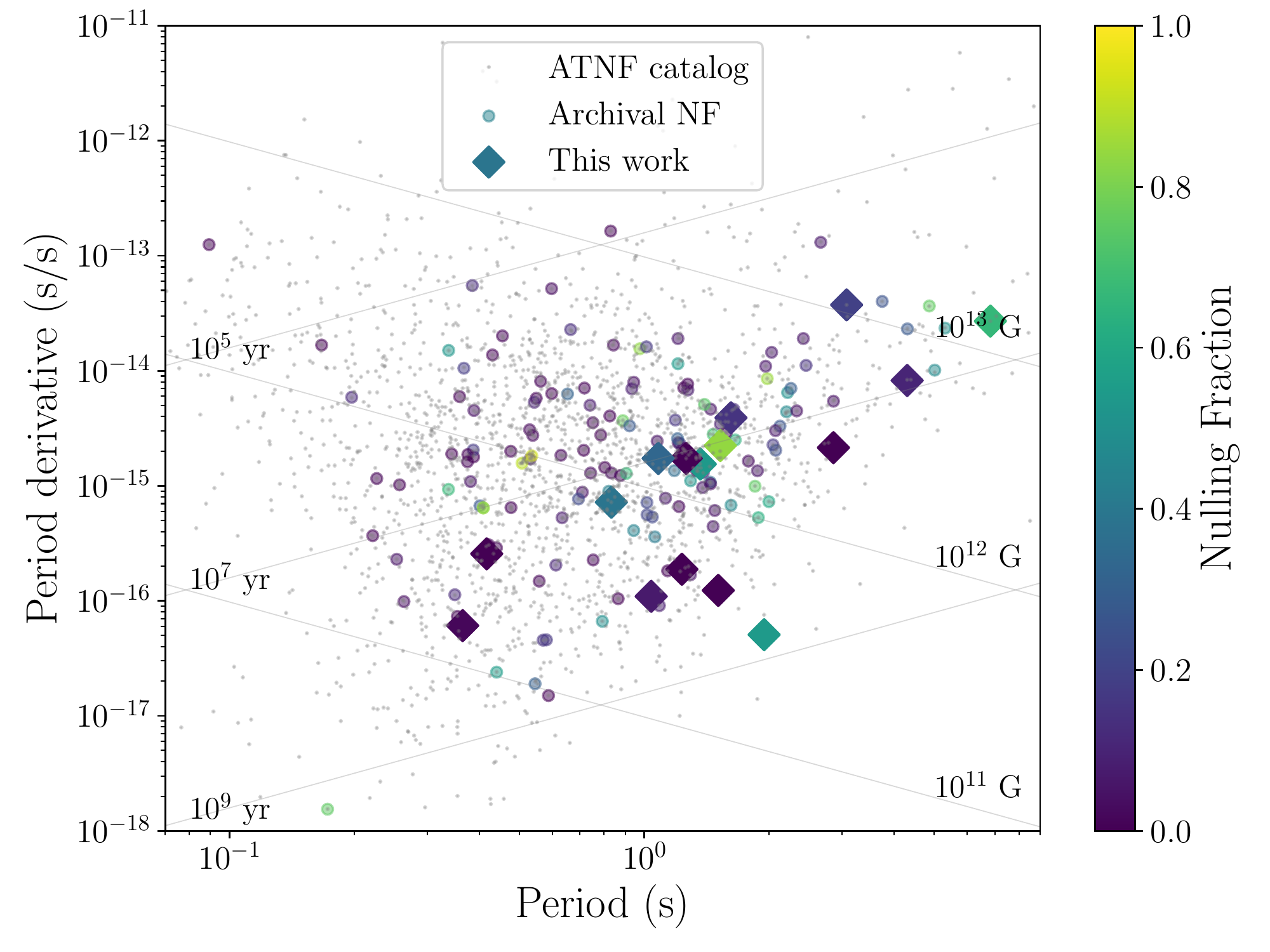}
    \caption{Period-period derivative ($P-\Dot{P}$) diagram highlighting   nulling pulsars. Shown in grey circles are all the pulsar from the ATNF catalog \citep{atnfcat}, in colored circles are the archival nulling pulsars  from \citet{Konar2019} and in diamonds are the pulsars from this study. The contours represent  lines of constant characteristic age $\tau_c$ and dipolar surface magnetic field ($B_{\rm surf}$). The color bar shows the nulling fraction which ranges from 0 to 1. No  clear discernible trend of \nf\ with any of $P$/$\Dot{P}$/$B_{\rm surf}$/$\tau_c$ is  visible.}
    \label{fig:ppdotnf_full}
\end{figure}

Comparing the nulling estimates from the mixture modelling and Ritchings' method in Table \ref{tab:nf_res}, it can be seen that there can be significant differences between these estimates. Such a scenario can lead to significant biases in population-wide studies that look for correlation between nulling fraction and pulsar properties. Figure \ref{fig:ppdotnf_full} shows the most complete list of  nulling pulsars, extended from \citet{Konar2019}, on the $P-\Dot{P}$ diagram.  We do not find any clear visual trends of \nf\, with respect to period ($P$), spin-down rate ($\Dot{P}$), characteristic age ($\tau_c$), or surface magnetic field ($B_{\rm surf}$), although we emphasize that most of the pulsars here (142/164) have their \nf\, estimates derived using some variant of the Ritchings method. 

Our sample size of 22 pulsars is too small to derive reliable correlations. However, we can test the similarity/disparity in the correlations obtained using nulling estimates derived with mixture models versus the Ritchings algorihtm. We use the Spearman correlation test, a non-parametric correlation test to quantify any correlations between the relevant parameters ($P$/$\Dot{P}$/$B_{\rm surf}$/$\tau_c$) and \nf. Table \ref{tab:nf_corr} shows the correlation coefficients of nulling fraction with parameters of interest ($P$, $\Dot{P}$, $B_{\rm surf}$, $\tau_c$). In no case do we see an evidence for strong correlations but we can see large differences between these coefficients obtained using the \nf\, derived using the two methods. We emphasize that the values of these have to be taken with a high degree of caution, given the relative sample size under study and the presence of   outliers. In particular we find that PSR~J2310+6706 turns out to be a strong outlier, especially in the $\tau_c$ and $B_{\rm surf}$ space and this significantly affects the results (see Table \ref{tab:nf_corr}), further illustrating the limitations of a small sample size.

Previously, 
using a sample size (23) comparable to ours, \citet{Wang2007} qualitatively found that  \nf\, is related to age with older population experiencing larger nulling fractions. \citet{Ritchings76} found a positive correlation both with the pulsar period and  age in a sample (32) slightly larger than the one in this study. However, as mentioned above those and most other nulling estimates in the literature are derived using some variant of Ritchings' algorithm.  Computing the Spearman coefficient for all of the archival sources we cannot confirm either correlation and suggest caution in interpreting results using Ritchings' algorithm.

However, we also note that the source of this disparity does not seem to be straightforward: For a sample of pulsars with a given SNR, the energy per single pulse will be lower for pulsars with shorter periods, which means that the \nf\, estimates for the short-period pulsars should experience larger biases and have higher nulling fractions measured with the Richtings' method. Under the (overly simplistic) assumption of a uniform distribution of luminosity with period \citep[cf.][]{faucher-giguere06,bates14}, the correlation of inferred nulling fraction with period will then be negative which is  contrary to the previous studies. This suggests that the source of this bias is not simple and needs careful understanding of the underlying distribution of \nf\, with pulsar properties and a larger sample of pulsars with more robust and unbiased \nf\, estimates.


\begin{figure*}
    \includegraphics[scale=0.475]{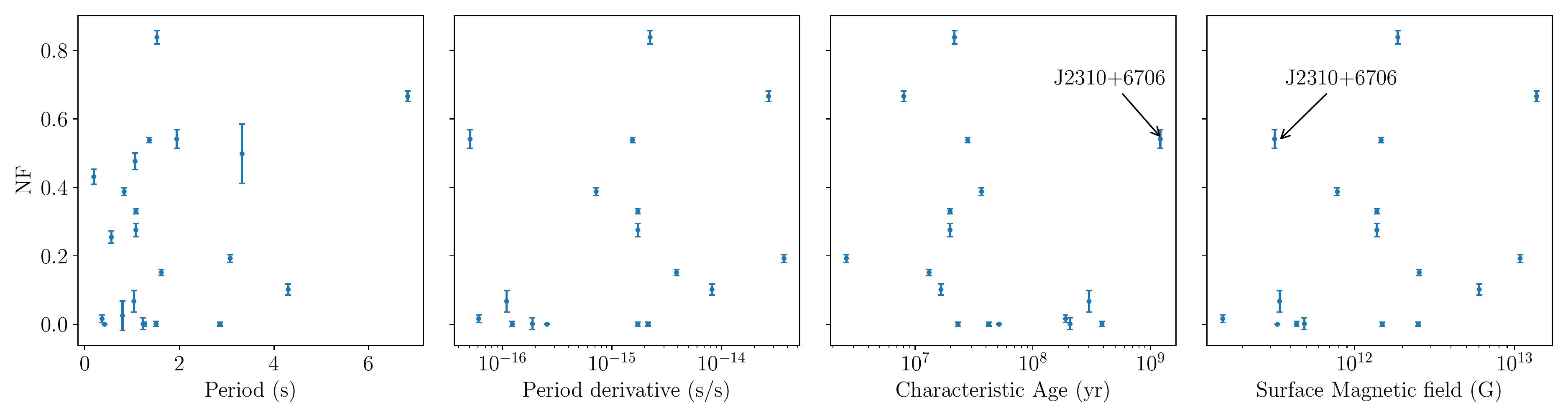}
    \caption{Scatter plot showing the \nf\, of the pulsars in this study vs their properties. It can be seen that the pulsars appear scattered in the $P$/$\Dot{P}$ space. However, with the exclusion of PSR J2310+6706 which appears as an outlier in the $\tau_c$/$B_{\rm surf}$ space, a rough trend can be seen that of \nf\, decreasing with the age $\tau_c$ and increasing with the surface magnetic field $B_{\rm surf}$. The correlation coefficients are given in Table \ref{tab:nf_corr}.}
    \label{fig:nf_corr_pul_params}
\end{figure*}

\begin{deluxetable}{cccc}[t]\label{tab:nf_corr}
\caption{Spearman rank correlation coefficients for our sample data set and archival data set.}
\tablehead{\colhead{Parameter} & \colhead{MM} & \colhead{Ritchings} & \colhead{Catalog}}
\startdata
$P$ & \phm{$-$}0.356 & \phm{$-$}0.008 & \phm{$-$}0.311 \\
 & \phm{$-$}0.314 & $-$0.064 & \nodata \\
$\lvert \Dot{P} \rvert$ & \phm{$-$}0.274 & \phm{$-$}0.035 & $-$0.013 \\
 & \phm{$-$}0.457 & \phm{$-$}0.057 & \nodata \\
$\tau_c$ & $-$0.353 & $-$0.088 & \phm{$-$}0.149 \\
 & $-$0.557 & $-$0.207 & \nodata \\
$B_{\rm surf}$ & \phm{$-$}0.291 & $-$0.006 & \phm{$-$}0.110 \\
 & \phm{$-$}0.450 & \phm{$-$}0.071 & \nodata \\
\enddata
\tablecomments{Not all the pulsars in the sample have $\Dot{P}$ measurements. Hence the sample size used for period is larger.  The two rows for each parameter correspond to the rank coefficients including and excluding PSR J2310+6706 (see Figure \ref{fig:nf_corr_pul_params}).}
\end{deluxetable}


\subsection{Is Nulling Periodic?}
As shown in Section~\ref{sec:nf_prob}, we find that  nulling  appears periodic/quasi-periodic in a subset of pulsars, with their periods noted in Table \ref{tab:nf_res}. \citet{herfindal07, herfindal09} also find evidence for such signatures and attributd this to the line of sight passing through a structured rotating carousel. In addition we also find that in PSR J0414+31, which was observed at two different frequencies with different instruments, this period is the same. It should be noted that the frequency resolution here is $\sim0.004\,{\rm pulse\,period}^{-1}$ (from the stacks of 256 pulses) and so we will be insensitive to any changes that are finer than this. Although significant correlations can not be drawn from these periodicities given our sample size and the number of pulsars that show periodic nulling, the occurrence of such a phenomenon in modest set of pulsars in our sample suggests that this might not be uncommon and should be searched for in future data. 

\section{Conclusions}
\label{sec:conc}
In this study, we have extended the Gaussian mixture model of \citet{Kaplan2018} to study nulling behavior in 22 pulsars, spanning a wider range of properties than in the initial paper but still not selected independent of nulling behavior.  We find that all pulsars can be well-represented by mixture model, but we find that a single Gaussian is not sufficient to describe the emission component in some pulsars\footnote{PSR J0054+6946 is better described by 2 different emission components, one at lower amplitude and the other at higher amplitude, as seen in Figure~\ref{fig:0054}.}. Similar to \citet{Kaplan2018}, we 
find that previous  methods used to estimate \nf\, can suffer significant biases when the pulsar emission is weak compared to the background noise. Such biases may lead to misinterpreting weak pulsars as nulling pulsars.  We also show that these biases may lead to spurious correlations between the \nf\, and pulsar properties in population-wide studies. 

Drawing on the more robust statistics that we calculate, we find that  nulling can appear periodic, with three pulsars in our sample showing this behavior.
Two pulsars in our sample, PSR J1822+02 and PSR J1829+25, shows clear signs of sub-pulse drifting, and they have an inferred nulling fraction consistent with 0.
In contrast, studies like \citet{Vishal2014, Davies1984} find sub-pulse drifting in pulsars that exhibit moderate nulling, indicating that sub-pulse drifting and nulling might be two independent manifestations of sub-pulse intensity variations. In all cases we look forward to using larger, less-biased samples to more robustly explore the nulling population and seeing if it is related to other phenomenology.

Two pulsars in our sample, PSR J0414+31 and PSR J1829+25, were observed at two different frequencies (430\,MHz and 820\,MHz), albeit not simultaneously. PSR J1829+25 has nulling estimates that agree at both  frequencies, consistent with 0, but we find that PSR J0414+31, has \nf\, estimates in tension at the $\sim 2\sigma$ level, with the \nf\, higher at lower frequencies. Although it is hard to draw definite conclusions from these two pulsars since the observations are not simultaneous, it emphasizes the need for simultaneous observations at multiple frequencies (or across a larger bandwidth). Observing at 4 different frequencies (325, 610, 1400, 4850 MHz), \cite{Vishal2014} find coherent nulling in three different pulsars whereas \citet{Bhat2007} find the evidence for null excess at lower frequencies in PSR~B1133+16 further emphasizing the need for multi-frequency observations in a larger sample to find whether nulling is universally broadband.  

One of the pulsars in our sample (PSR J2310+6706) has a two-component profile with a faint leading peak in addition to the primary peak. The very low SNR of the leading component limits our ability to find a stringent estimate of the \nf\, independent of the primary component, but we find that the \nf\, values obtained from each component is consistent. Analyzing nulling characteristics in pulsars with multi-component pulse profiles with a robust method like mixture modeling can provide insights into the simultaneous nulling in the different regions of the pulsar's magnetosphere.

So far we have only analyzed normal, non-recycled pulsars.  Current sensitivity limitations restrict the sample of nulling pulsars  to normal pulsars (as is evident from Figure \ref{fig:ppdotnf_full}), while MSPs are largely unexplored. Initial single pulse studies done by \cite{Kaustubh2014} do not find any compelling evidence for nulling in MSPs. Using the mixture model technique, which does not suffer from the same biases at low signal-to-noise, for MSPs, together with newer higher-sensitivity facilities may help explore whether the nulling phenomenon affects all pulsars, or is limited to a sub-population.

\begin{acknowledgments}
We thank an anonymous referee for helpful suggestions that clarified this work.
AA, JS, and DK receive support from National Science Foundation (NSF) Physics Frontiers Center award numbers 1430284 and 2020265. AA thanks Alex McEwen for helpful discussions  during the data reduction stage. The Arecibo Observatory is a facility of the NSF operated under cooperative agreement (\#AST-1744119) by the University of Central Florida (UCF) in alliance with Universidad Ana G. Méndez (UAGM) and Yang Enterprises (YEI), Inc. The Green Bank Observatory is a facility of the NSF operated under cooperative agreement by Associated Universities, Inc.
\end{acknowledgments}

\facilities{GBT (GUPPI), Arecibo (PUPPI)}
\software{\texttt{PINT} \citep{pint}, \texttt{PSRCHIVE} \citep{psrchive}, \texttt{dspsr} \citep{dspsr}, \texttt{NumPy} \citep{numpy}, \texttt{Matplotlib} \citep{matplotlib}, \texttt{AstroPy} \citep{astropy1, astropy2}, \texttt{emcee} \citep{emcee}}

\appendix

\section{Nulling Results for All Pulsars}
\label{app:nulling}
We show pulse profiles, MCMC corner plot results, and nulling histograms for all of the pulsars in our sample.

\bibliography{nulling}{}
\bibliographystyle{aasjournal}

\input{all_figures}
\end{document}

%% file: GBT_obs_schedule.tex
\begin{deluxetable}{lcc}[!h]
\tablecaption{Times and durations of GBT observations \label{tab:gbt_data}}
\tablehead{\colhead{Pulsar} & \colhead{Observations} & \colhead{Total Time}\\
\colhead{} & \colhead{MJD (hr)} & \colhead{(hr)}}
\startdata
J0054+6946 & 58163 (2.00) & 2.00 \\
J0111+6624 & 58163 (2.24) & 2.24 \\
J0325+6744 & 58163 (1.52) & 2.00 \\
 & 58164 (0.48) & \nodata \\
J0414+31\tablenotemark{a} & 58164 (1.50) & 1.50 \\
J0614+83 & 58164 (1.90) & 1.90 \\
J0738+6904 & 58209 (2.00) & 2.00 \\
J1529$-$26 & 58209 (1.50) & 1.50 \\
J1536$-$30 & 58209 (1.50) & 1.50 \\
J1629+33 & 58209 (1.50) & 1.50 \\
J1821+4147 & 58209 (1.69) & 1.69 \\
J1829+25\tablenotemark{a} & 58246 (1.50) & 1.50 \\
J1901$-$04 & 58246 (1.50) & 1.50 \\
J2040$-$21 & 58246 (1.50) & 1.50 \\
J2131$-$31 & 58246 (0.33) & 0.33 \\
J2310+6706 & 58246 (1.75) & 1.75 \\
\enddata
\tablecomments{For each pulsar we give the individual Modified Julian Date (MJD) and duration of each session, as well as the total observing time.}
\tablenotetext{a}{This pulsar was observed at both AO and GBT}
\end{deluxetable}

%% file: AO_obs_schedule.tex
\begin{deluxetable}{lcc}[!ht]
\tablecaption{Times and durations of Arecibo observations \label{tab:ao_data}}
\tablehead{\colhead{Pulsar} & \colhead{Observations} & \colhead{Total Time}\\
\colhead{} & \colhead{MJD (hr)} & \colhead{(hr)}}
\startdata
J0355+28 & 58890 (0.25), 58922 (0.33) & 2.95 \\
 & 58924 (0.42), 58928 (0.39) & \nodata \\
 & 58936 (0.39), 58951 (0.39) & \nodata \\
 & 58982 (0.39), 59013 (0.39) & \nodata \\
J0414+31\tablenotemark{a} & 58890 (0.50), 58922 (0.38) & 3.46 \\
 & 58924 (0.50), 58928 (0.35) & \nodata \\
 & 58936 (0.30), 58951 (0.40) & \nodata \\
 & 58982 (0.63), 59013 (0.40) & \nodata \\
J1822+02 & 58941 (0.22), 58968 (0.17) & 1.55 \\
 & 58970 (0.17), 58974 (0.17) & \nodata \\
 & 58981 (0.17), 59000 (0.33) & \nodata \\
 & 59029 (0.17), 59063 (0.17) & \nodata \\
J1829+25\tablenotemark{a} & 58852 (0.17), 58941 (0.17) & 1.03 \\
 & 58968 (0.14), 58970 (0.11) & \nodata \\
 & 58974 (0.11), 58981 (0.11) & \nodata \\
 & 59029 (0.11), 59063 (0.11) & \nodata \\
J1904+33 & 58852 (0.17), 58882 (0.17) & 1.34 \\
 & 58941 (0.17), 58968 (0.14) & \nodata \\
 & 58970 (0.14), 58974 (0.14) & \nodata \\
 & 58981 (0.14), 59029 (0.14) & \nodata \\
 & 59063 (0.14) & \nodata \\
J1928+28 & 58852 (0.17), 58882 (0.17) & 1.98 \\
 & 58941 (0.17), 58968 (0.14) & \nodata \\
 & 58970 (0.17), 58974 (0.17) & \nodata \\
 & 58981 (0.17), 59000 (0.50) & \nodata \\
 & 59029 (0.17), 59063 (0.17) & \nodata \\
J1941+02 & 58852 (0.17), 58882 (0.17) & 1.5 \\
 & 58912 (0.14), 58941 (0.17) & \nodata \\
 & 58968 (0.14), 58970 (0.14) & \nodata \\
 & 58974 (0.14), 58981 (0.10) & \nodata \\
 & 59029 (0.17), 59063 (0.17) & \nodata \\
J2000+29 & 58852 (0.39), 58882 (0.17) & 1.83 \\
 & 58941 (0.10), 58968 (0.14) & \nodata \\
 & 58970 (0.14), 58974 (0.14) & \nodata \\
 & 58981 (0.14), 59000 (0.33) & \nodata \\
 & 59029 (0.14), 59063 (0.14) & \nodata \\
J2044+28 & 58852 (0.17), 58882 (0.17) & 1.18 \\
 & 58968 (0.07), 58970 (0.14) & \nodata \\
 & 58974 (0.14), 58981 (0.14) & \nodata \\
 & 59000 (0.07), 59029 (0.14) & \nodata \\
 & 59063 (0.14) & \nodata \\
\enddata
\tablenotetext{a}{This pulsar was observed at both AO and GBT}
\end{deluxetable}

%% file: timing_params.tex
\begin{deluxetable*}{lcccclll}
\tablehead{
\colhead{Pulsar} & \multicolumn{4}{c}{Position (J2000)} & \colhead{Period} & \colhead{Period derivative} & \colhead{DM} \\
\tableline\\
\colhead{} & \colhead{RA} & \colhead{RA error} & \colhead{DEC} & \colhead{DEC error} & \colhead{} & \colhead{} & \colhead{} \\
\cline{2-5}\\
\colhead{} & \colhead{} & \colhead{(\arcsec)} & \colhead{} & \colhead{(\arcsec)} & \colhead{(s)} & \colhead{($10^{-15}$\,s/s)} & \colhead{(pc/$\rm cm^3$)}
}
\tablecaption{Timing Parameters for the GBNCC pulsars used to study nulling \label{tab:tim_param}}
\startdata
\sidehead{GBT sample}
J0054+6946\tablenotemark{a} & 00$^{\rm h}$\;54$^{\rm m}$\;59$\fs$1 & 00.1 & +69\arcdeg\;46\arcmin\;16$\farcs$8 & 00.0(3) & 0.832911328744(4) & $-$0.7194(8) & 116.52(5) \\
J0111+6624\tablenotemark{a} & 01$^{\rm h}$\;11$^{\rm m}$\;21$\fs$9 & 01.7 & +66\arcdeg\;24\arcmin\;10$\farcs$9 & 00.6\phm{(0)} & 4.3018721007(3) & $-$8.4(2) & 111.20(3) \\
J0325+6744\tablenotemark{a} & 03$^{\rm h}$\;25$^{\rm m}$\;05$\fs$1 & 00.3 & +67\arcdeg\;44\arcmin\;59$\farcs$4 & 00.1\phm{(0)} & 1.36467876728(1) & $-$1.553(9) & 65.28(5) \\
J0414+31\tablenotemark{b} & 04$^{\rm h}$\;14$^{\rm m}$\;35$\fs$6 & 02.6 & +31\arcdeg\;38\arcmin\;35$\farcs$4 & 25.3\phm{(0)} & 1.0805116(1) & $-$3.6(5) & 64.64(3) \\
J0614+83\tablenotemark{c} & 06$^{\rm h}$\;14$^{\rm m}$\;03$\fs$4 & 34.6 & +83\arcdeg\;13\arcmin\;46$\farcs$2 & 34.6\phm{(0)} & 1.03918794(5) & \nodata & 44.2(1) \\
J0738+6904\tablenotemark{a} & 07$^{\rm h}$\;38$^{\rm m}$\;22$\fs$6 & 00.5 & +69\arcdeg\;04\arcmin\;20$\farcs$0 & 00.3\phm{(0)} & 6.8276928023(5) & $-$26.97(4) & 17.22(2) \\
J1529$-$26\tablenotemark{c} & 15$^{\rm h}$\;29$^{\rm m}$\;07$\fs$2 & 38.9 & $-$26\arcdeg\;26\arcmin\;35$\farcs$5 & 38.9\phm{(0)} & 0.79857094(5) & \nodata & 44.7(1) \\
J1536$-$30\tablenotemark{c} & 15$^{\rm h}$\;36$^{\rm m}$\;33$\fs$4 & 17.3 & $-$30\arcdeg\;06\arcmin\;14$\farcs$4 & 17.3\phm{(0)} & 0.190084143(9) & \nodata & 63.40(7) \\
J1629+33\tablenotemark{c} & 16$^{\rm h}$\;29$^{\rm m}$\;22$\fs$6 & 99.2 & +33\arcdeg\;23\arcmin\;35$\farcs$9 & 99.2\phm{(0)} & 1.5247311(3) & \nodata & 34.8(5) \\
J1821+4147\tablenotemark{a} & 18$^{\rm h}$\;21$^{\rm m}$\;52$\fs$3 & 00.1 & +41\arcdeg\;47\arcmin\;02$\farcs$6 & 00.0(4) & 1.26185719(3) & $-$1.7292(9) & 40.63(5) \\
J1829+25\tablenotemark{b} & 18$^{\rm h}$\;30$^{\rm m}$\;31$\fs$8 & 01.8 & +25\arcdeg\;08\arcmin\;00$\farcs$4 & 01.4\phm{(0)} & 2.85769207(9) & $-$1.9(4) & 73.64(9) \\
J1901$-$04\tablenotemark{c} & 19$^{\rm h}$\;01$^{\rm m}$\;37$\fs$1 & 62.0 & $-$04\arcdeg\;54\arcmin\;44$\farcs$9 & 62.0\phm{(0)} & 1.8255459(8) & \nodata & 105.4(9) \\
J2040$-$21\tablenotemark{c} & 20$^{\rm h}$\;40$^{\rm m}$\;40$\fs$6 & 09.7 & +21\arcdeg\;52\arcmin\;51$\farcs$6 & 09.7\phm{(0)} & 0.562564125(4) & \nodata & 23.77(1) \\
J2131$-$31\tablenotemark{c} & 21$^{\rm h}$\;31$^{\rm m}$\;30$\fs$9 & 65.9 & $-$31\arcdeg\;32\arcmin\;53$\farcs$4 & 65.9\phm{(0)} & 3.32537(3) & \nodata & 31.753 \\
J2310+6706\tablenotemark{a} & 23$^{\rm h}$\;10$^{\rm m}$\;42$\fs$1 & 02.9 & +67\arcdeg\;06\arcmin\;52$\farcs$1 & 00.9\phm{(0)} & 1.944788973(1) & $-$0.06(5) & 97.7(2) \\
\tableline
\sidehead{AO sample}
J0355+28 & 03$^{\rm h}$\;55$^{\rm m}$\;22$\fs$8 & 00.4 & +28\arcdeg\;38\arcmin\;50$\farcs$1 & 00.8\phm{(0)} & 0.36492919909(3) & \nodata & 48.788 \\
J0414+31\tablenotemark{b} & 04$^{\rm h}$\;14$^{\rm m}$\;35$\fs$6 & 02.6 & +31\arcdeg\;38\arcmin\;35$\farcs$4 & 25.3\phm{(0)} & 1.0805116(1) & $-$3.6(5) & 64.64(3) \\
J1822+02 & 18$^{\rm h}$\;22$^{\rm m}$\;43$\fs$6 & 01.4 & +02\arcdeg\;28\arcmin\;53$\farcs$8 & 01.2\phm{(0)} & 1.5081132778(9) & \nodata & 103.22 \\
J1829+25\tablenotemark{b} & 18$^{\rm h}$\;30$^{\rm m}$\;31$\fs$8 & 01.8 & +25\arcdeg\;08\arcmin\;00$\farcs$4 & 01.4\phm{(0)} & 2.85769207(9) & $-$1.9(4) & 73.64(9) \\
J1904+33 & 19$^{\rm h}$\;04$^{\rm m}$\;40$\fs$2 & 00.2 & +33\arcdeg\;58\arcmin\;25$\farcs$9 & 00.1\phm{(0)} & 0.417032327(1) & $-$0.247(5) & 81.139 \\
J1928+28 & 19$^{\rm h}$\;27$^{\rm m}$\;58$\fs$4 & 01.1 & +28\arcdeg\;59\arcmin\;12$\farcs$4 & 01.0\phm{(0)} & 1.0630373062(5) & \nodata & 79.34 \\
J1941+02 & 19$^{\rm h}$\;40$^{\rm m}$\;34$\fs$1 & 00.8 & +02\arcdeg\;39\arcmin\;21$\farcs$7 & 01.0\phm{(0)} & 1.23229077(1) & $-$0.18(9) & 87.478 \\
J2000+29 & 20$^{\rm h}$\;00$^{\rm m}$\;16$\fs$5 & 00.4 & +29\arcdeg\;20\arcmin\;07$\farcs$6 & 00.1\phm{(0)} & 3.07377646(2) & $-$37.37(8) & 132.62 \\
J2044+28 & 20$^{\rm h}$\;43$^{\rm m}$\;36$\fs$9 & 00.4 & +28\arcdeg\;28\arcmin\;37$\farcs$3 & 00.2\phm{(0)} & 1.61816650(1) & $-$3.99(4) & 90.169 \\
\enddata
\tablecomments{Quantities in parentheses are 1$\sigma$ uncertainties on the last digit.}
\tablenotetext{a}{Coherent timing solutions are given in \citet{Lynch2018}}
\tablenotetext{b}{Timing solution is obtained by combining AO and GBT data.}
\tablenotetext{c}{Astrometric positions are estimated from gridding and the positional uncertainties are estimated from the beam size (15\arcmin) and the Signal to Noise Ratio (SNR)}
\end{deluxetable*}

%% file: period_plots.tex
\begin{figure*}[!htb]
    \gridline{\fig{period_plots/J0414+31_GBT/dspsr_fold/nf_periodic.pdf}{0.45\textwidth}{(a) PSR J0414+31 (GBT) \label{fig:nf_periodic}}
              \fig{period_plots/J0414+31_arecibo/dspsr_fold/nf_periodic.pdf}{0.45\textwidth}{(b) PSR J0414+31 (AO)}
                }
    \gridline{
              \fig{period_plots/J2040-21/dspsr_fold/nf_periodic.pdf}{0.45\textwidth}{(e) PSR J2040$-$21}
              \fig{period_plots/J0738+6904/dspsr_fold/nf_periodic.pdf}{0.45\textwidth}{(f) PSR J0738+6904}
                }
    \caption{Fourier transform of the null probability for the pulsars in our sample that  show periodicity. Power combined incoherently from multiple stacks of 256 pulses is shown at 129 discrete frequencies (in the units of 1/pulse period) in the blue line. The orange curve shows the same for the OFF component (background noise) which can be used to eliminate any instrumental variations/artifacts and/or RFI. The black dotted line shows the upper limit that allows for 1 false positive in 1000 trails, corresponding to a 99.9\% confidence limit. The gray curves are the normalized power from the individual stacks (not to scale) that are used to look for quasi-periodicity. The value of the periodicities are given in Table \ref{tab:nf_res}}
    \label{fig:nfperiod}
\end{figure*}

%% file: all_figures.tex
\begin{figure*}
      \includegraphics[width=0.5\textwidth]{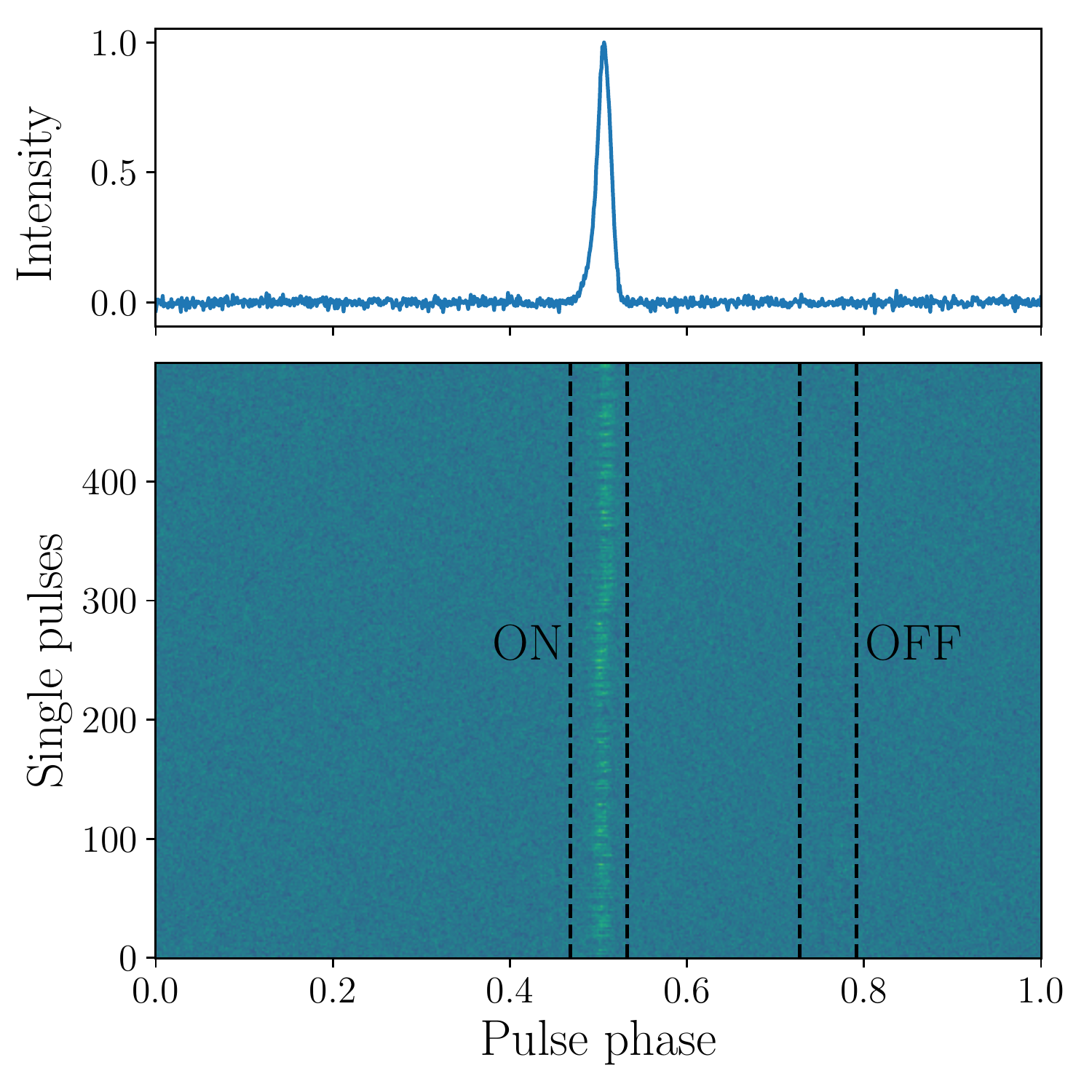}
      \includegraphics[width=0.5\textwidth]{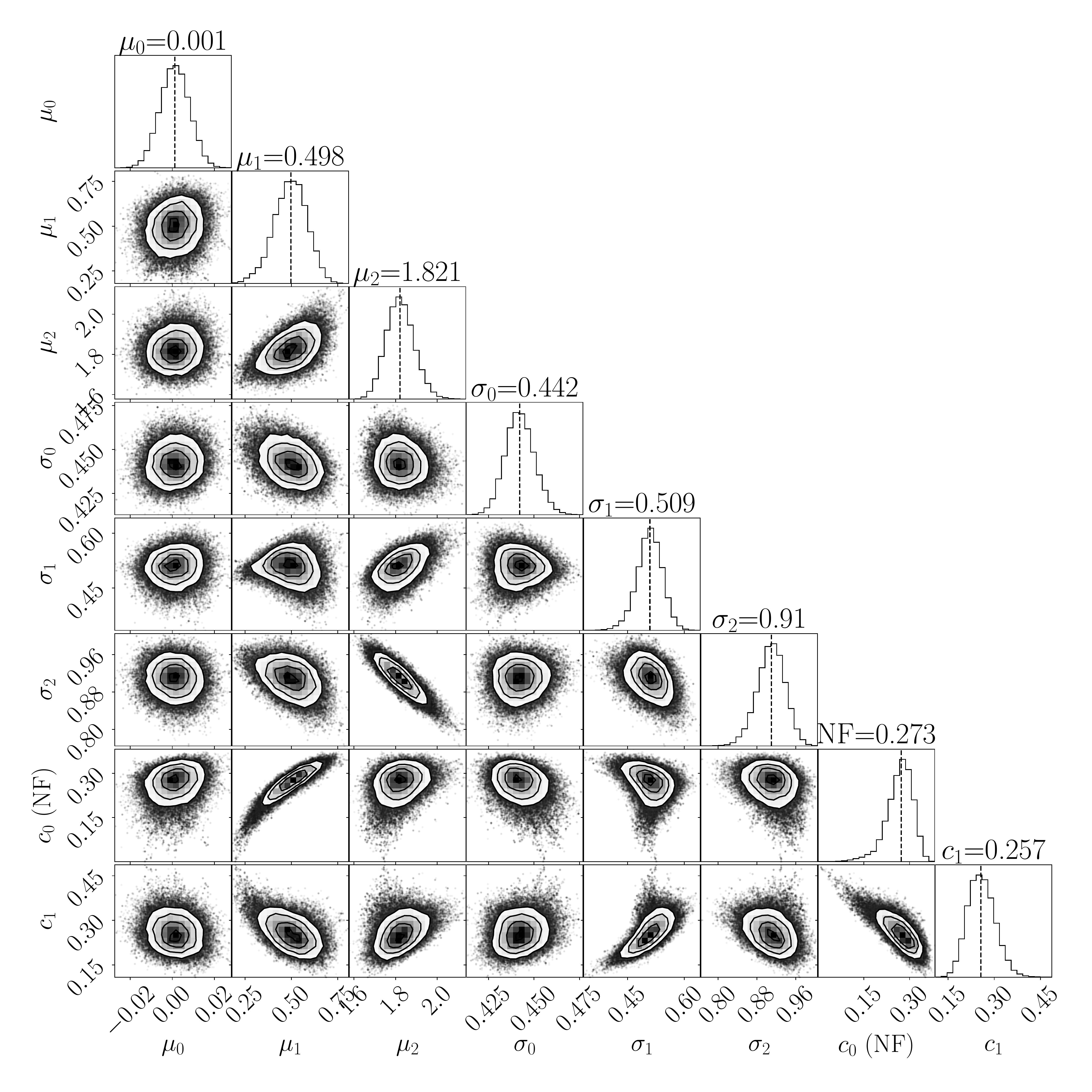}
      \centerline{\includegraphics[width=0.67\textwidth]{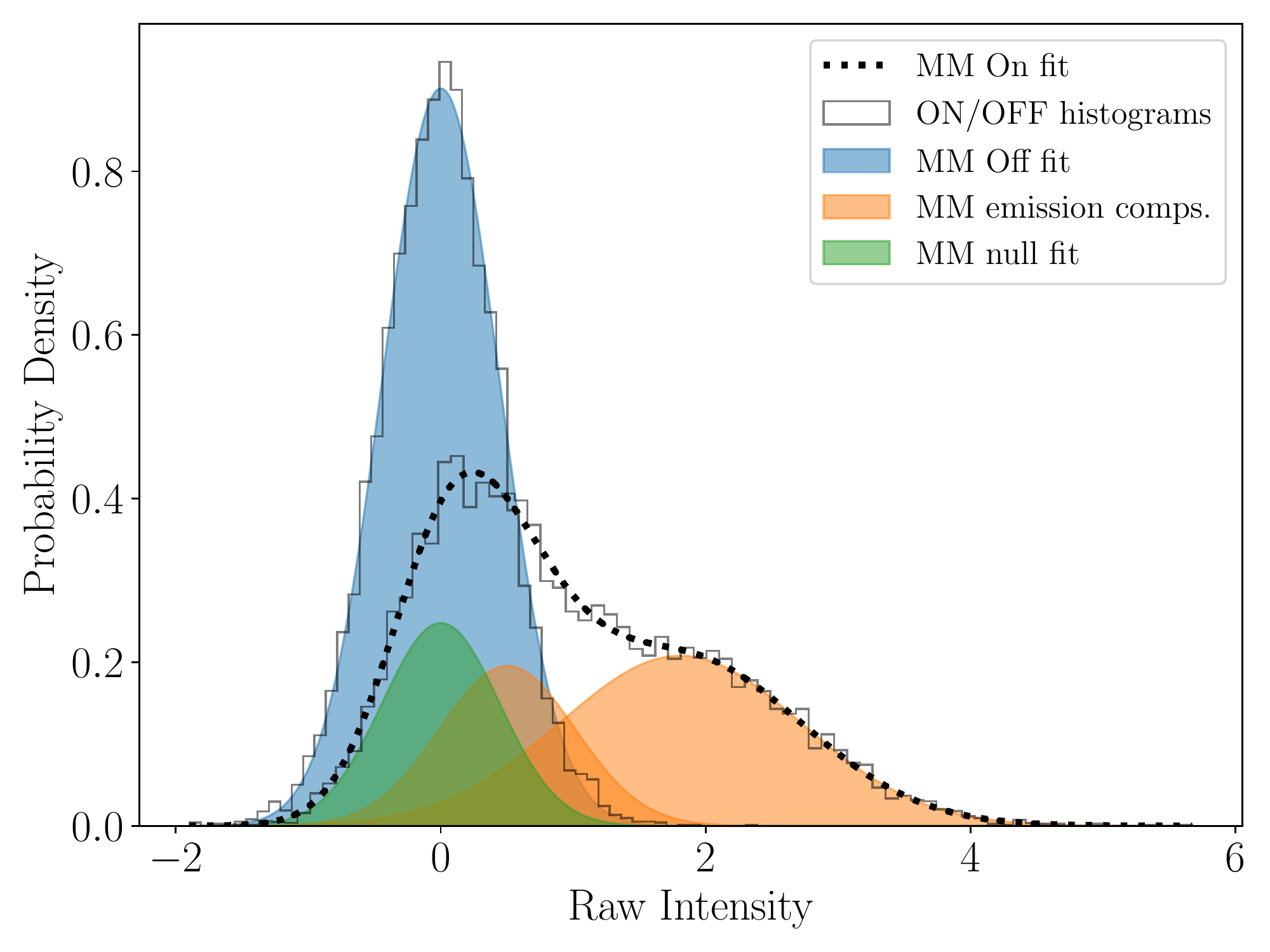}}          
      \caption{Single pulse stack (upper left), MCMC corner plot (bottom), and pulse intensity histogram (upper right) for PSR J0054+6946. In this case the best fit model is a 3-component Gaussian mixture}
      \label{fig:0054}
 \end{figure*}

\begin{figure*}
      \includegraphics[width=0.5\textwidth]{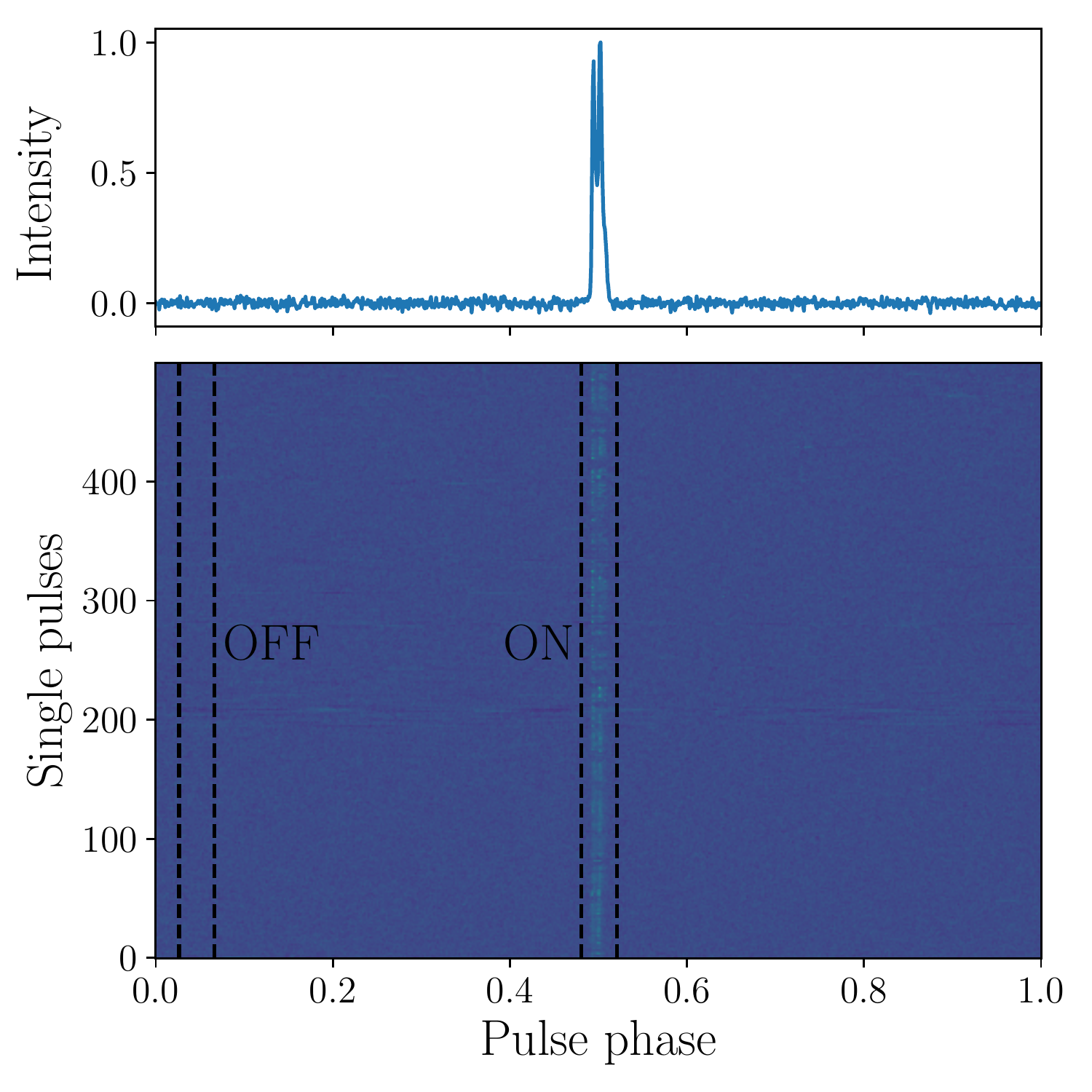}
      \includegraphics[width=0.5\textwidth]{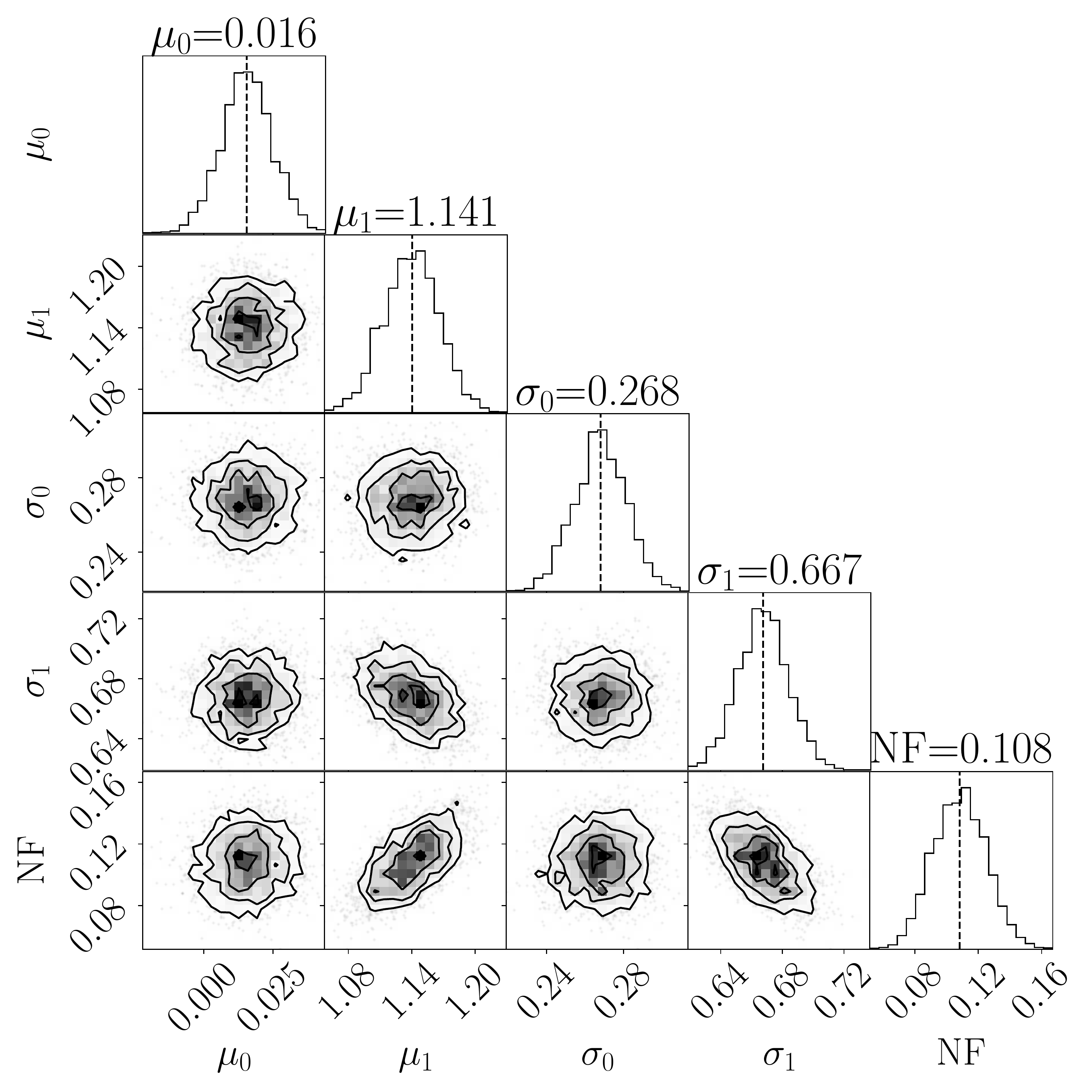}
      \centerline{\includegraphics[width=0.67\textwidth]{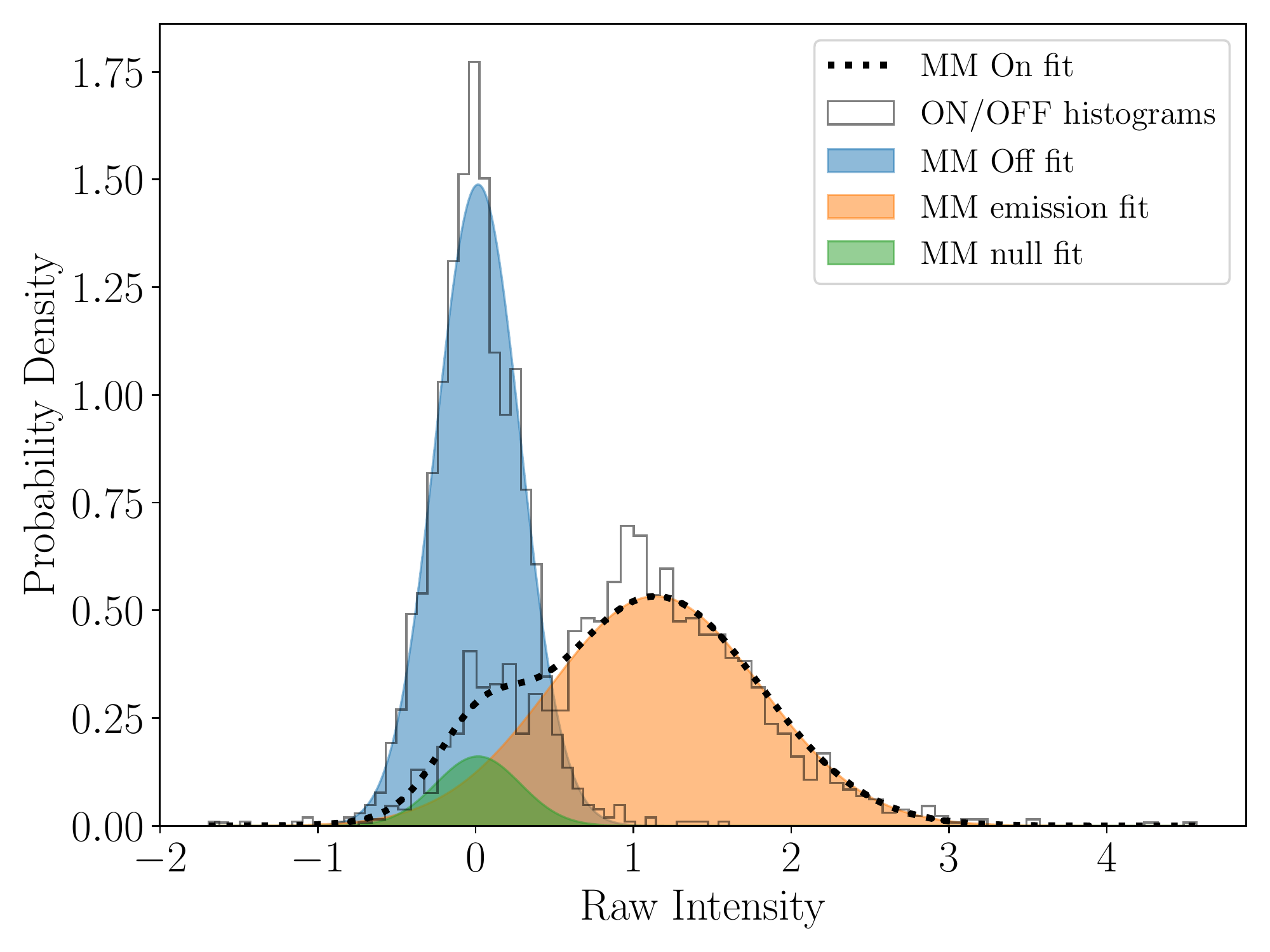}}          
      \caption{Single pulse stack (upper left), MCMC corner plot (bottom), and pulse intensity histogram (upper right) for PSR J0111+6624. In this case the best fit model is a 2-component Gaussian mixture}
 \end{figure*}

\begin{figure*}
      \includegraphics[width=0.5\textwidth]{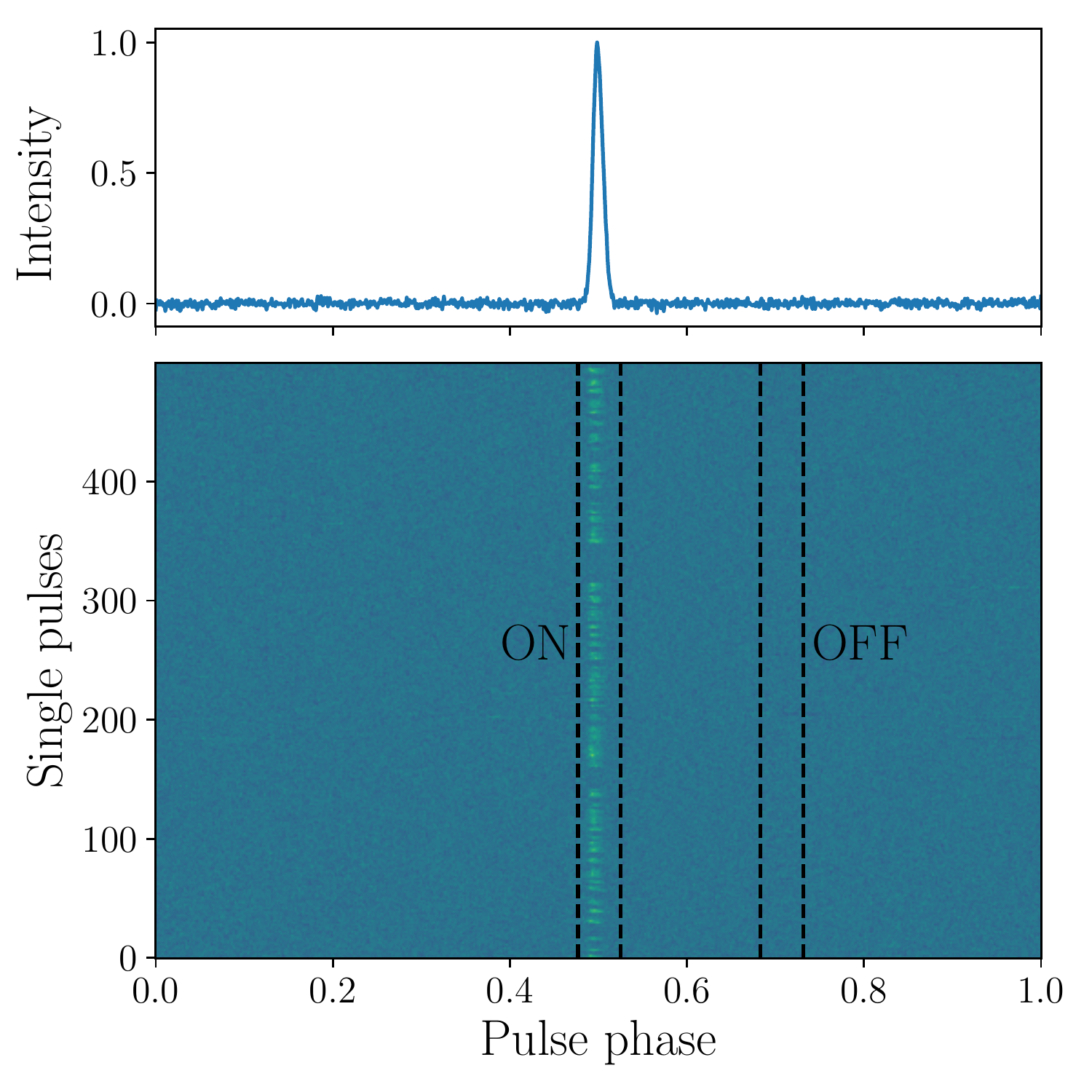}
      \includegraphics[width=0.5\textwidth]{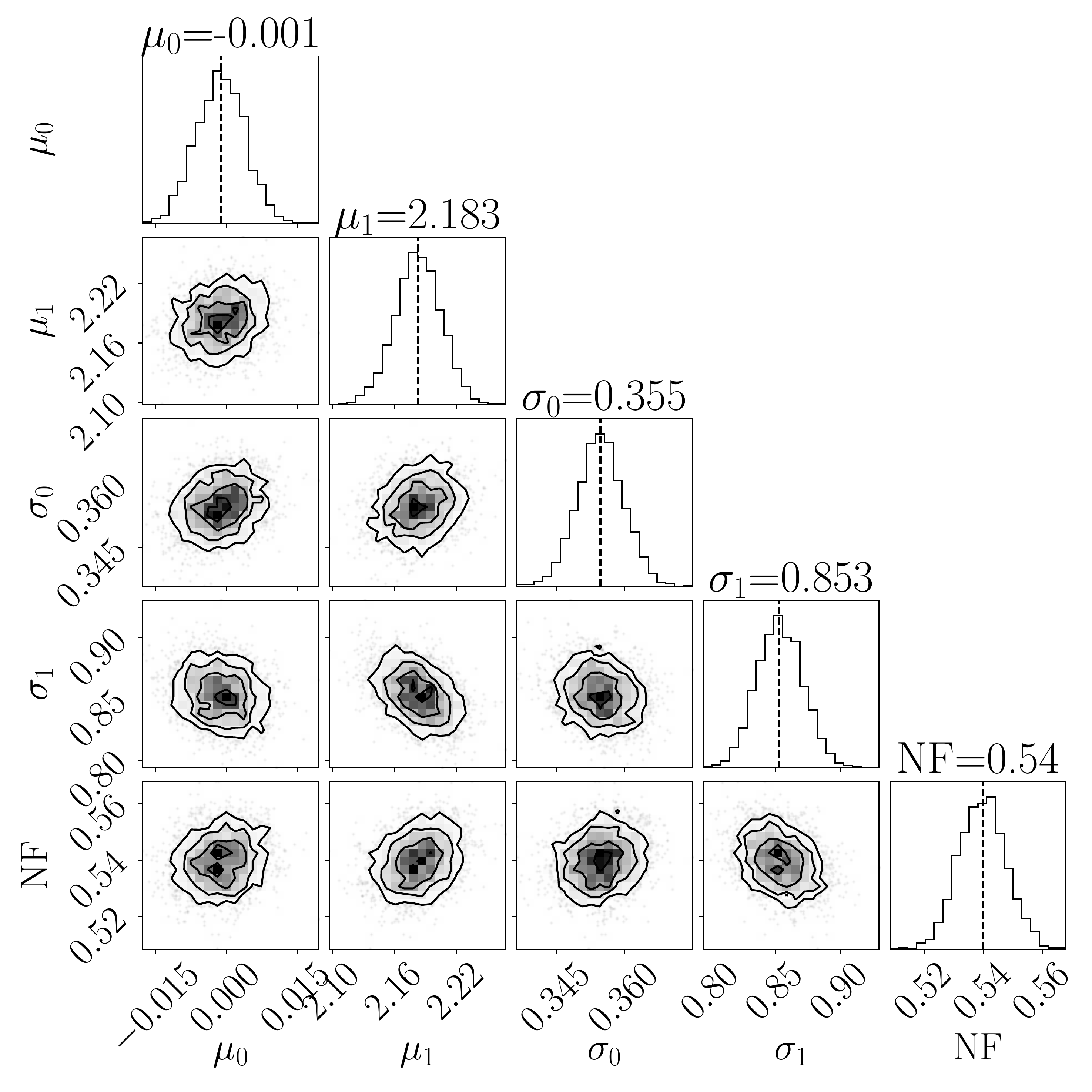}
      \centerline{\includegraphics[width=0.67\textwidth]{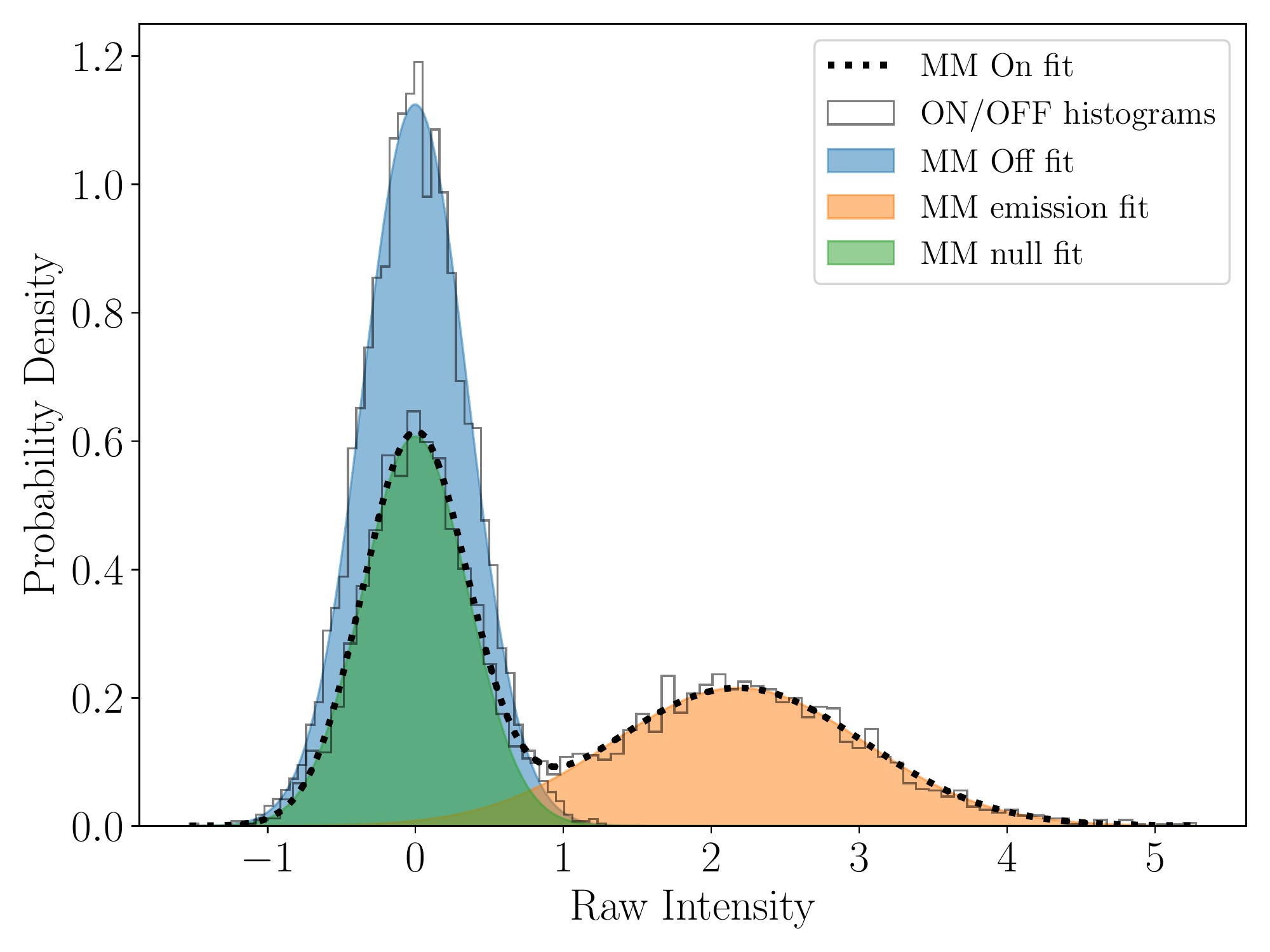}}          
      \caption{Single pulse stack (upper left), MCMC corner plot (bottom), and pulse intensity histogram (upper right) for PSR J0325+6744. In this case the best fit model is a 2-component Gaussian mixture}
 \end{figure*}

\begin{figure*}
      \includegraphics[width=0.5\textwidth]{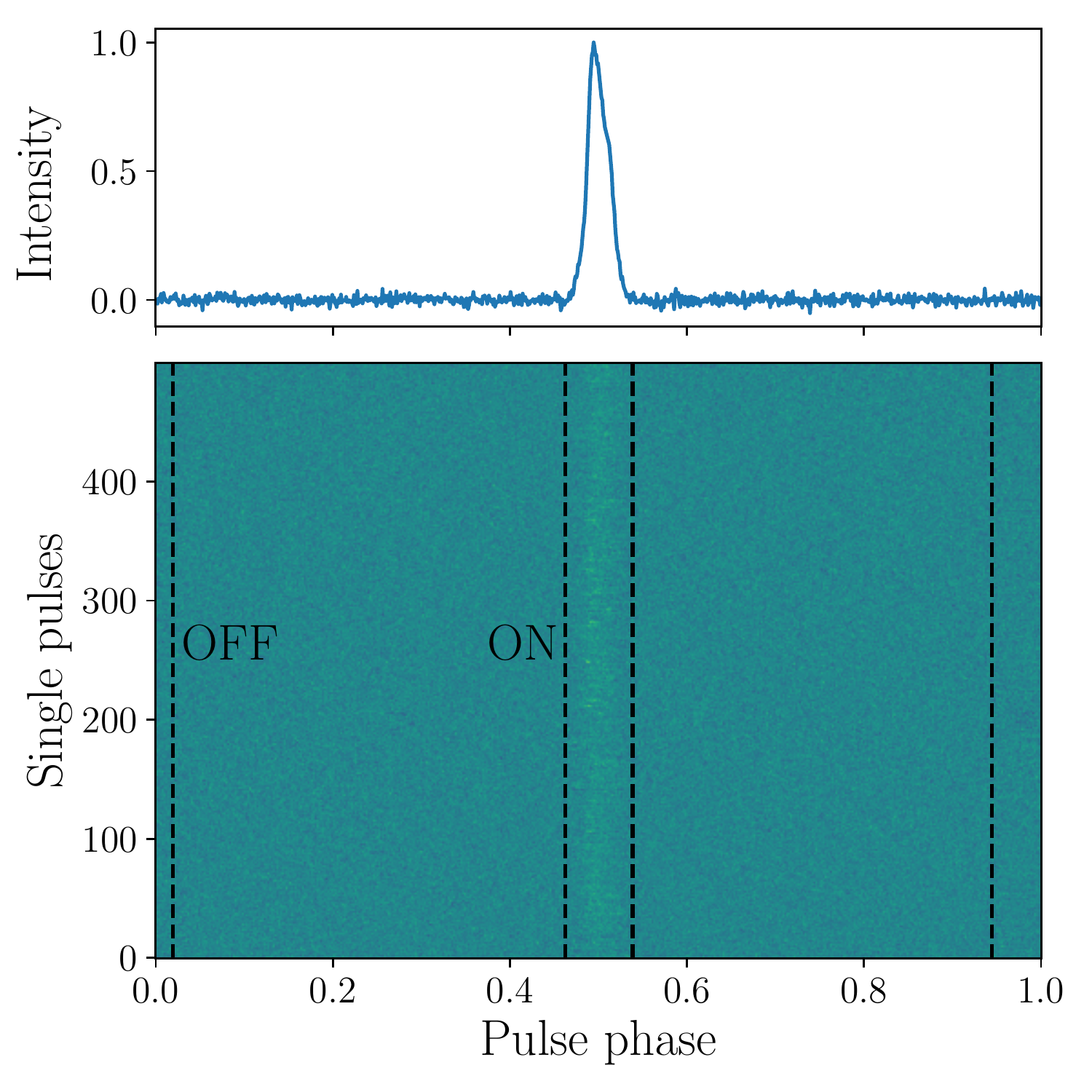}
      \includegraphics[width=0.5\textwidth]{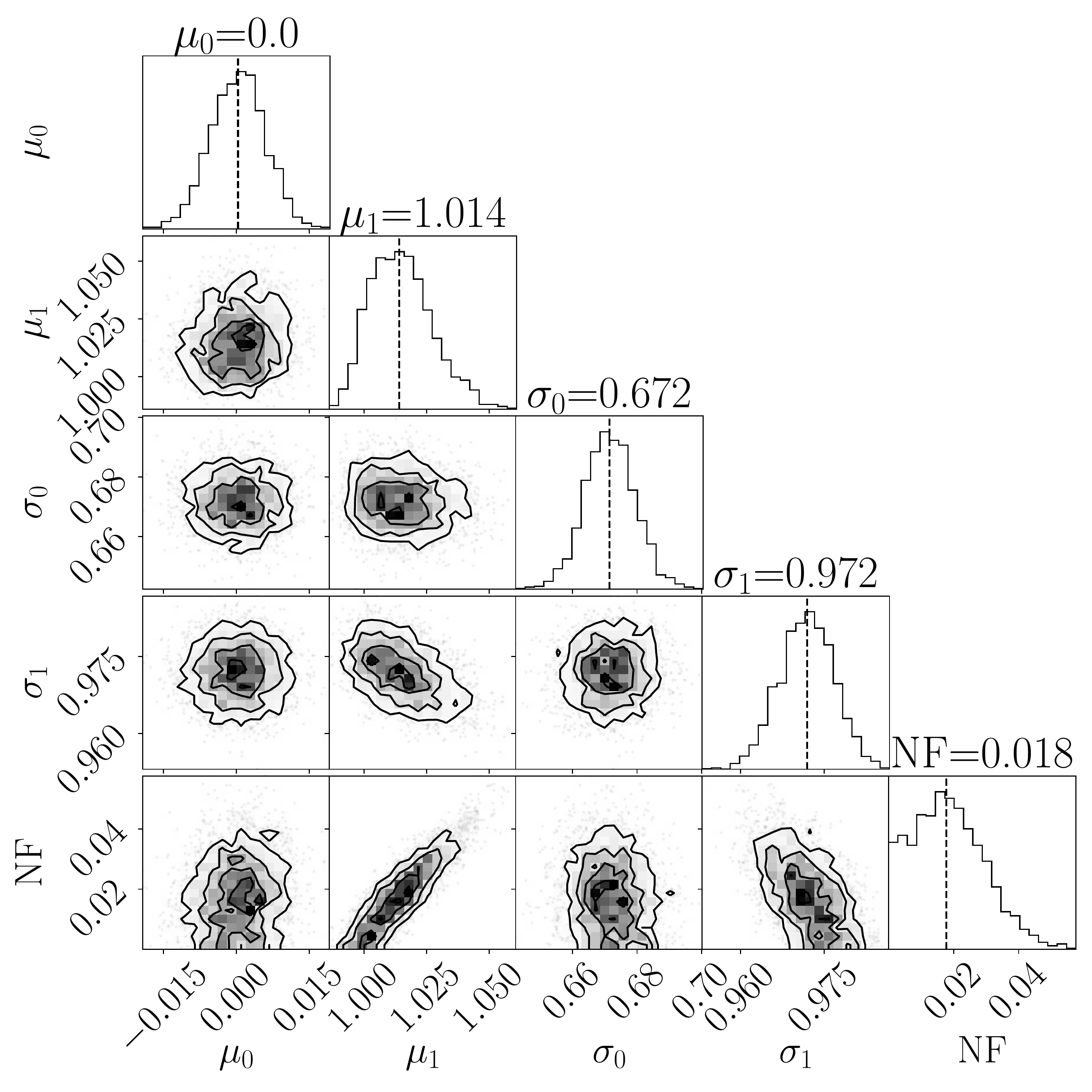}
      \centerline{\includegraphics[width=0.67\textwidth]{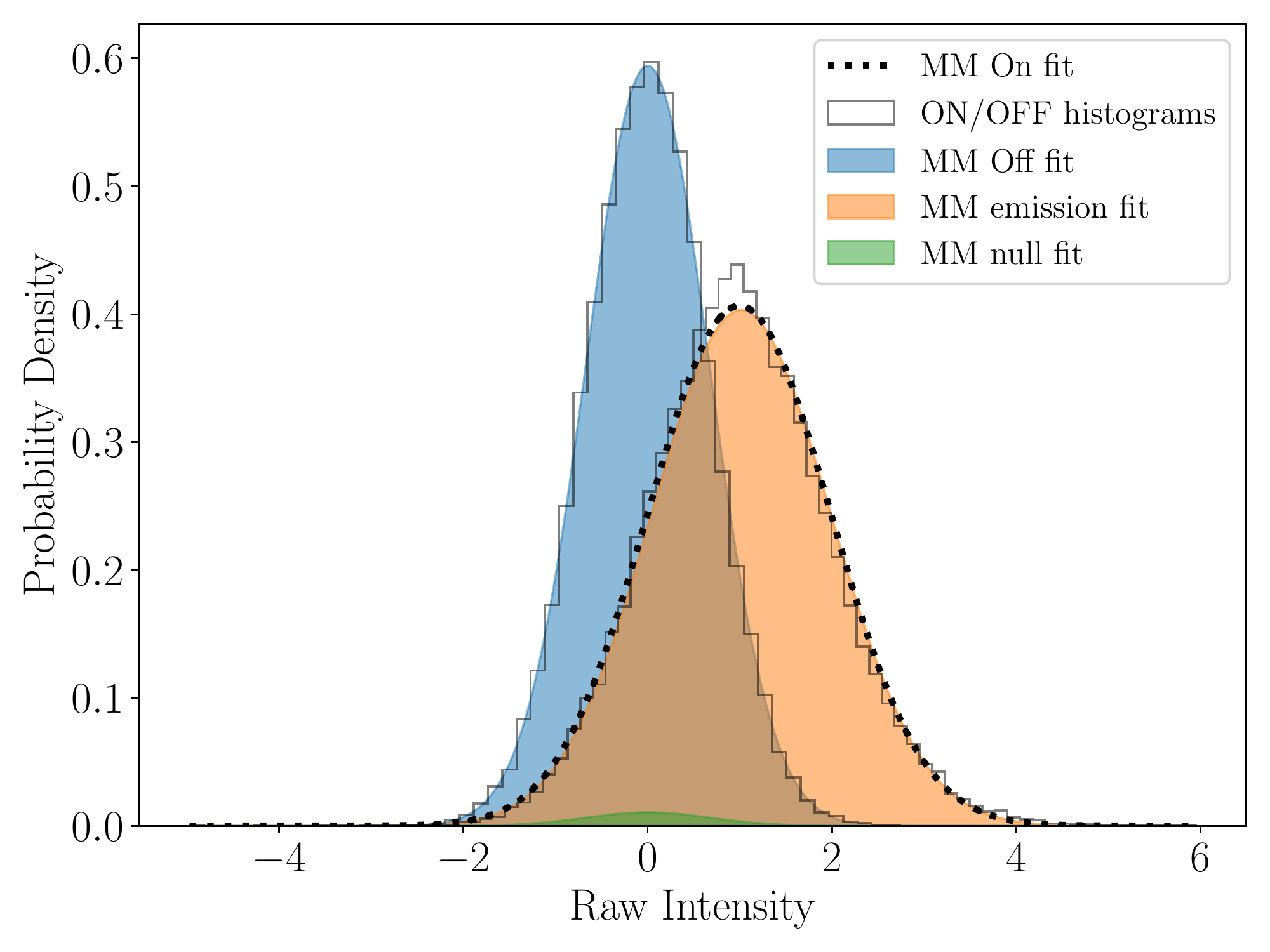}}          
      \caption{Single pulse stack (upper left), MCMC corner plot (bottom), and pulse intensity histogram (upper right) for PSR J0355+28. In this case the best fit model is a 2-component Gaussian mixture}
 \end{figure*}

\begin{figure*}
      \includegraphics[width=0.5\textwidth]{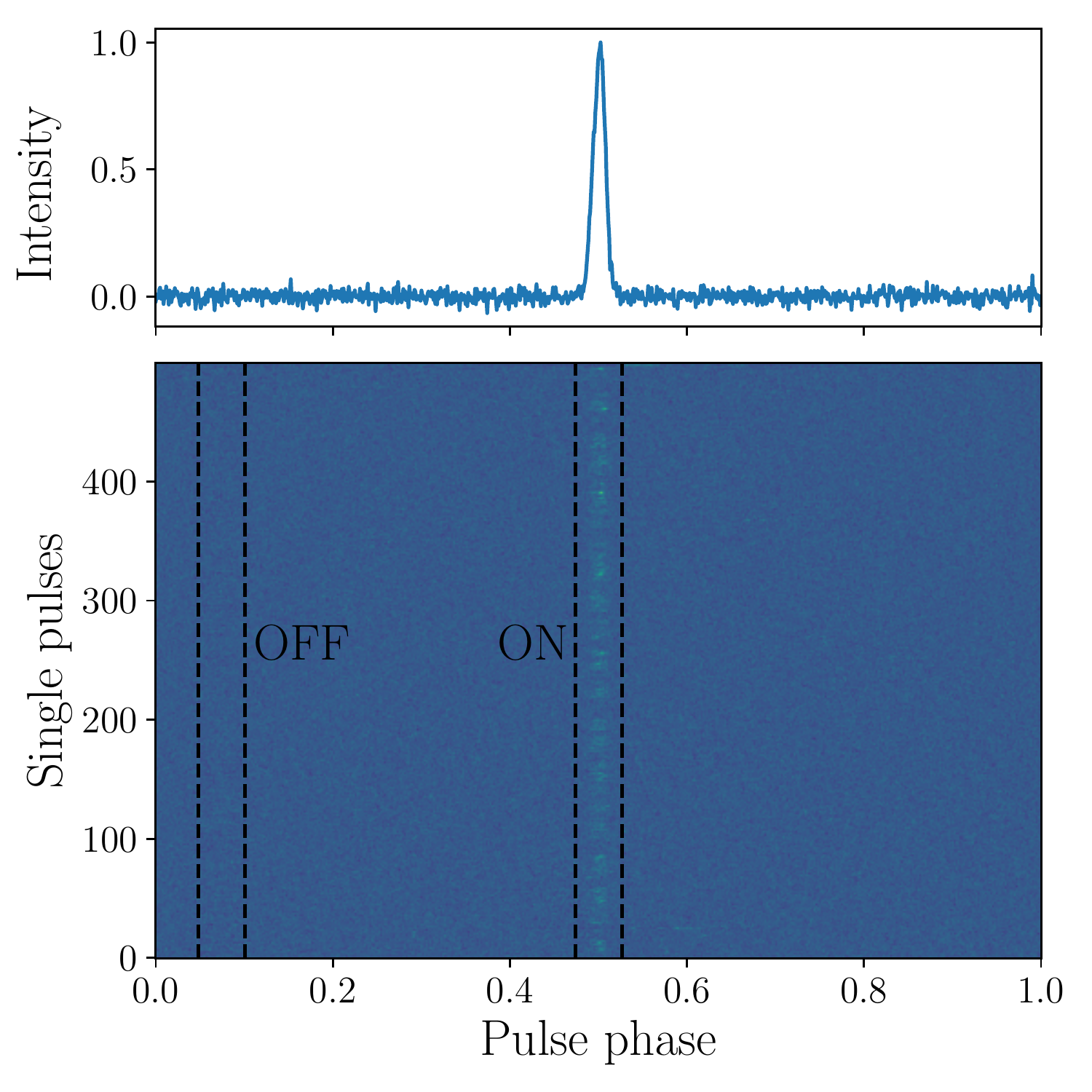}
      \includegraphics[width=0.5\textwidth]{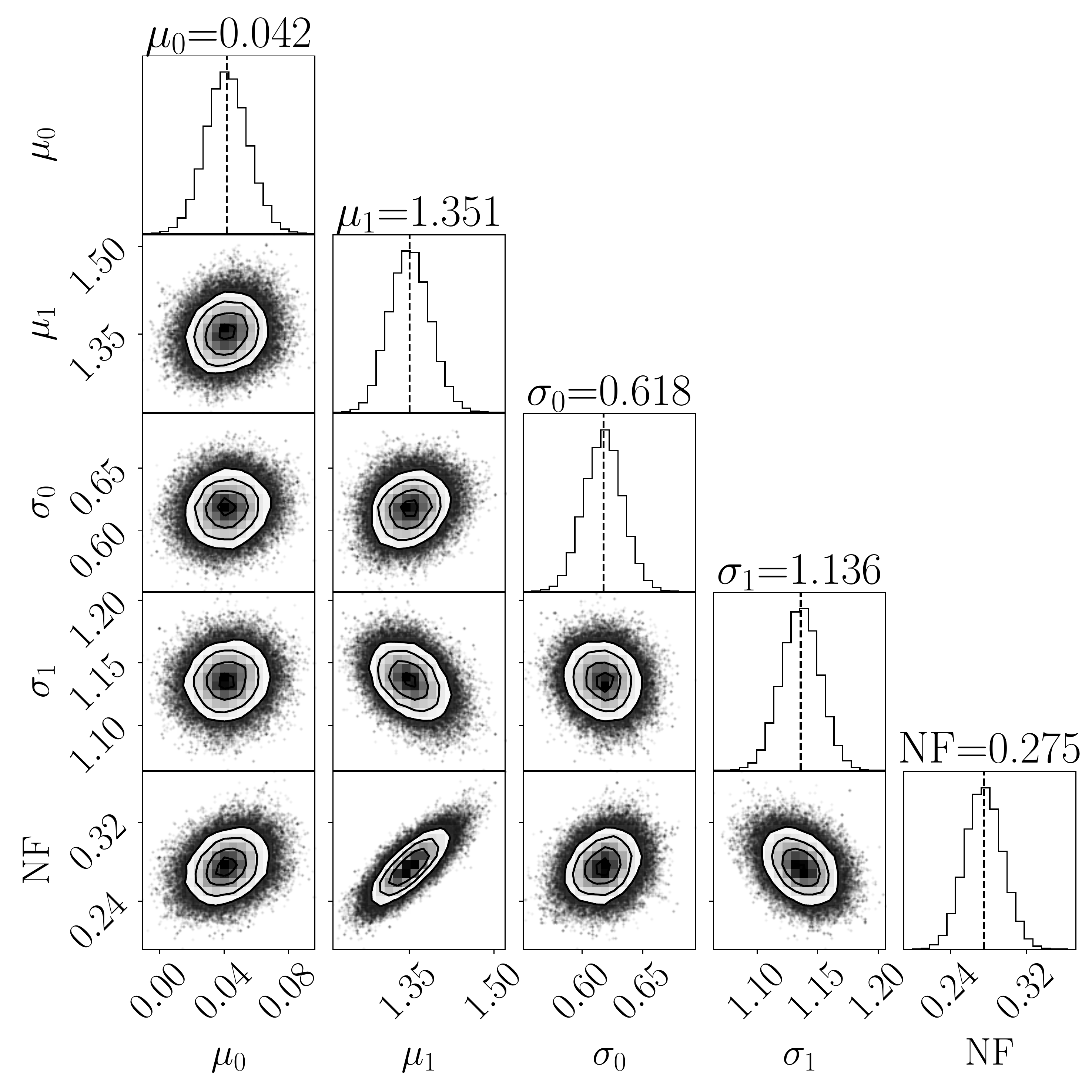}
      \centerline{\includegraphics[width=0.67\textwidth]{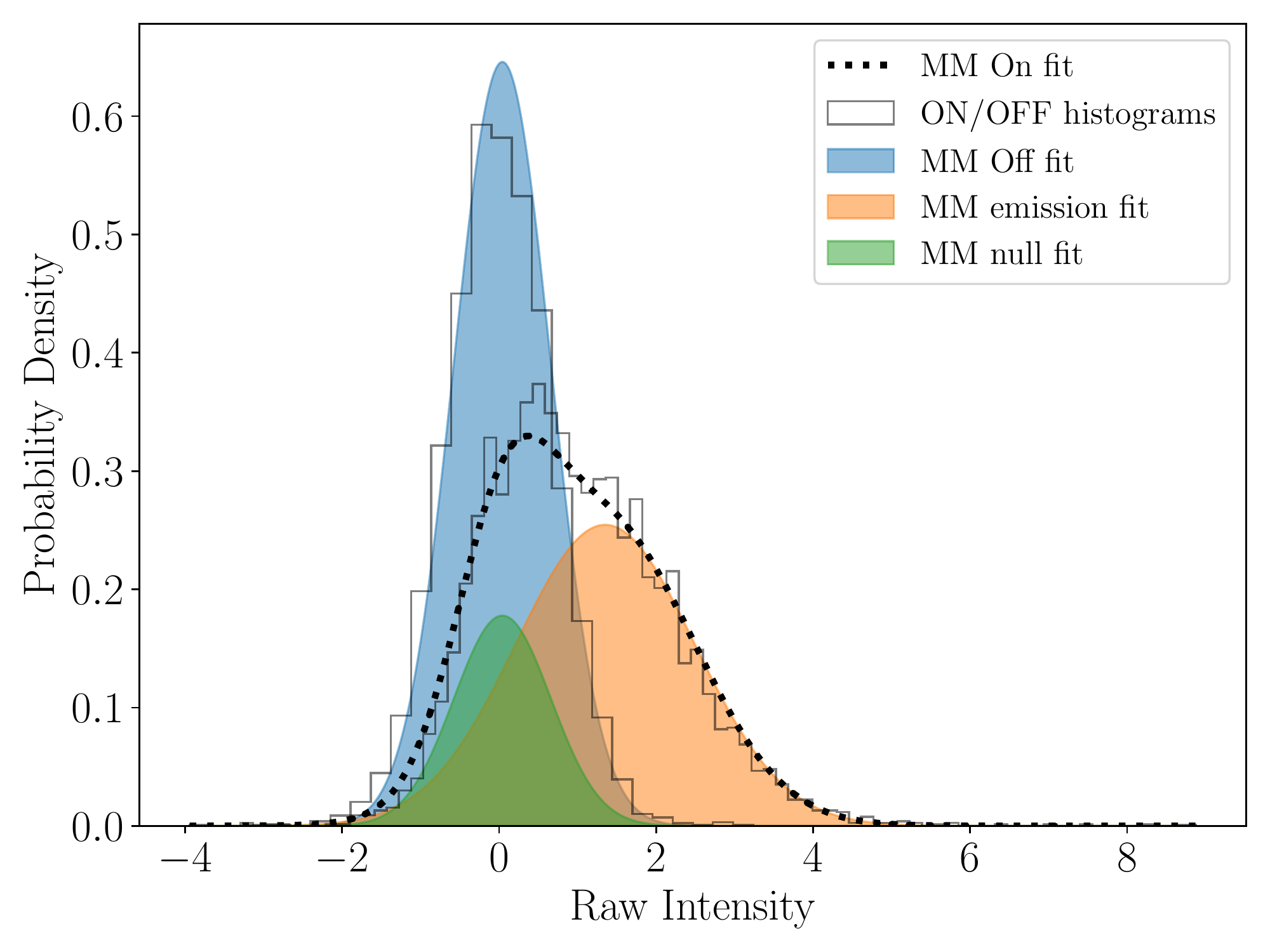}}          
      \caption{Single pulse stack (upper left), MCMC corner plot (bottom), and pulse intensity histogram (upper right) for PSR J0414+31 (GBT). In this case the best fit model is a 2-component Gaussian mixture}
 \end{figure*}

\begin{figure*}
      \includegraphics[width=0.5\textwidth]{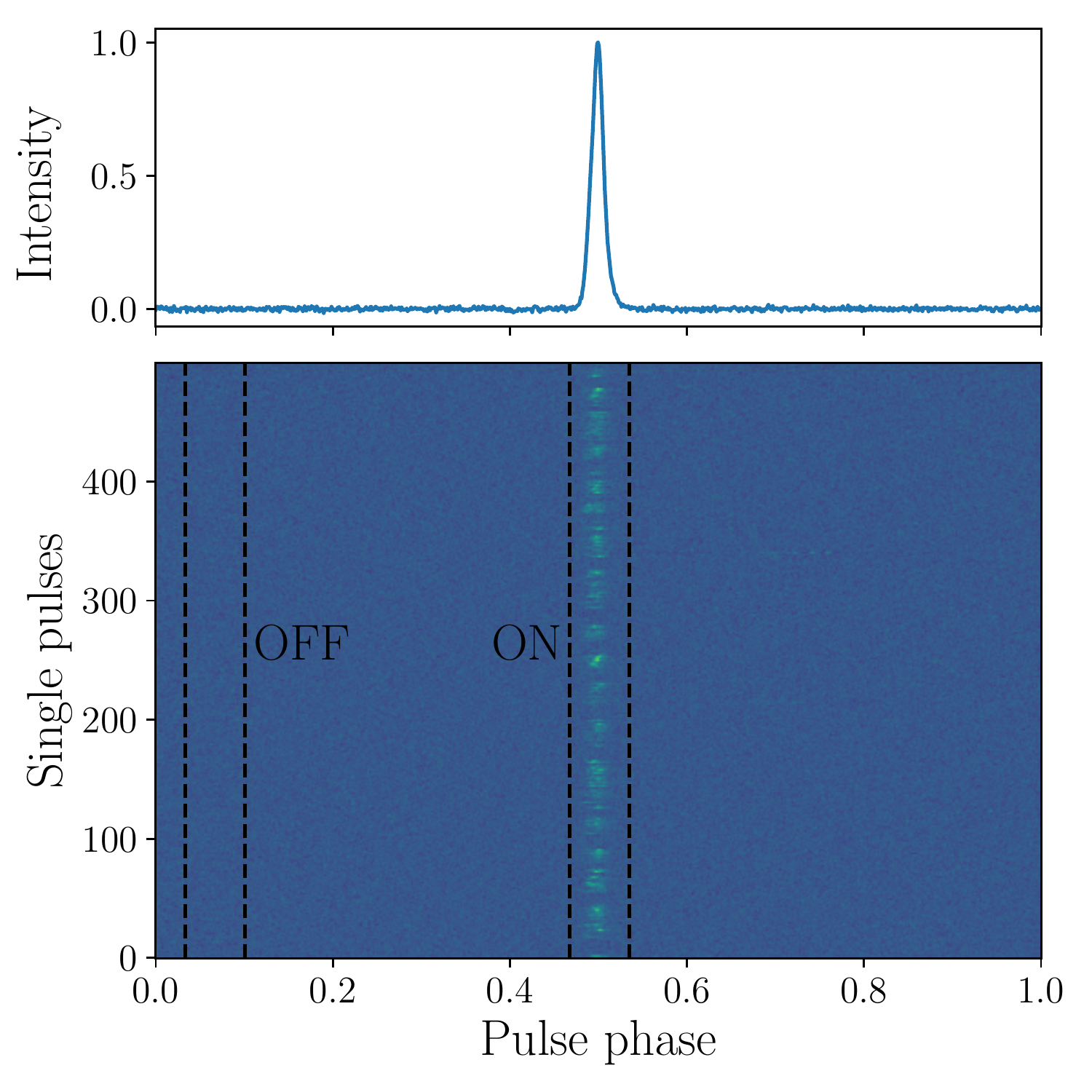}
      \includegraphics[width=0.5\textwidth]{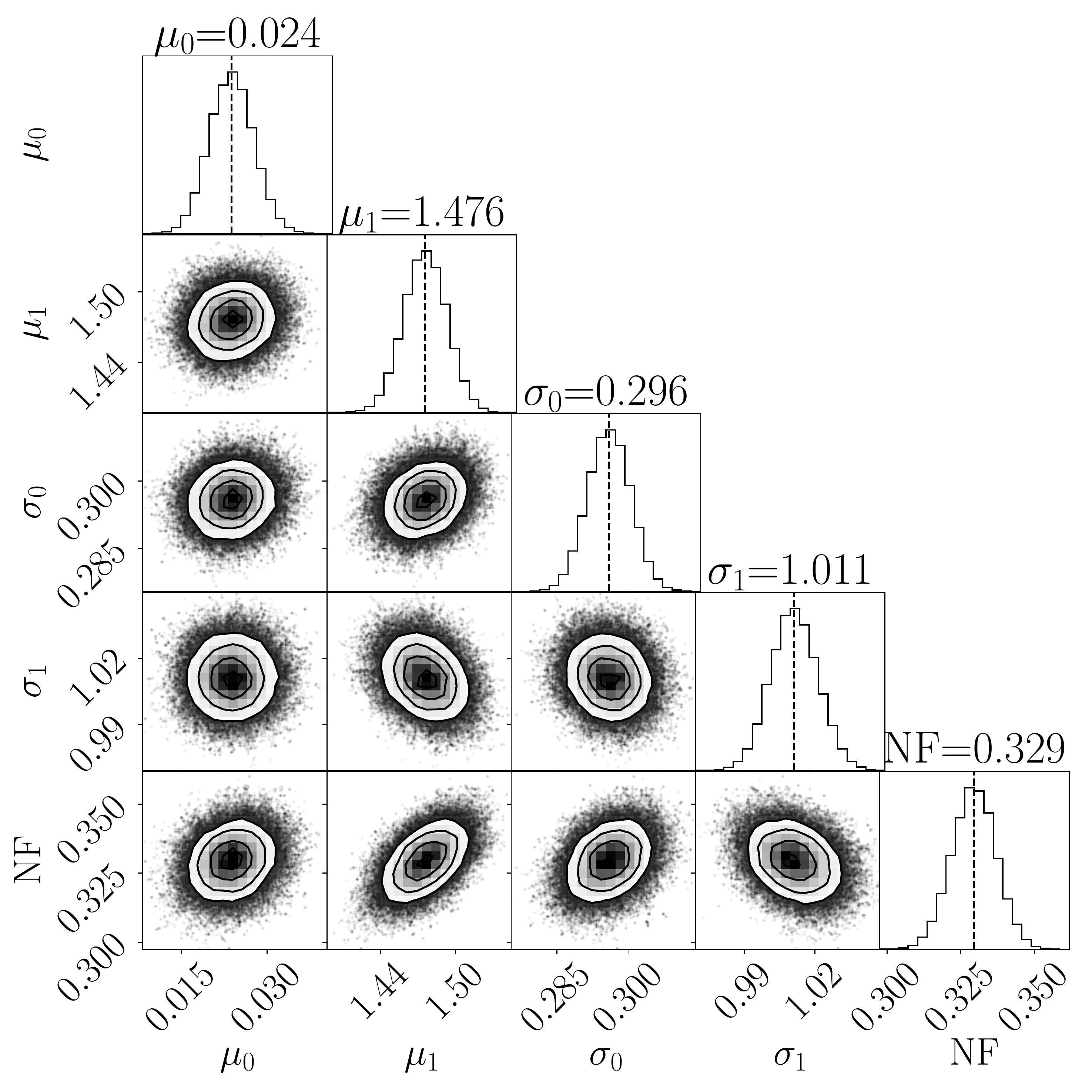}
      \centerline{\includegraphics[width=0.67\textwidth]{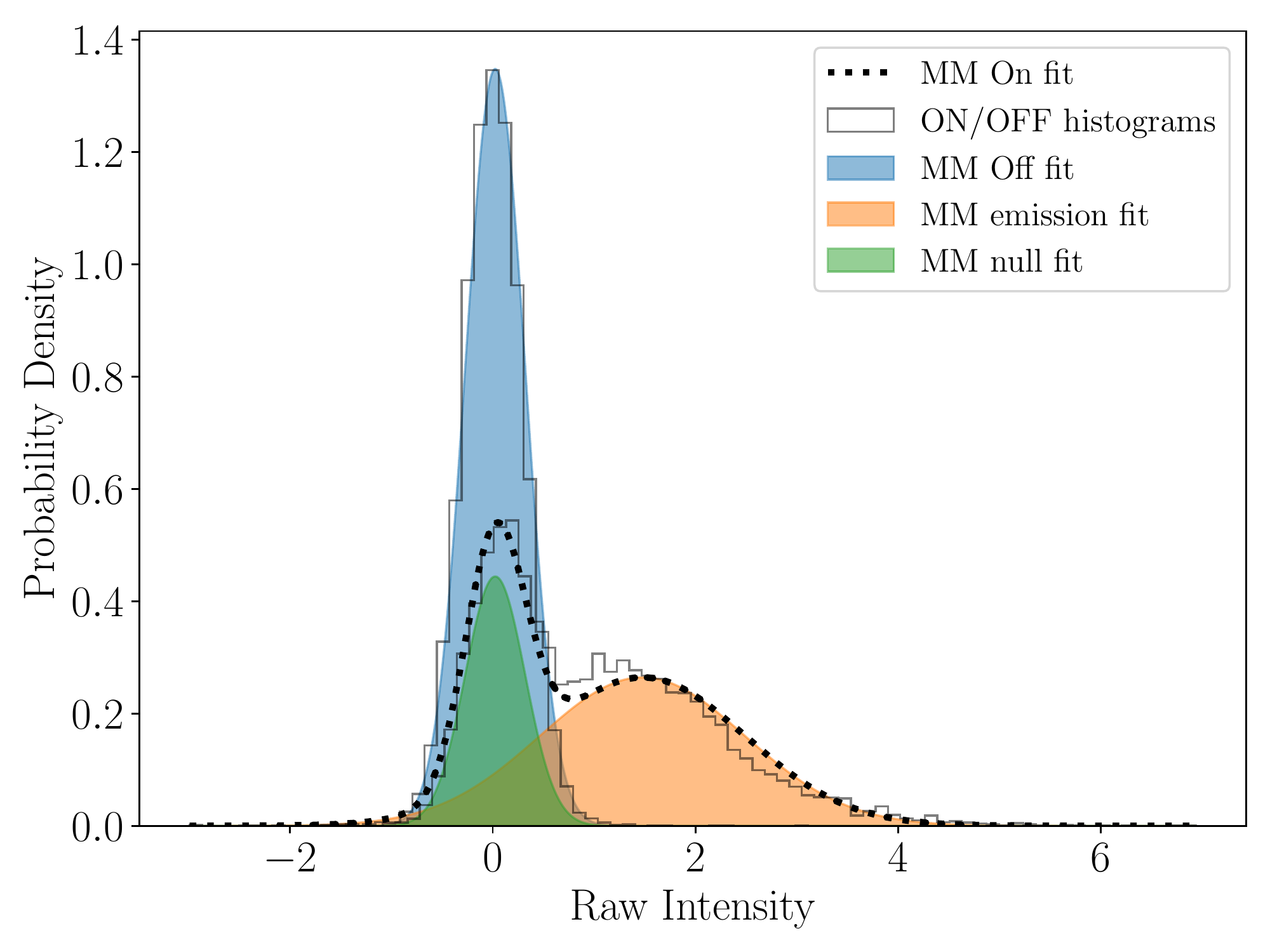}}          
      \caption{Single pulse stack (upper left), MCMC corner plot (bottom), and pulse intensity histogram (upper right) for PSR J0414+31 (arecibo). In this case the best fit model is a 2-component Gaussian mixture}
 \end{figure*}

\begin{figure*}
      \includegraphics[width=0.5\textwidth]{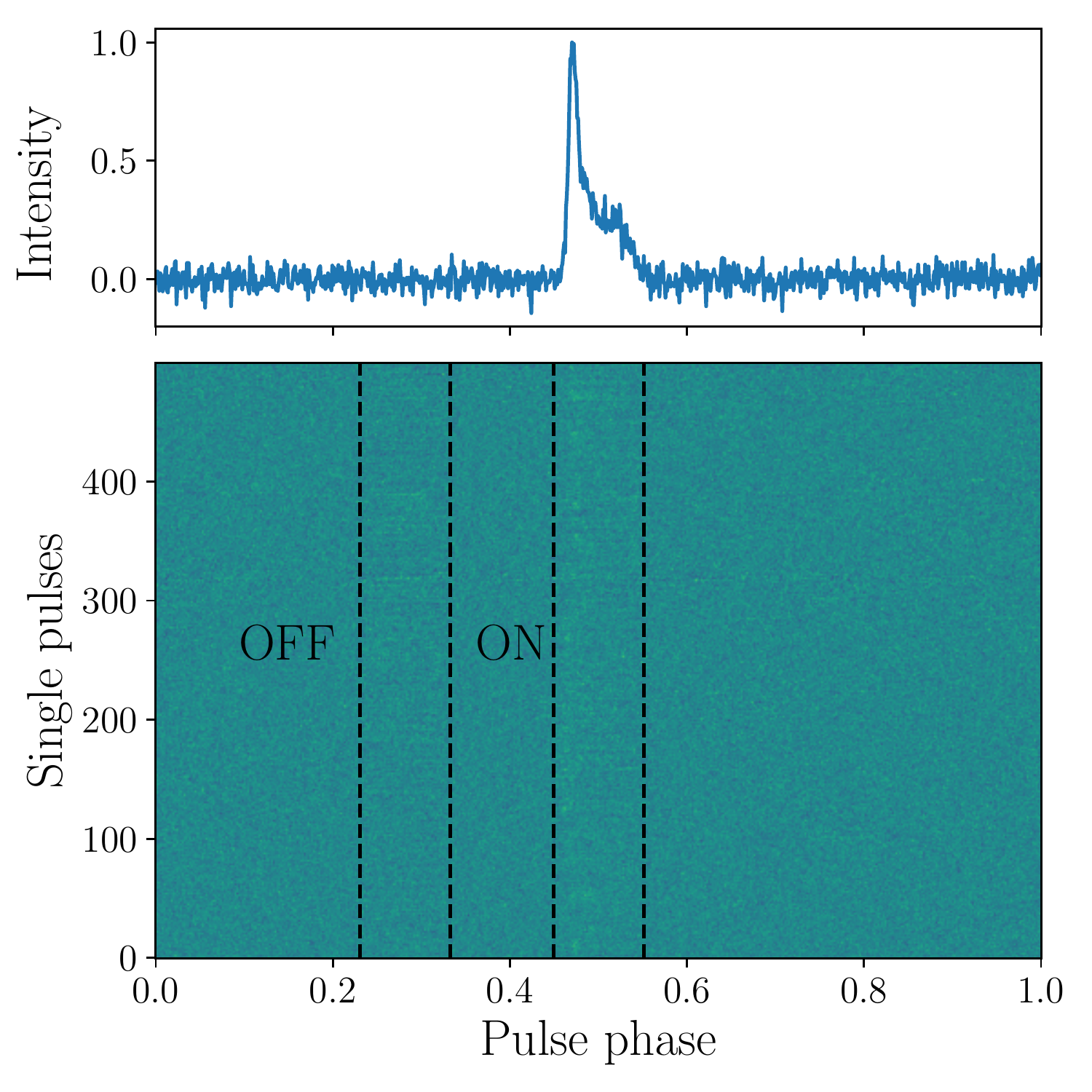}
      \includegraphics[width=0.5\textwidth]{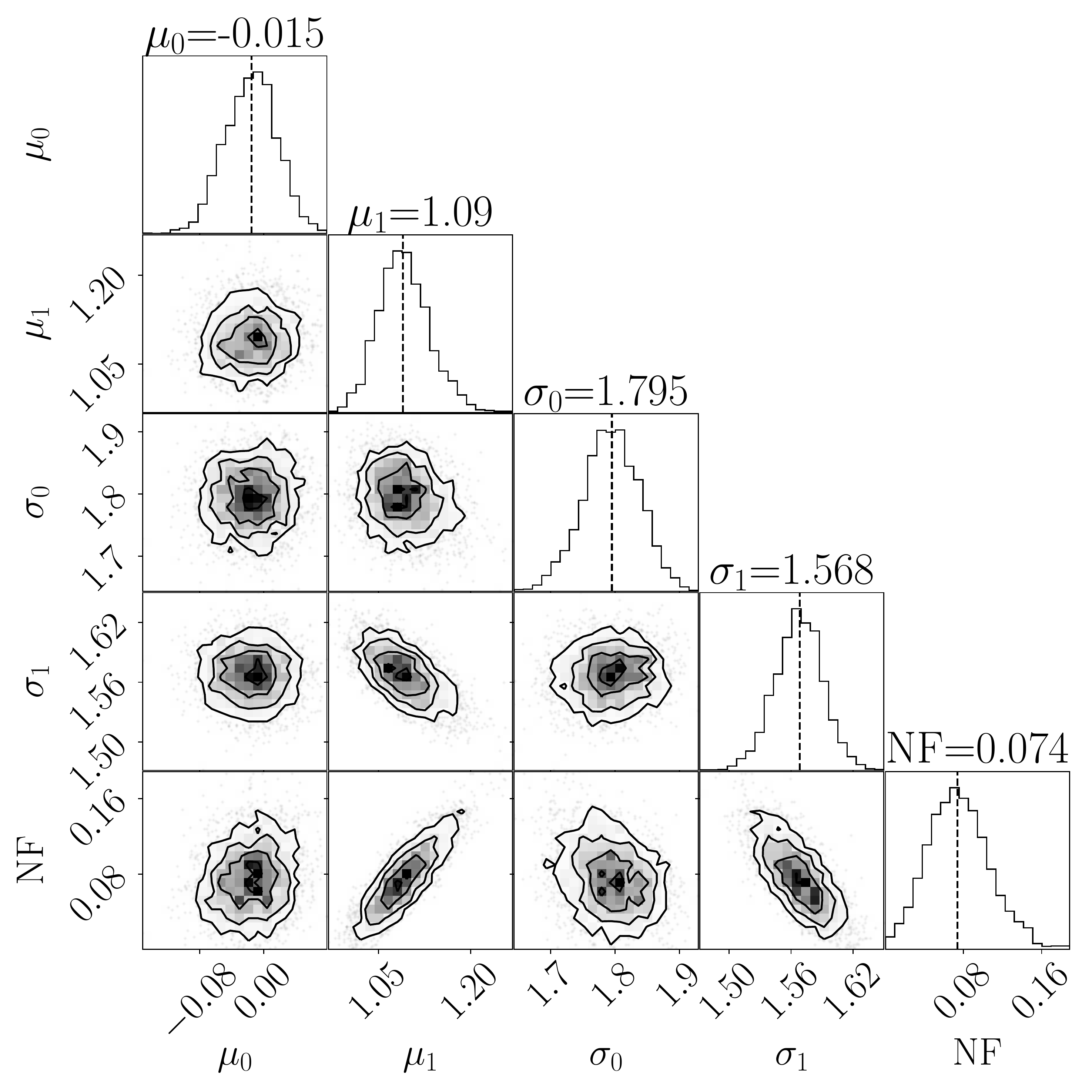}
      \centerline{\includegraphics[width=0.67\textwidth]{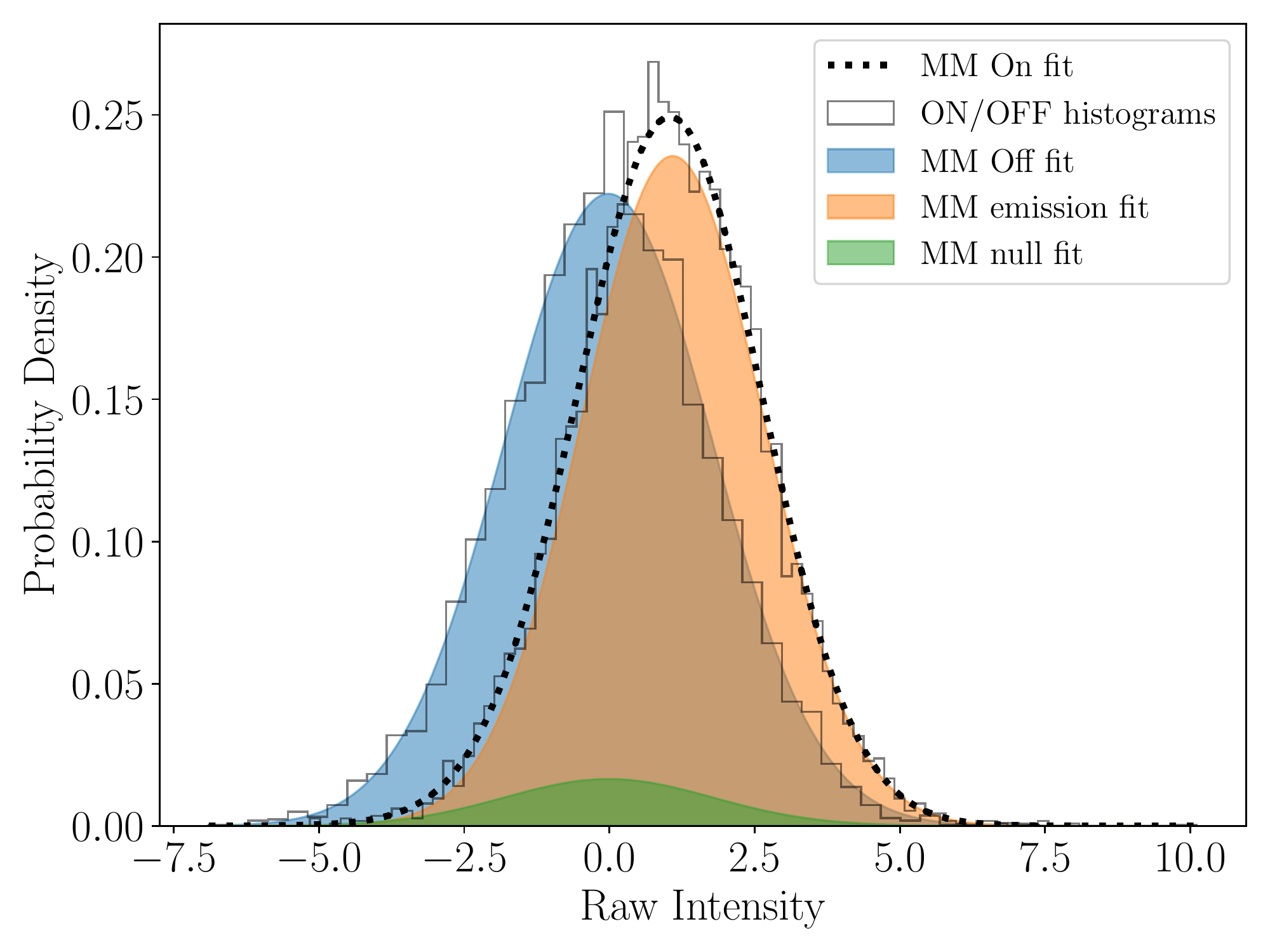}}          
      \caption{Single pulse stack (upper left), MCMC corner plot (bottom), and pulse intensity histogram (upper right) for PSR J0614+83. In this case the best fit model is a 2-component Gaussian mixture}
 \end{figure*}

\begin{figure*}
      \includegraphics[width=0.5\textwidth]{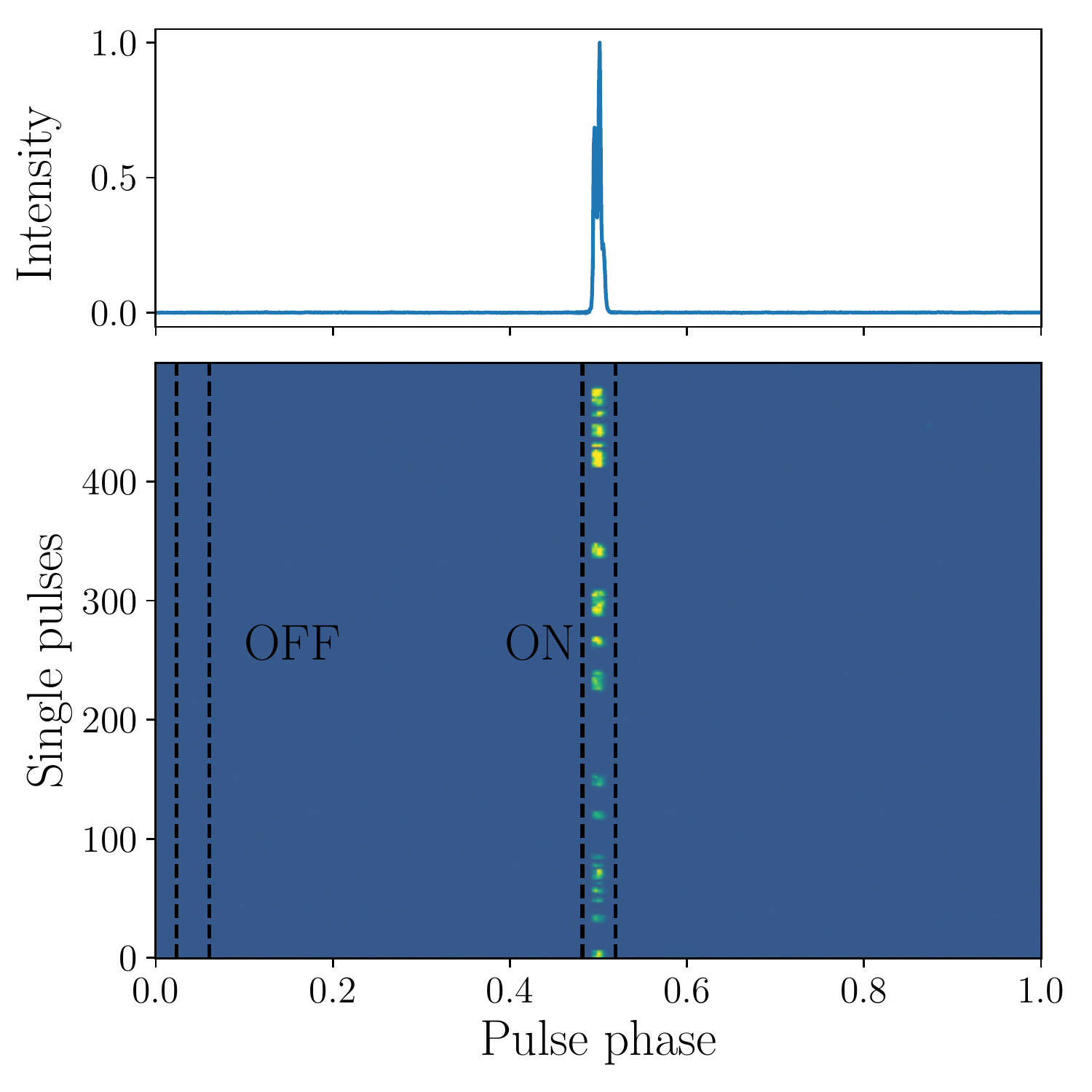}
      \includegraphics[width=0.5\textwidth]{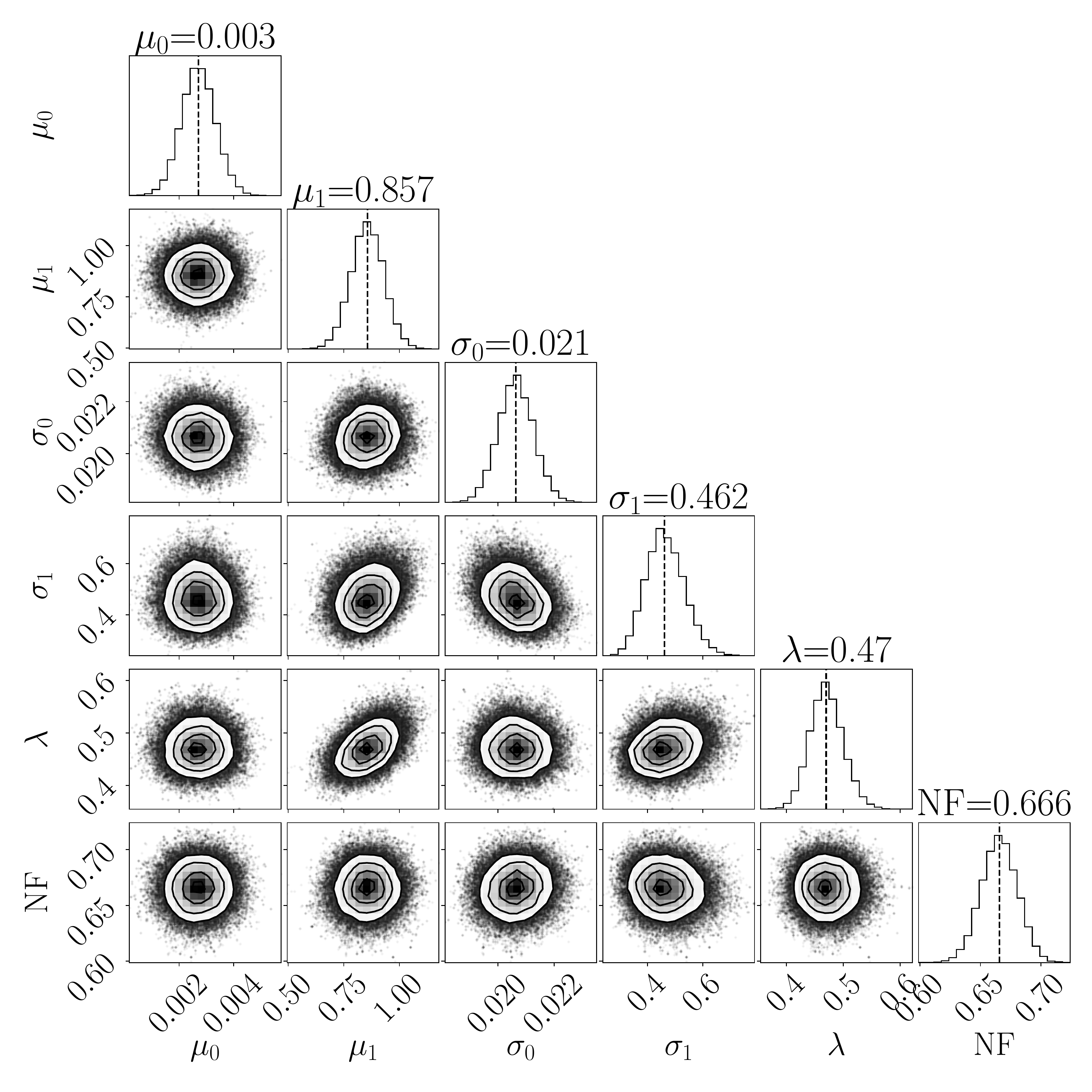}
      \centerline{\includegraphics[width=0.67\textwidth]{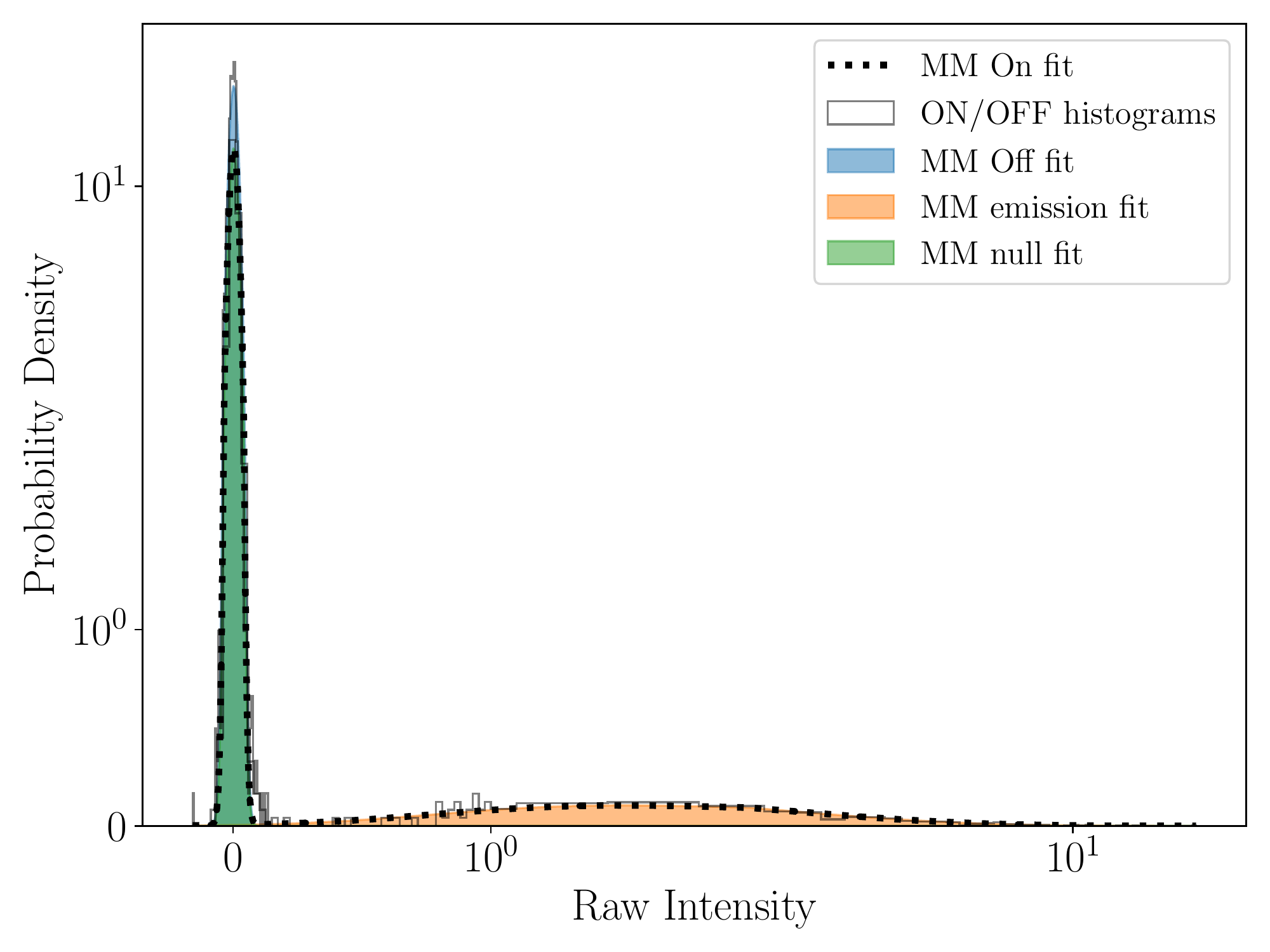}}          
      \caption{Single pulse stack (upper left), MCMC corner plot (bottom), and pulse intensity histogram (upper right) for PSR J0738+6904. In this case the best fit model is a 2-component Exponential convolved Gaussian mixture}
 \end{figure*}

\begin{figure*}
      \includegraphics[width=0.5\textwidth]{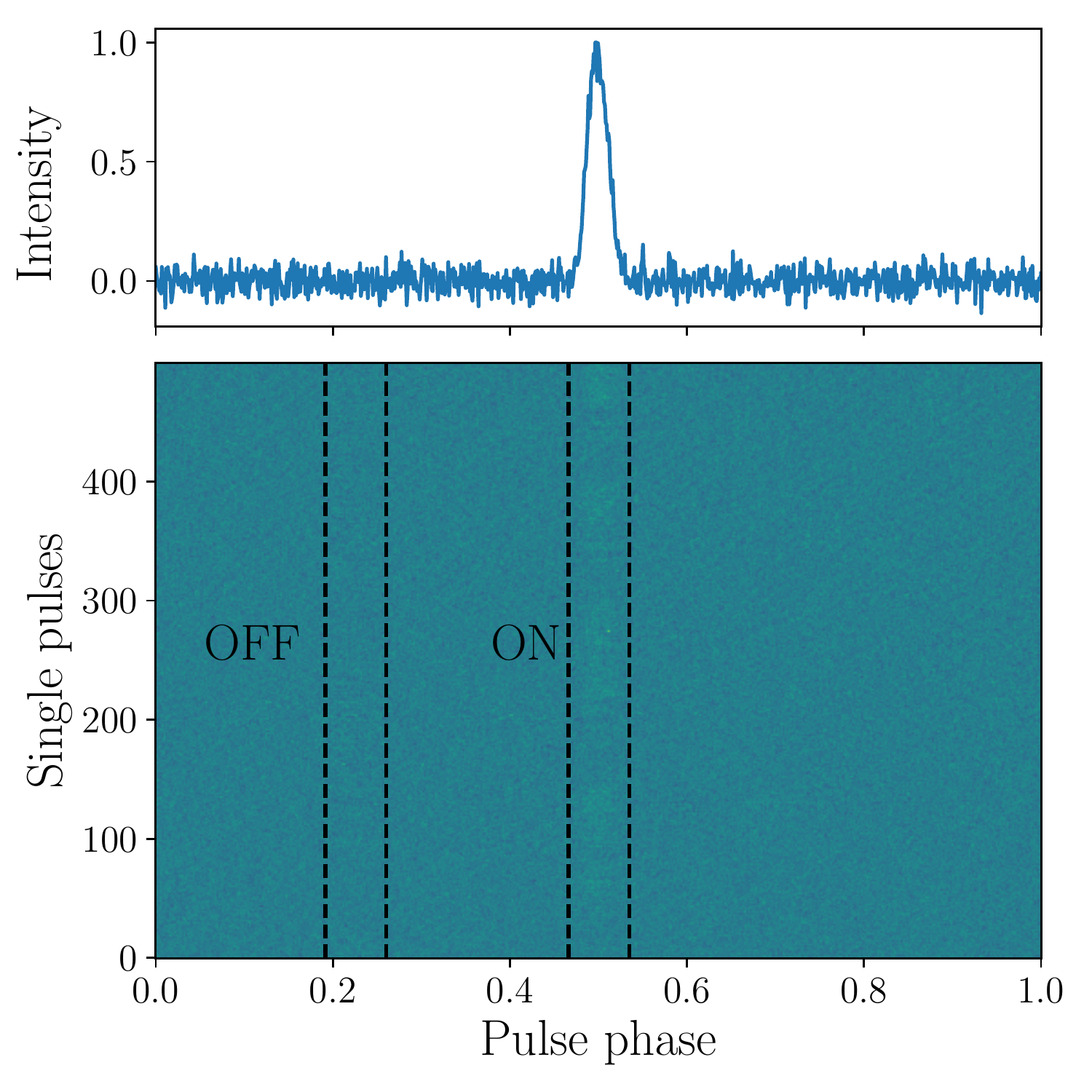}
      \includegraphics[width=0.5\textwidth]{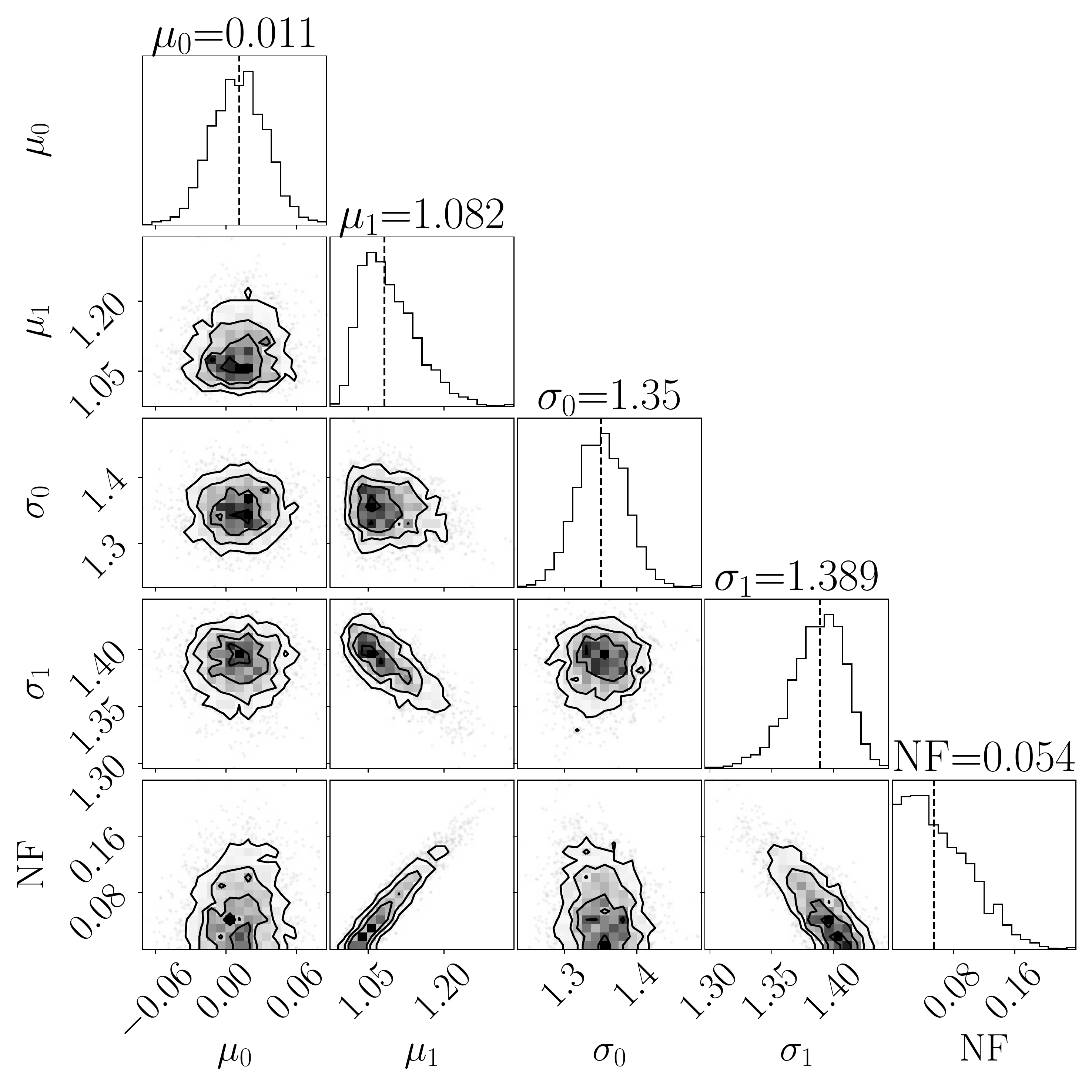}
      \centerline{\includegraphics[width=0.67\textwidth]{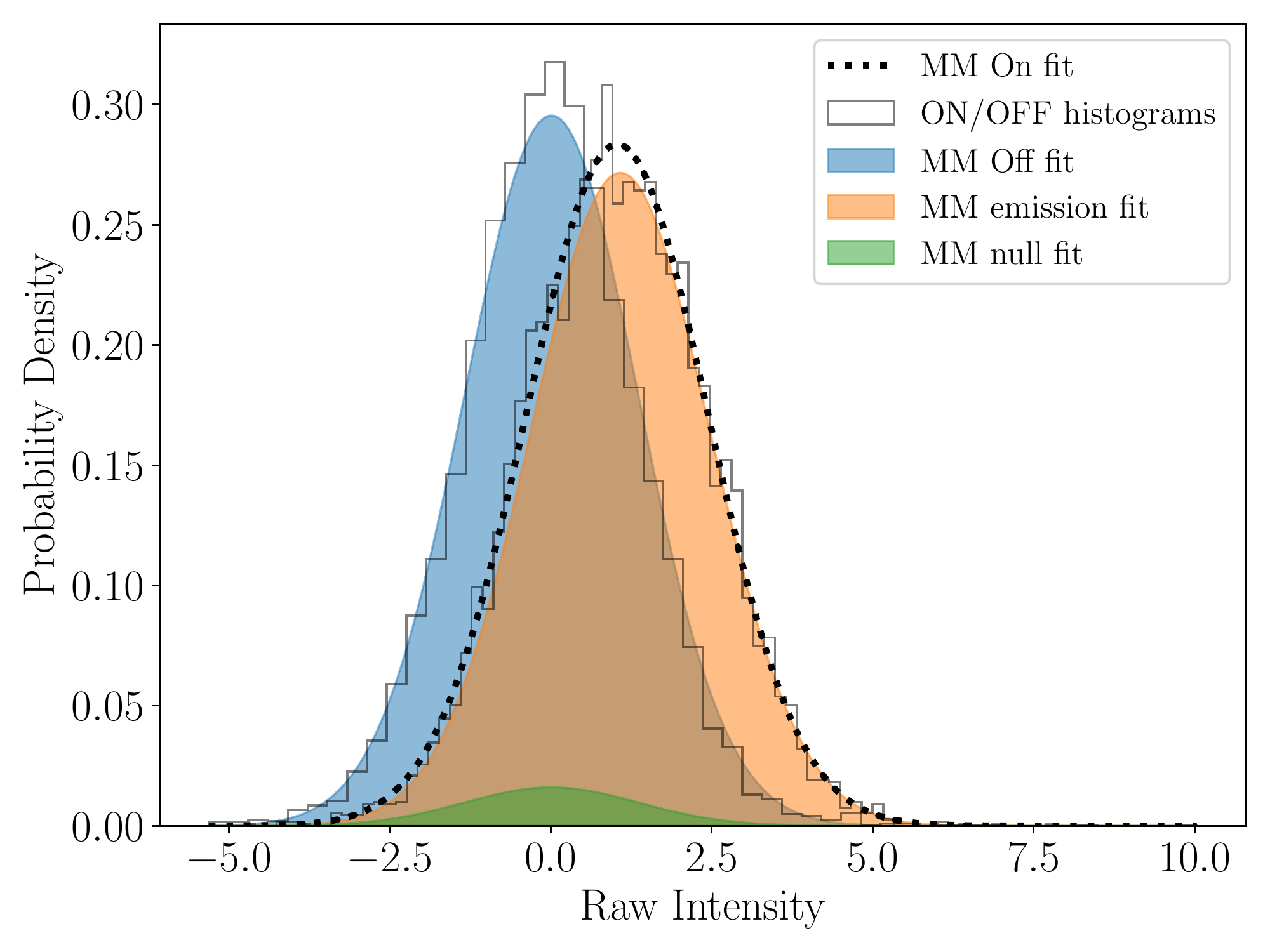}}          
      \caption{Single pulse stack (upper left), MCMC corner plot (bottom), and pulse intensity histogram (upper right) for PSR J1529-26. In this case the best fit model is a 2-component Gaussian mixture}
 \end{figure*}

\begin{figure*}
      \includegraphics[width=0.5\textwidth]{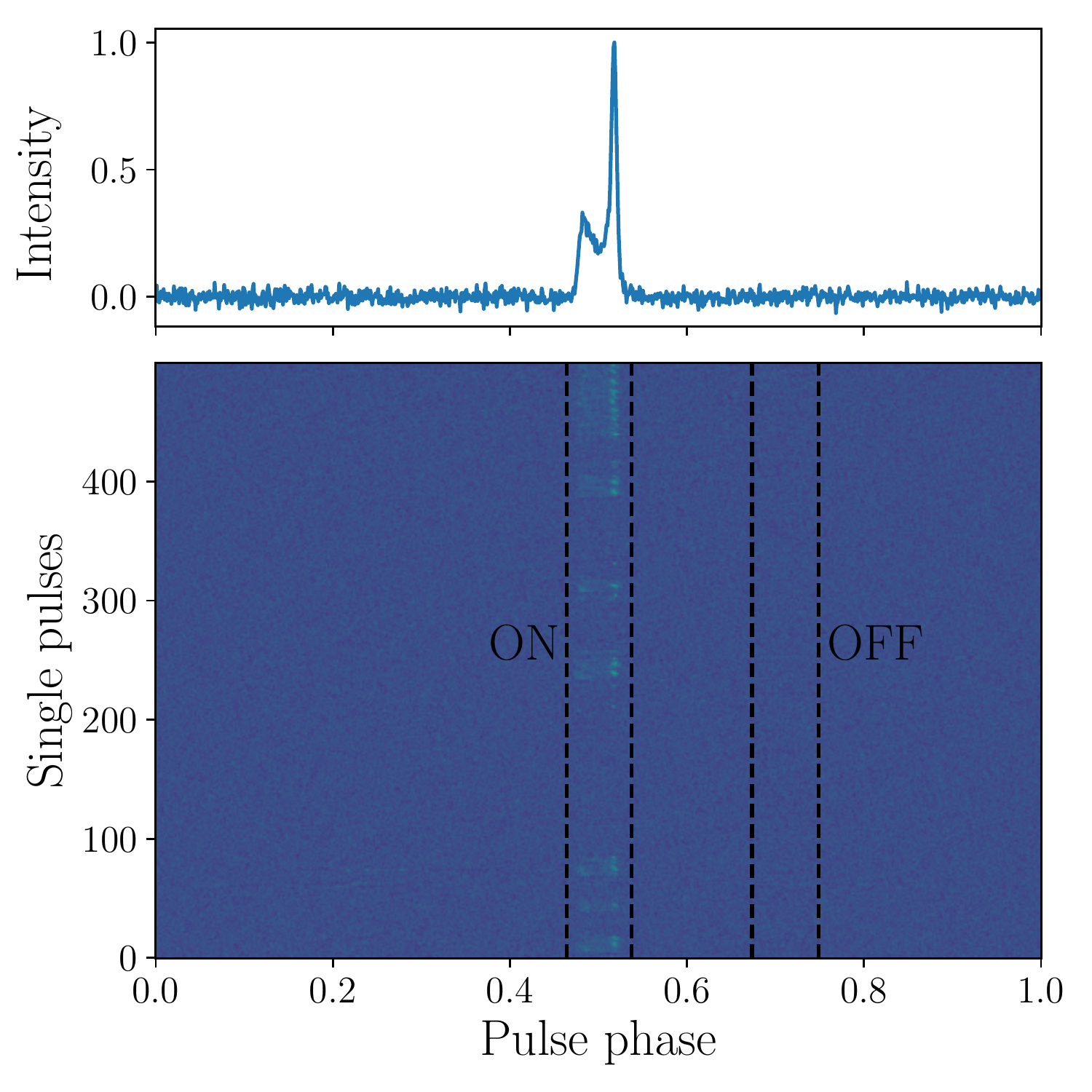}
      \includegraphics[width=0.5\textwidth]{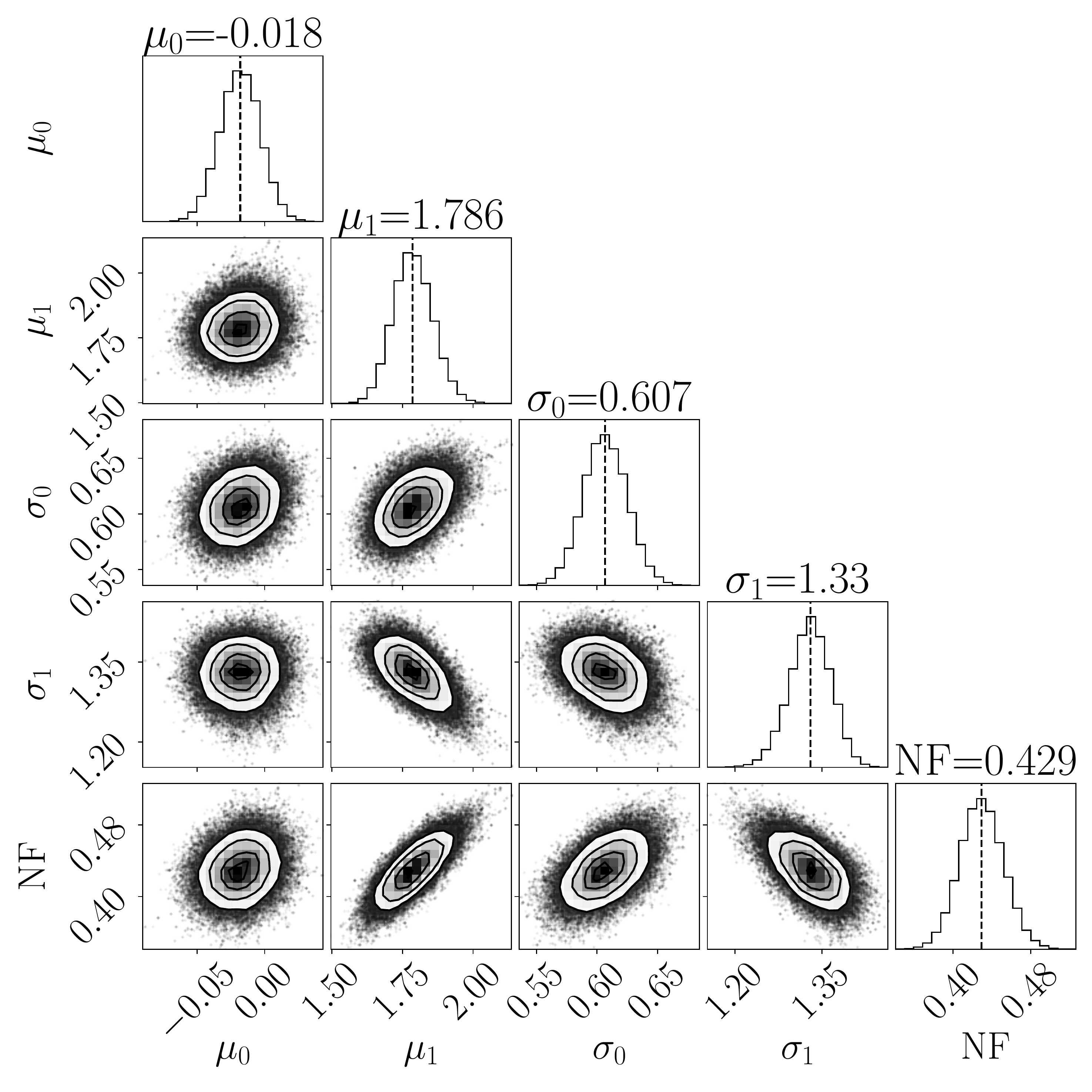}
      \centerline{\includegraphics[width=0.67\textwidth]{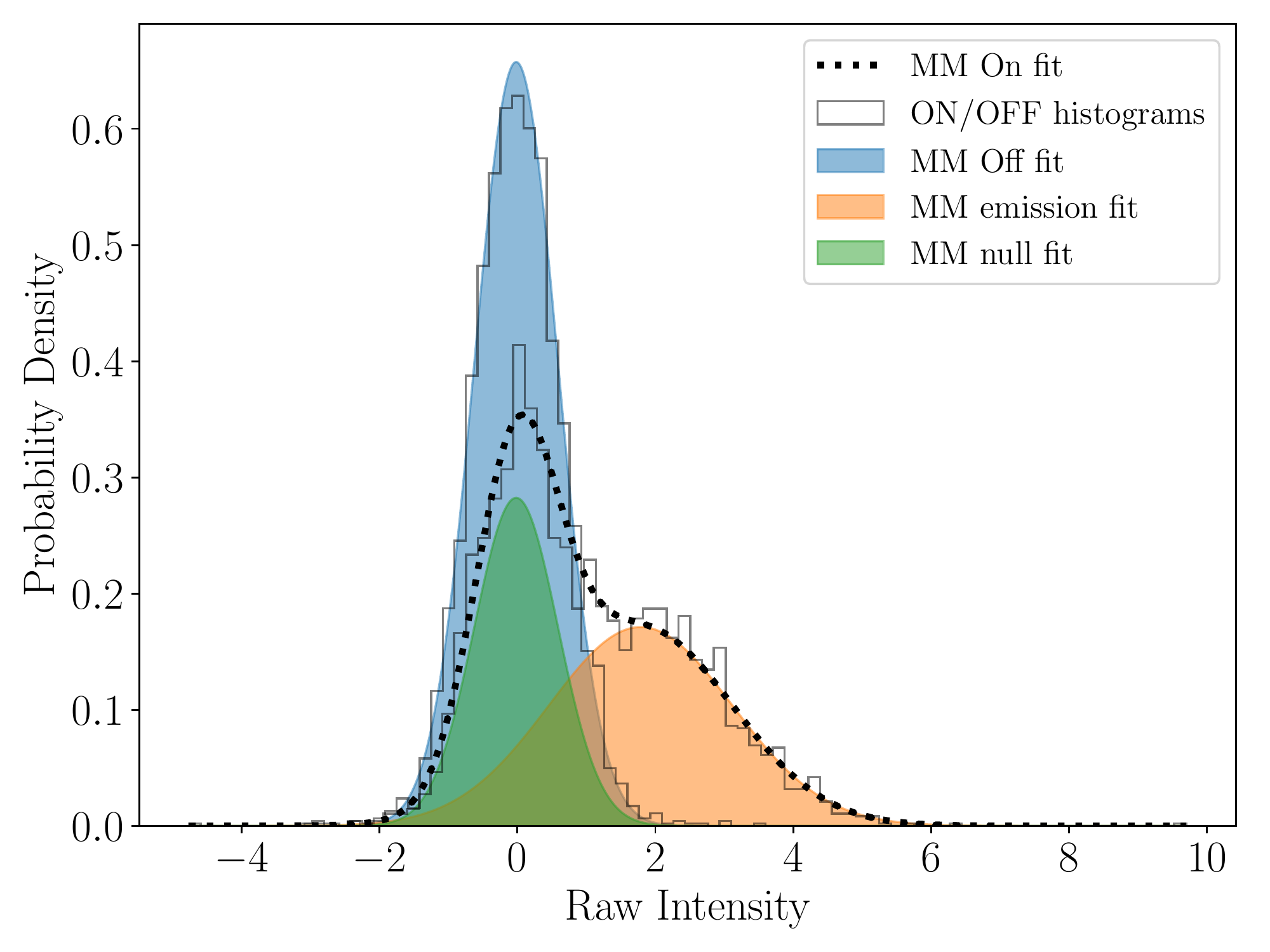}}          
      \caption{Single pulse stack (upper left), MCMC corner plot (bottom), and pulse intensity histogram (upper right) for PSR J1536-30. In this case the best fit model is a 2-component Gaussian mixture}
 \end{figure*}

\begin{figure*}
      \includegraphics[width=0.5\textwidth]{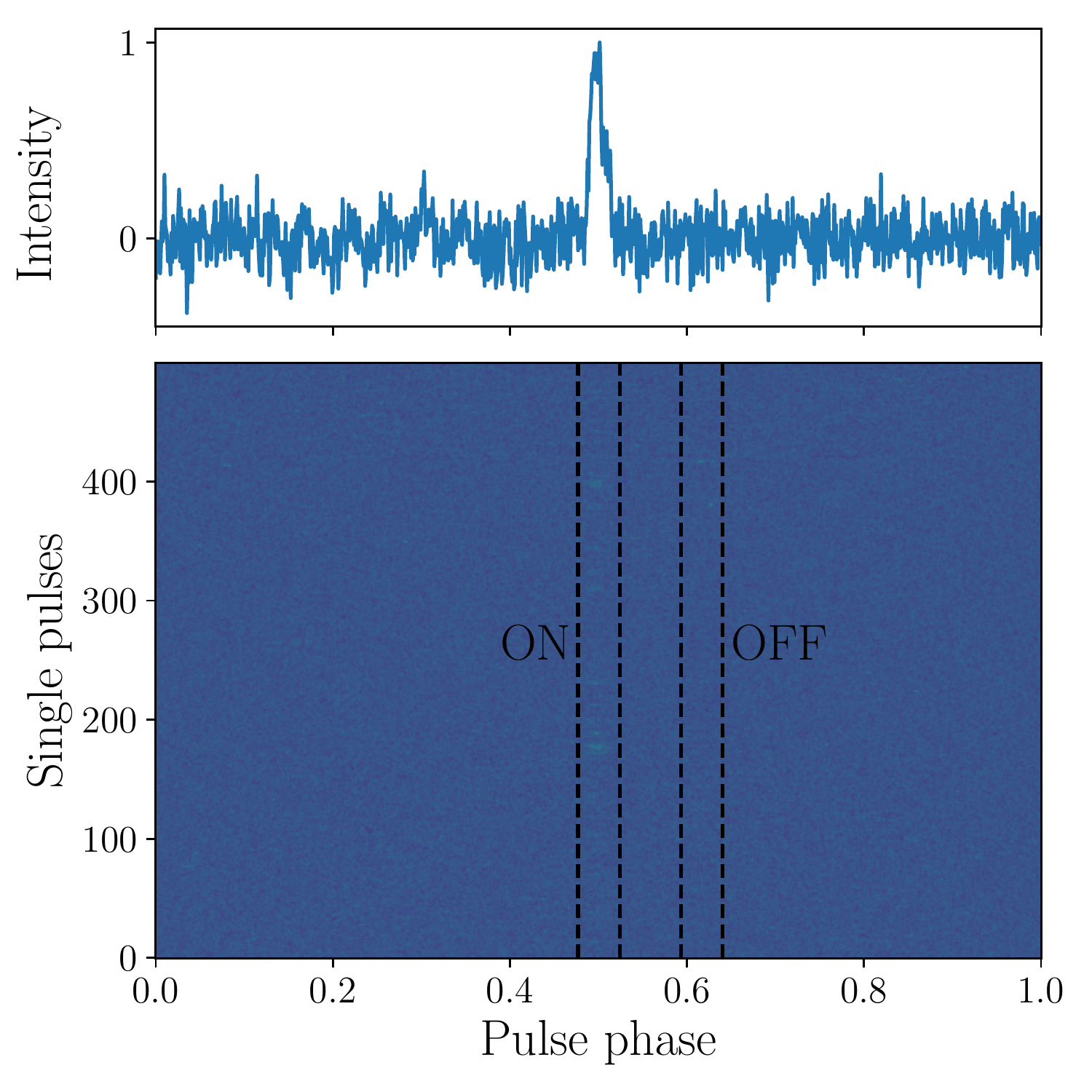}
      \includegraphics[width=0.5\textwidth]{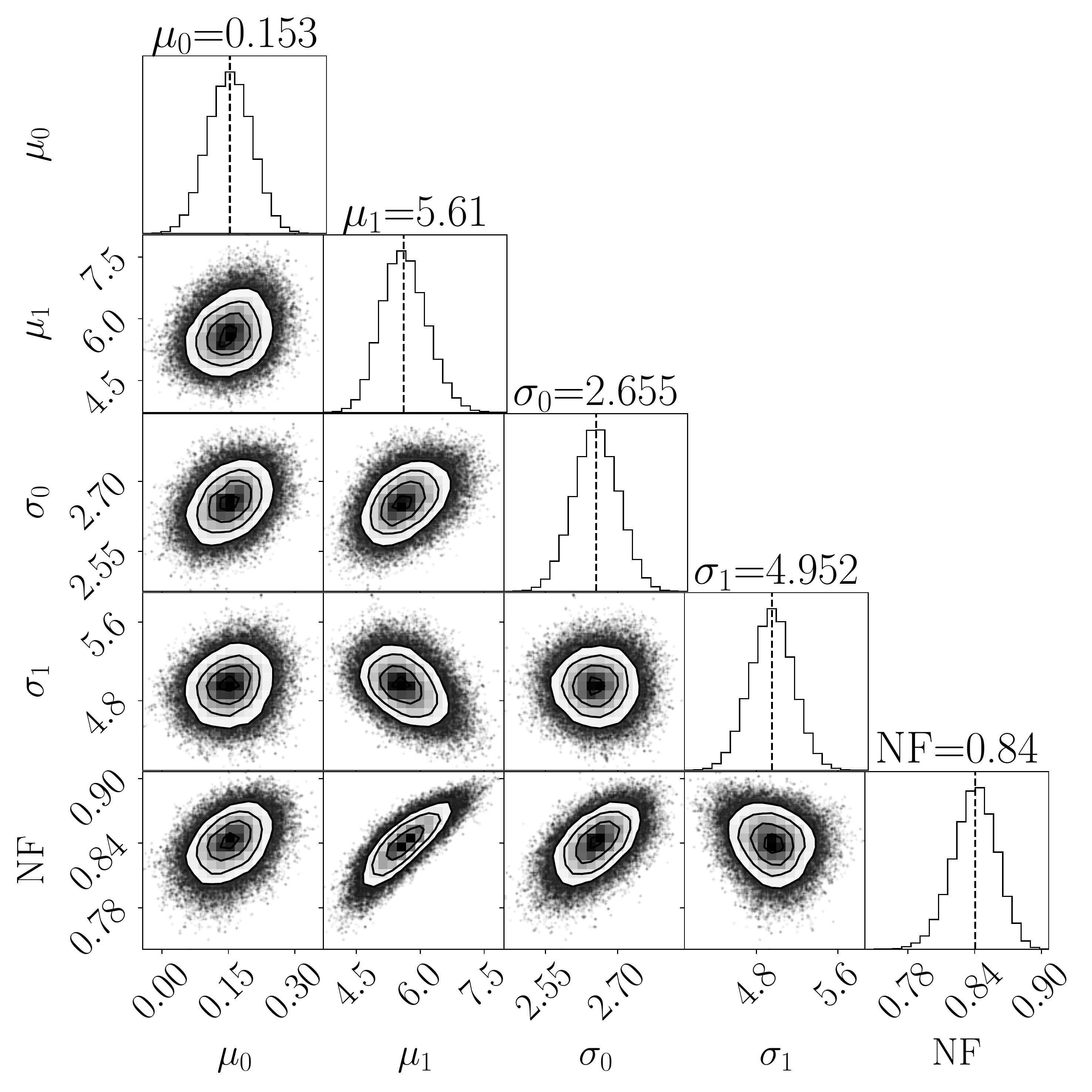}
      \centerline{\includegraphics[width=0.67\textwidth]{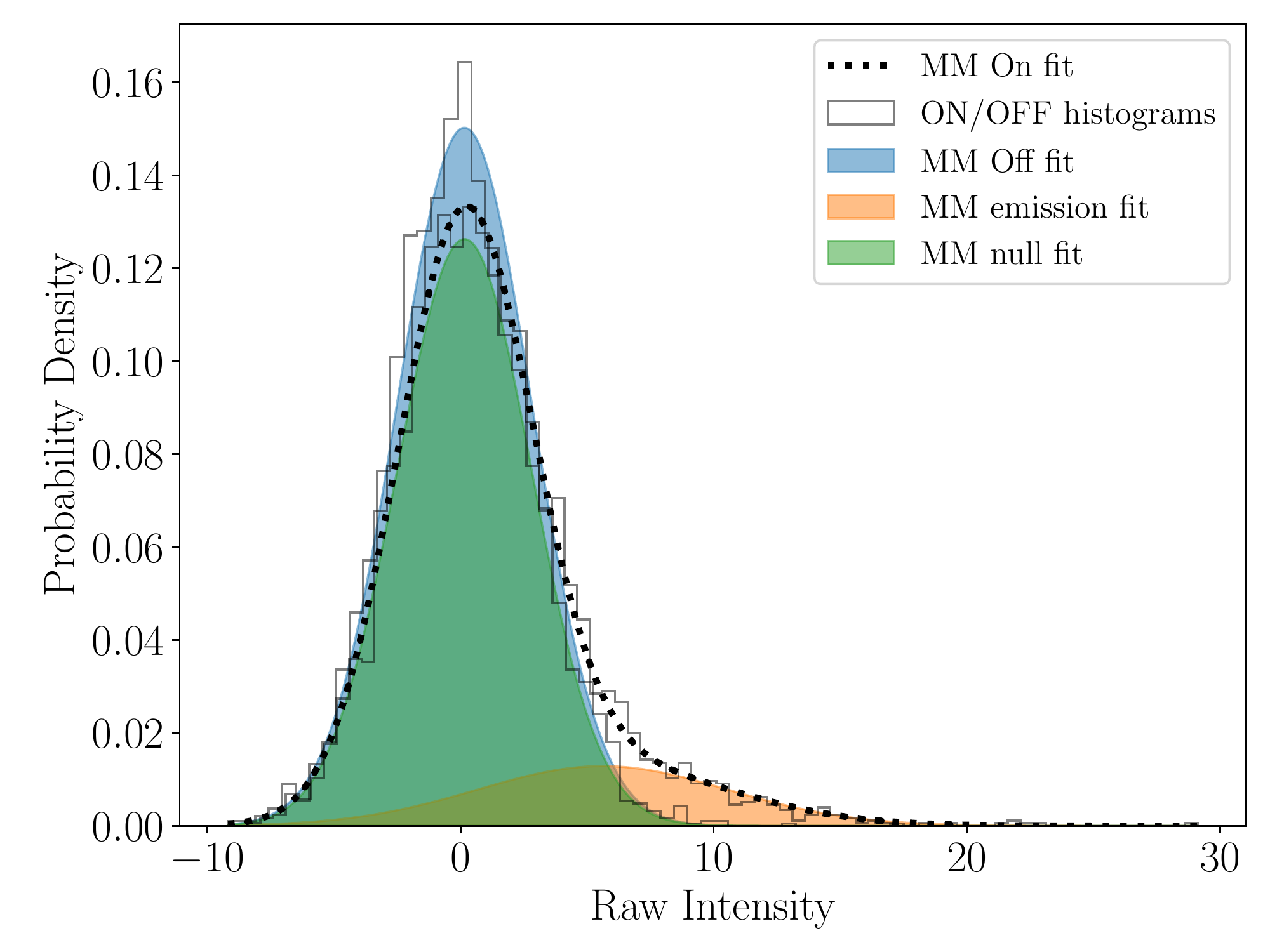}}          
      \caption{Single pulse stack (upper left), MCMC corner plot (bottom), and pulse intensity histogram (upper right) for PSR J1629+33. In this case the best fit model is a 2-component Gaussian mixture}
 \end{figure*}

\begin{figure*}
      \includegraphics[width=0.5\textwidth]{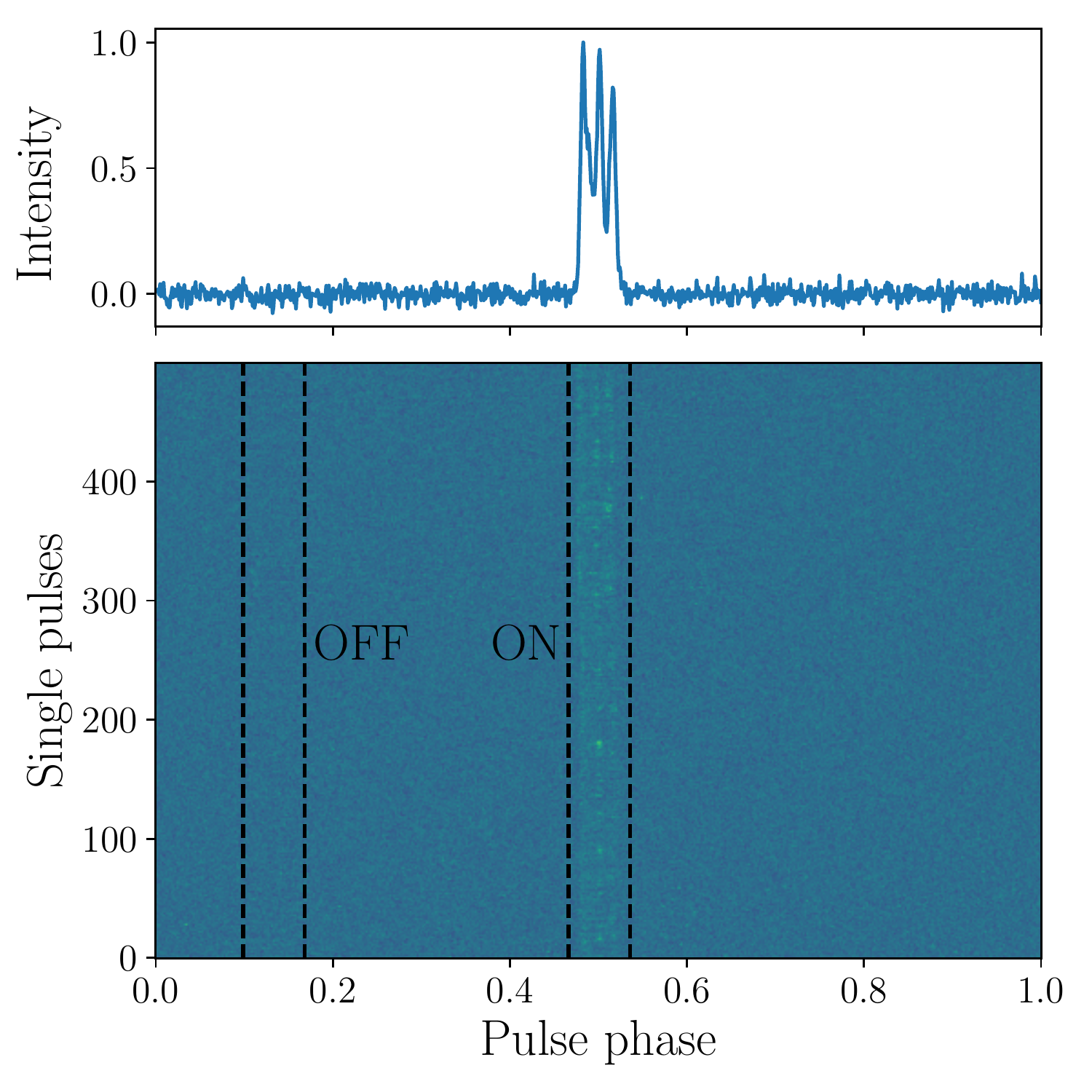}
      \includegraphics[width=0.5\textwidth]{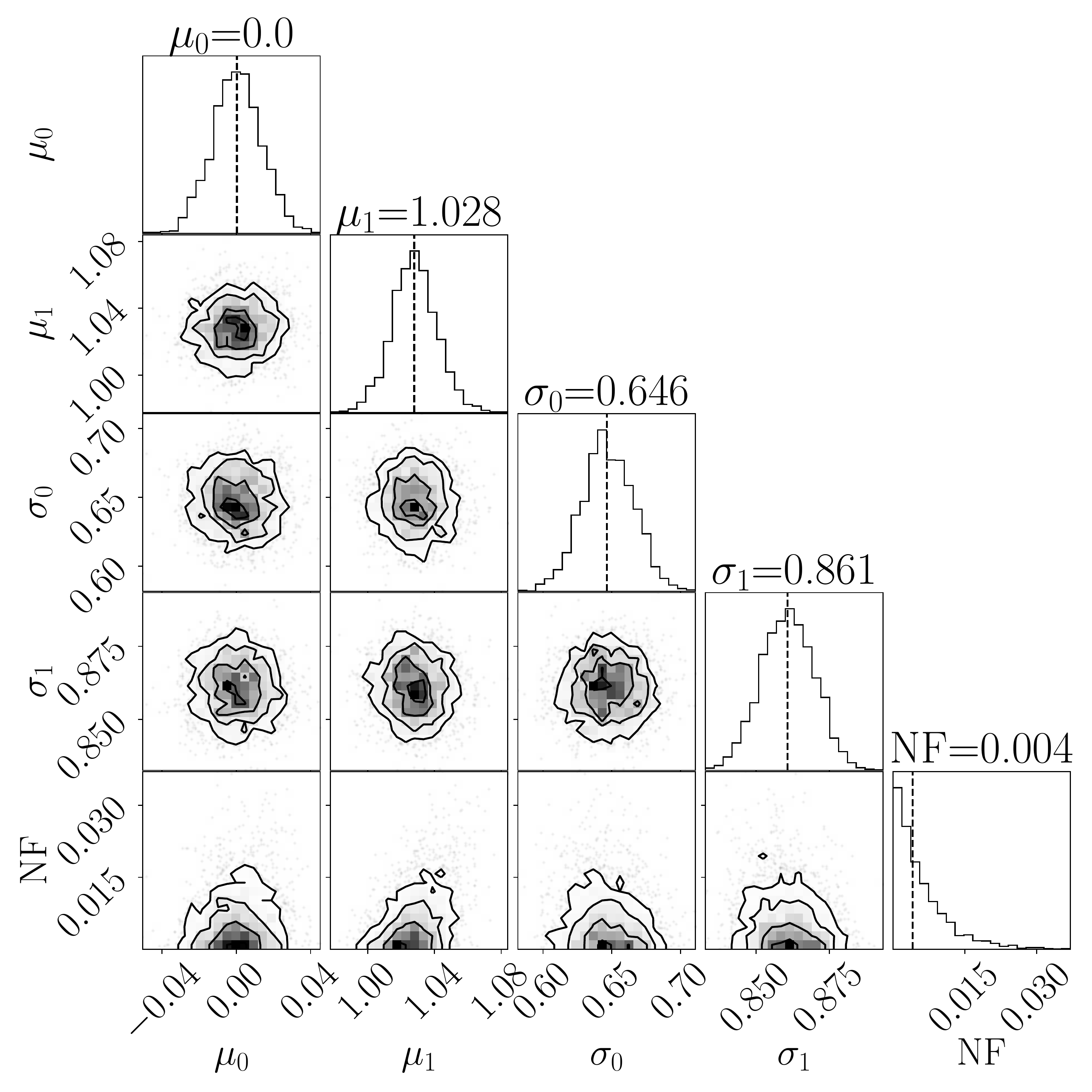}
      \centerline{\includegraphics[width=0.67\textwidth]{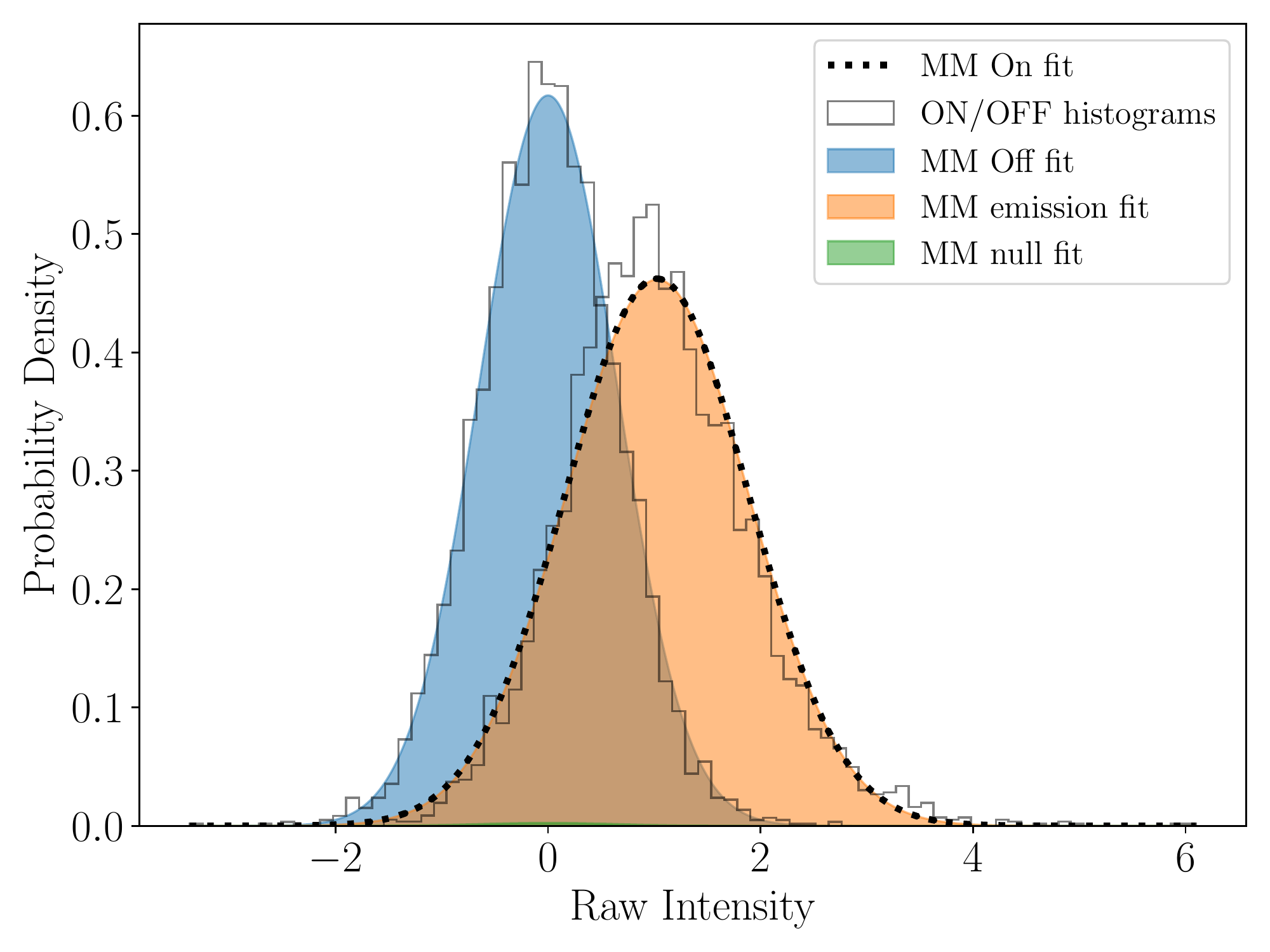}}          
      \caption{Single pulse stack (upper left), MCMC corner plot (bottom), and pulse intensity histogram (upper right) for PSR J1821+4147. In this case the best fit model is a 2-component Gaussian mixture}
 \end{figure*}

\begin{figure*}
      \includegraphics[width=0.5\textwidth]{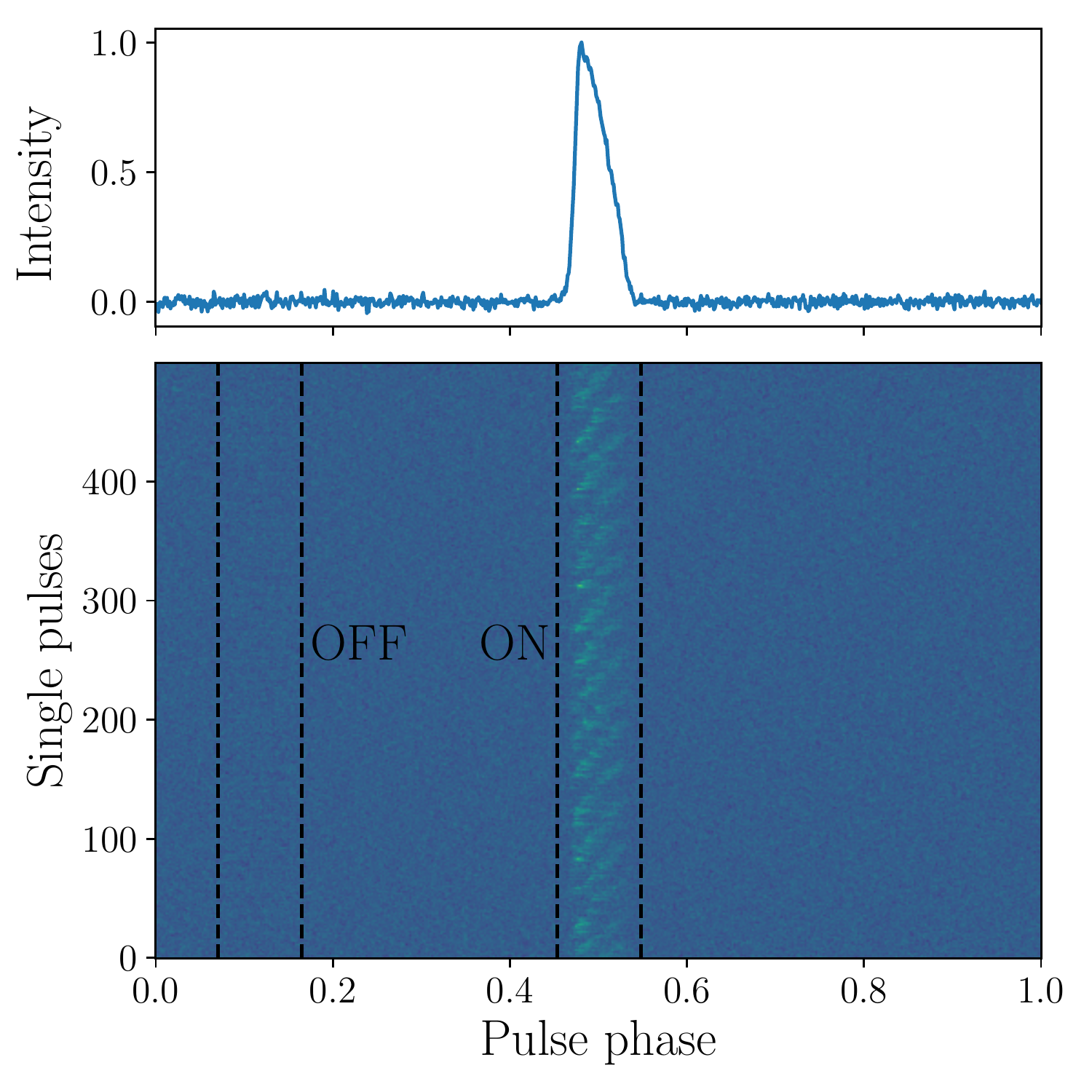}
      \includegraphics[width=0.5\textwidth]{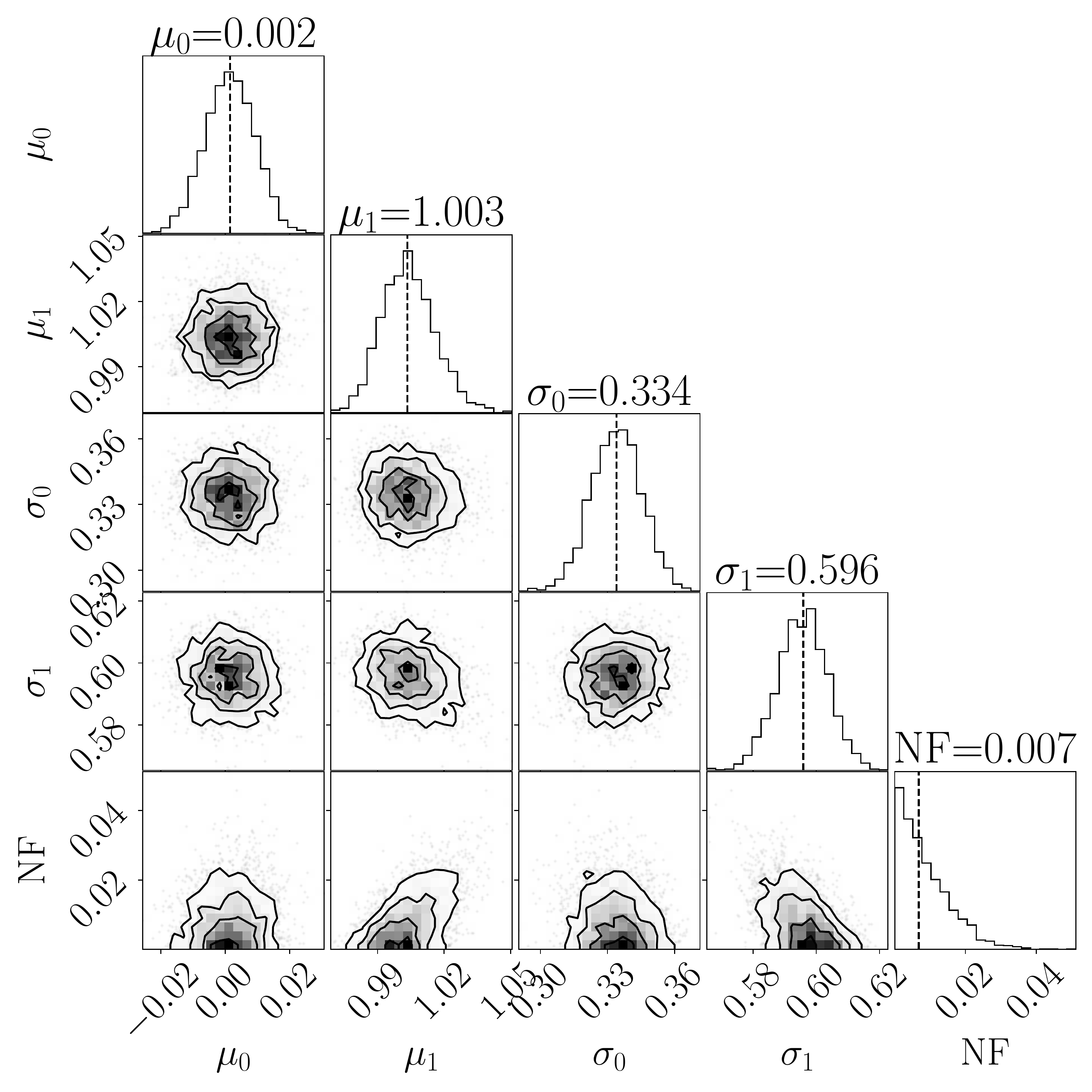}
      \centerline{\includegraphics[width=0.67\textwidth]{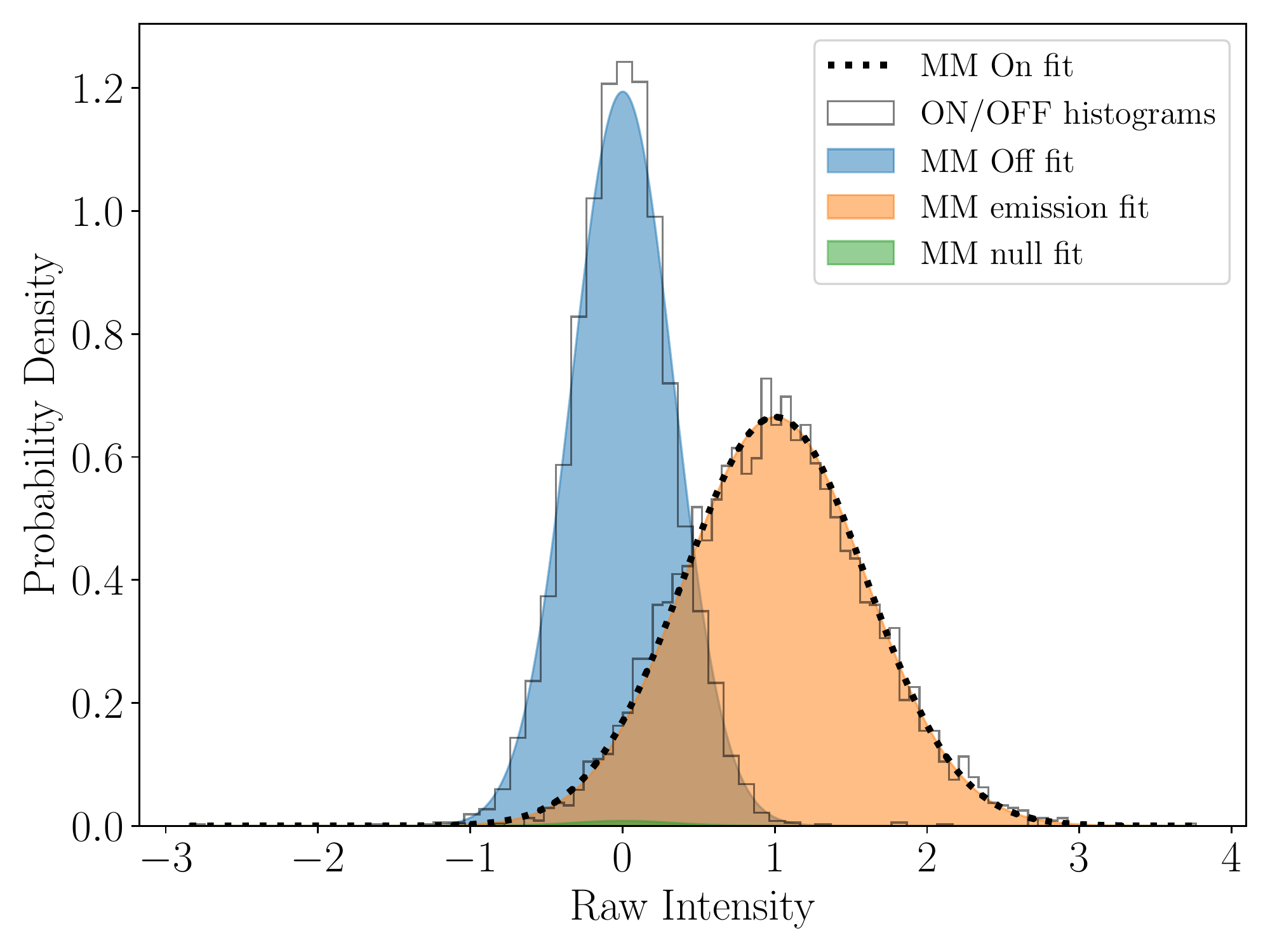}}          
      \caption{Single pulse stack (upper left), MCMC corner plot (bottom), and pulse intensity histogram (upper right) for PSR J1822+02. In this case the best fit model is a 2-component Gaussian mixture}
 \end{figure*}

\begin{figure*}
      \includegraphics[width=0.5\textwidth]{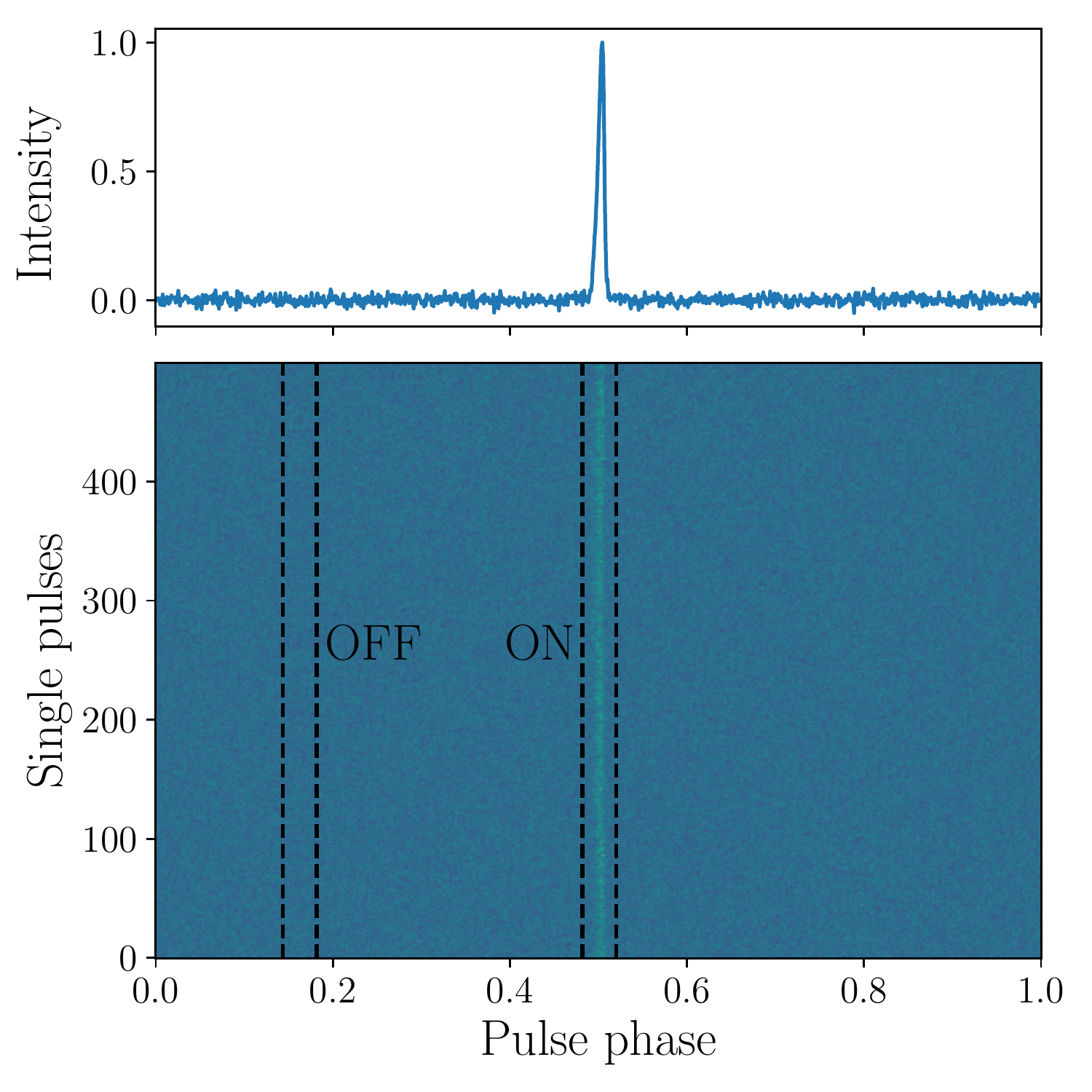}
      \includegraphics[width=0.5\textwidth]{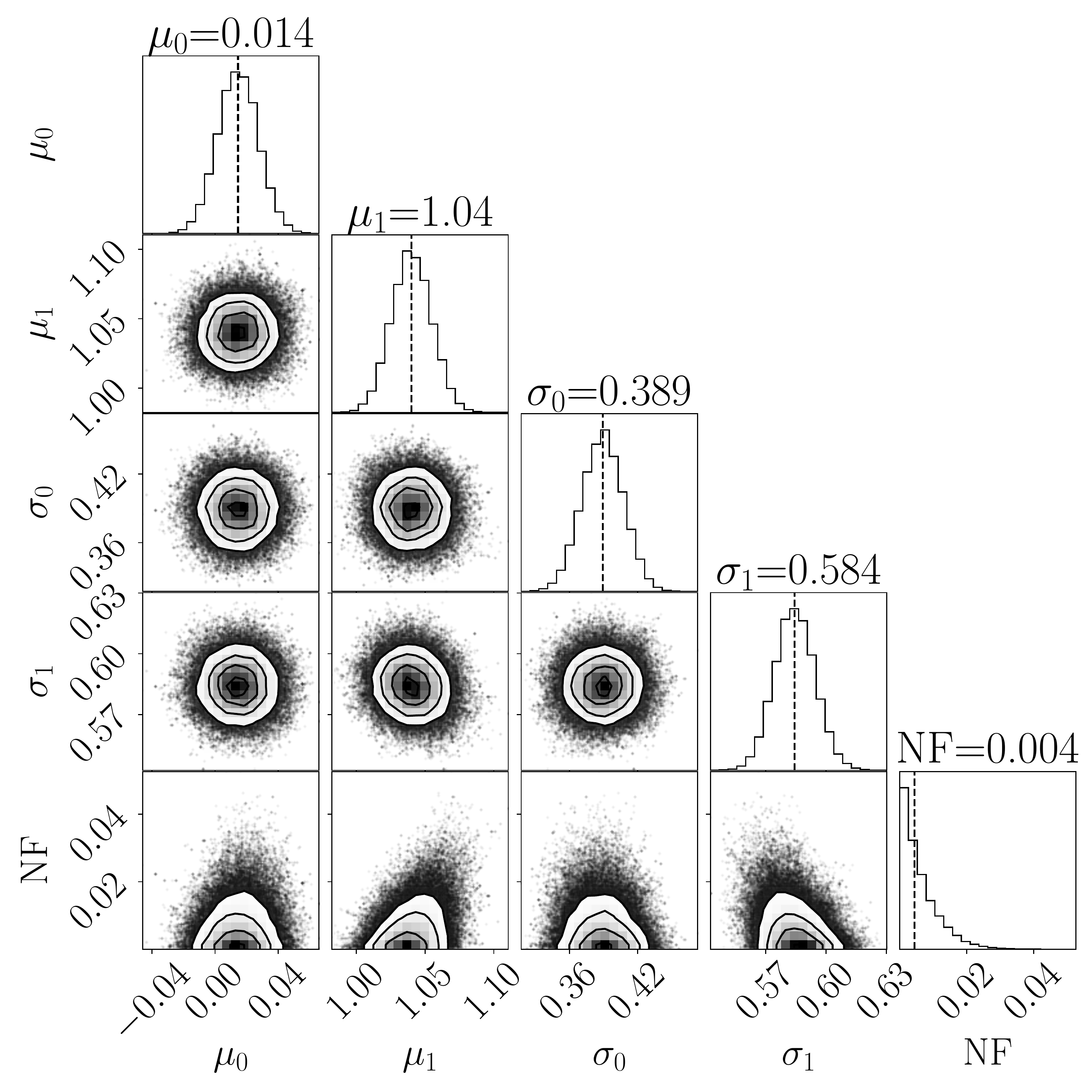}
      \centerline{\includegraphics[width=0.67\textwidth]{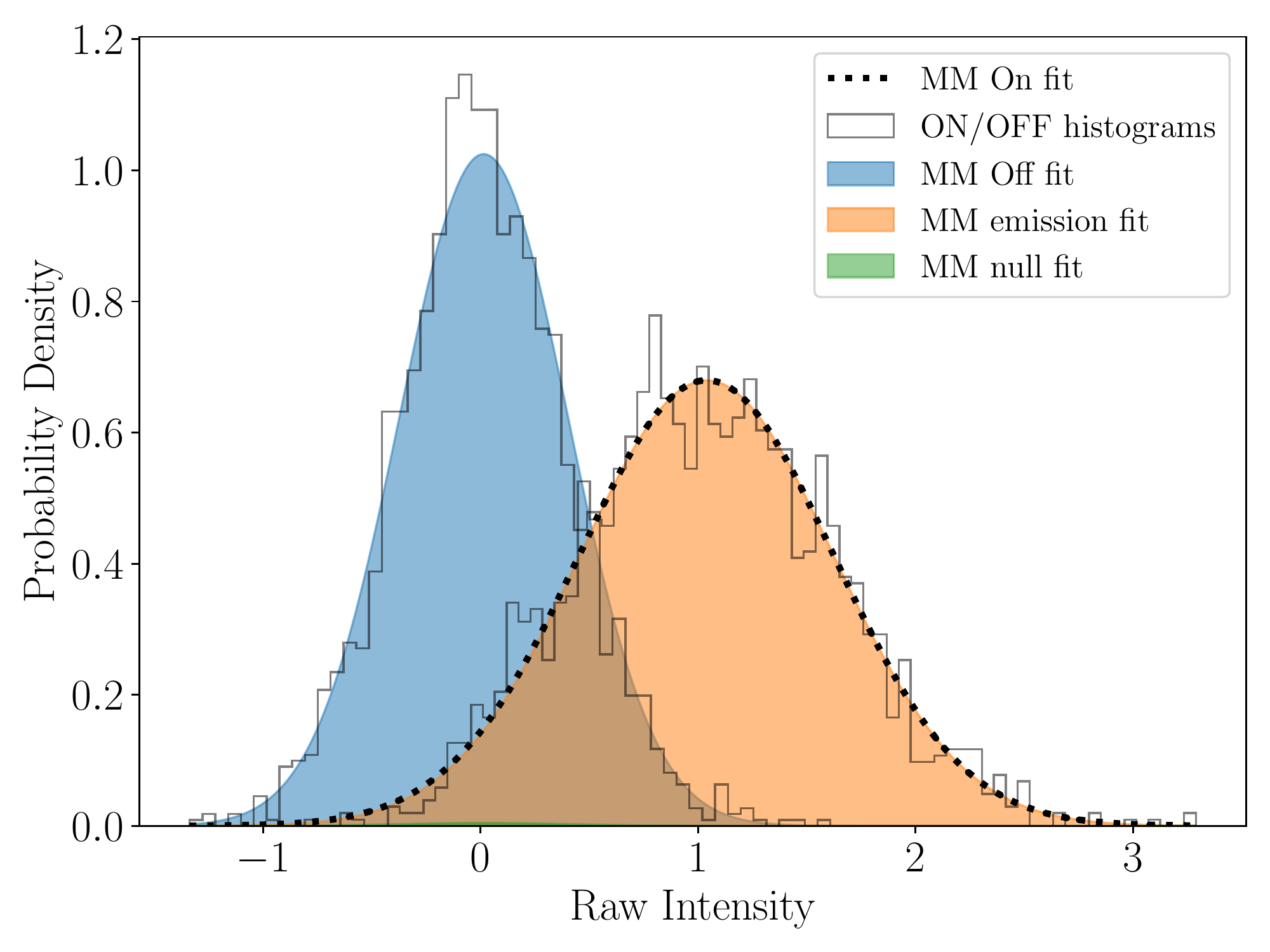}}          
      \caption{Single pulse stack (upper left), MCMC corner plot (bottom), and pulse intensity histogram (upper right) for PSR J1829+25 (GBT). In this case the best fit model is a 2-component Gaussian mixture}
 \end{figure*}

\begin{figure*}
      \includegraphics[width=0.5\textwidth]{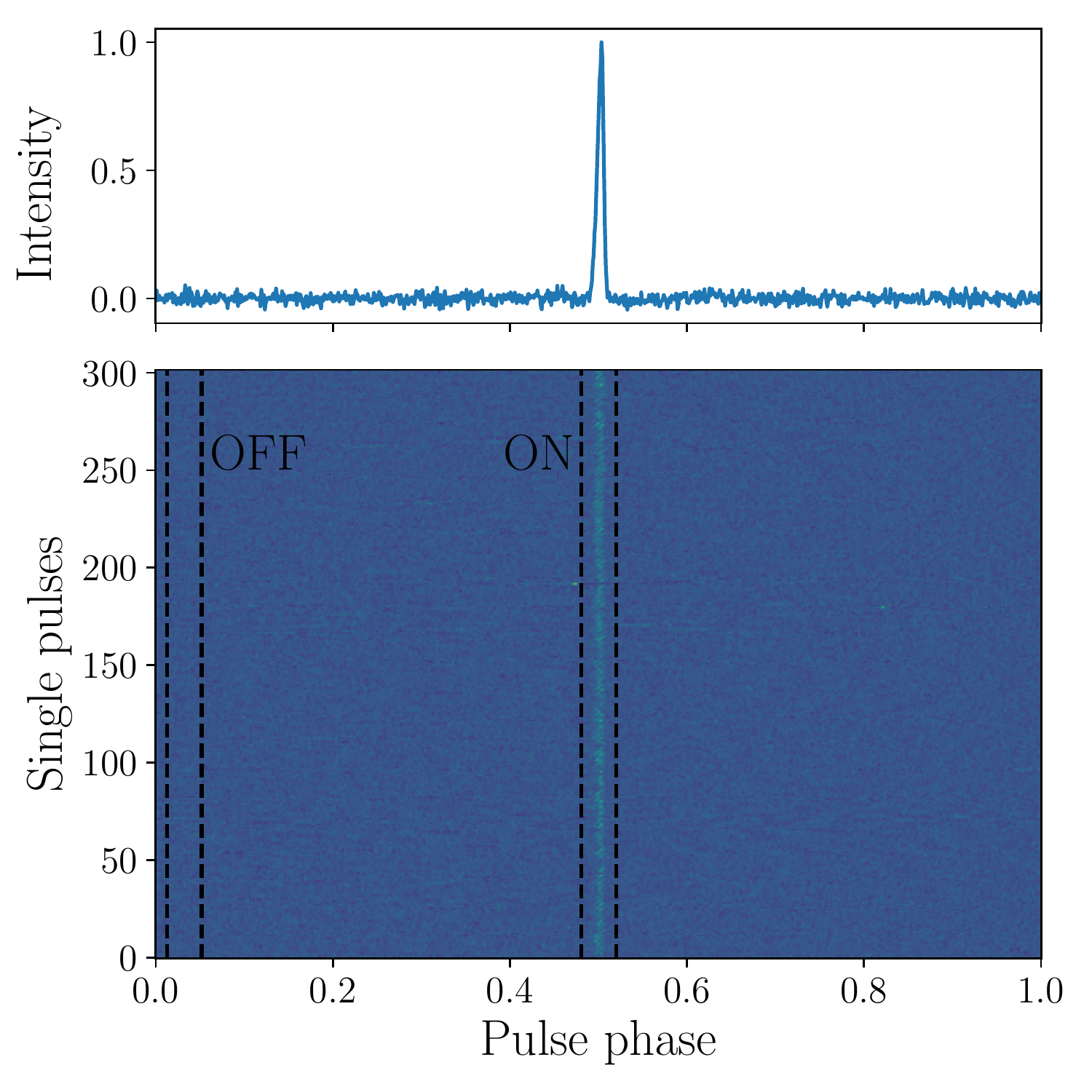}
      \includegraphics[width=0.5\textwidth]{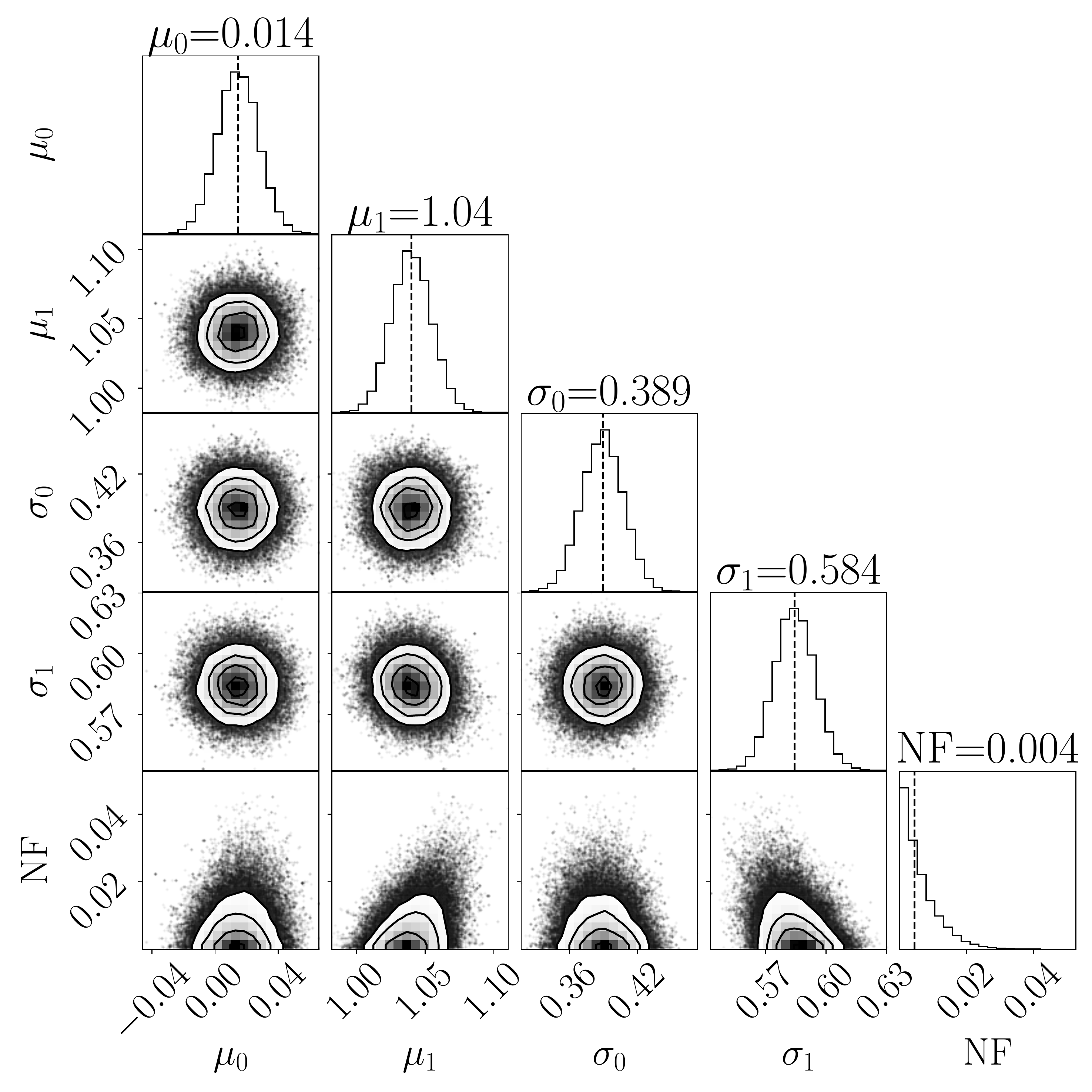}
      \centerline{\includegraphics[width=0.67\textwidth]{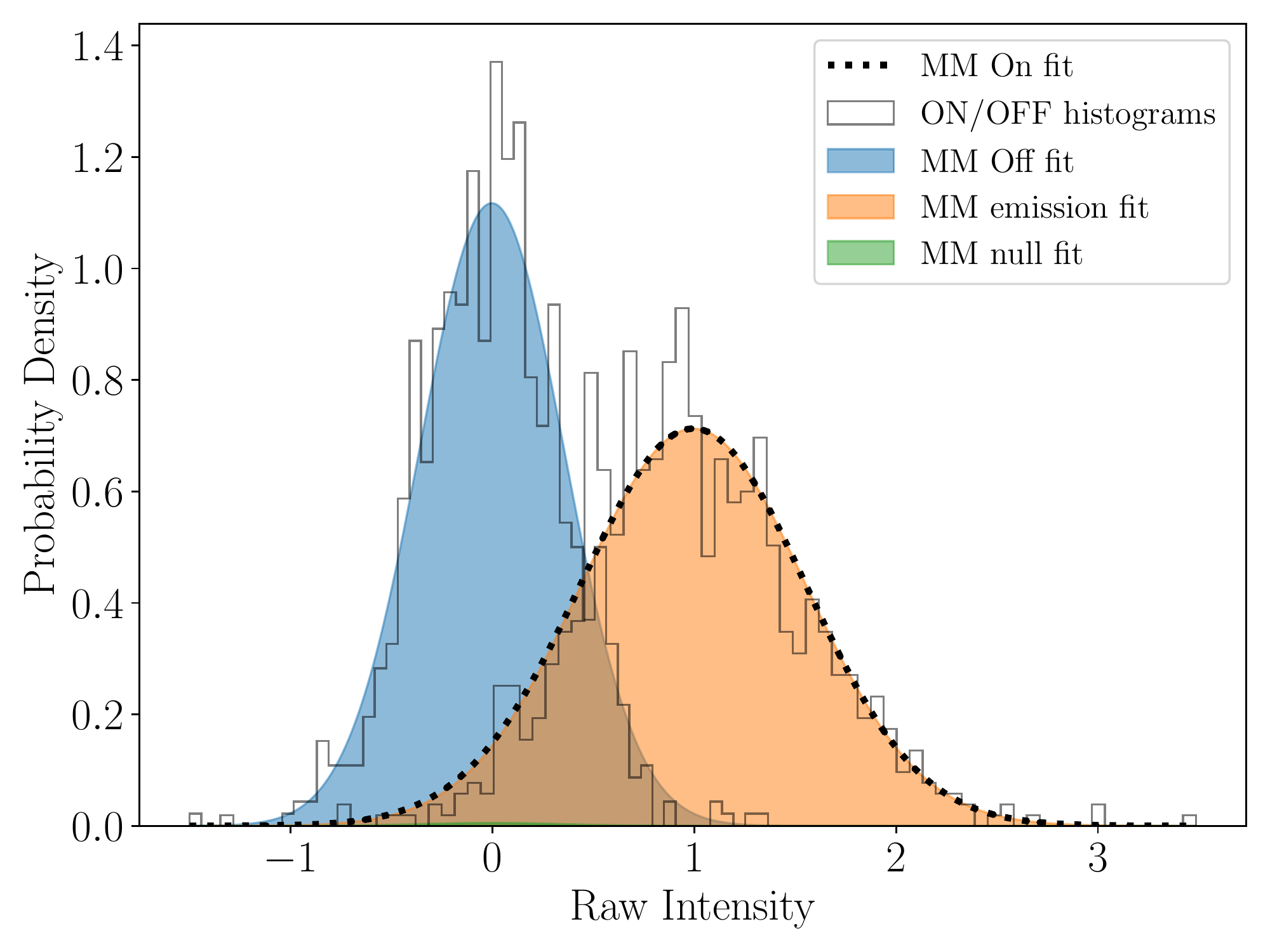}}          
      \caption{Single pulse stack (upper left), MCMC corner plot (bottom), and pulse intensity histogram (upper right) for PSR J1829+25 (AO). In this case the best fit model is a 2-component Gaussian mixture}
 \end{figure*}

\begin{figure*}
      \includegraphics[width=0.5\textwidth]{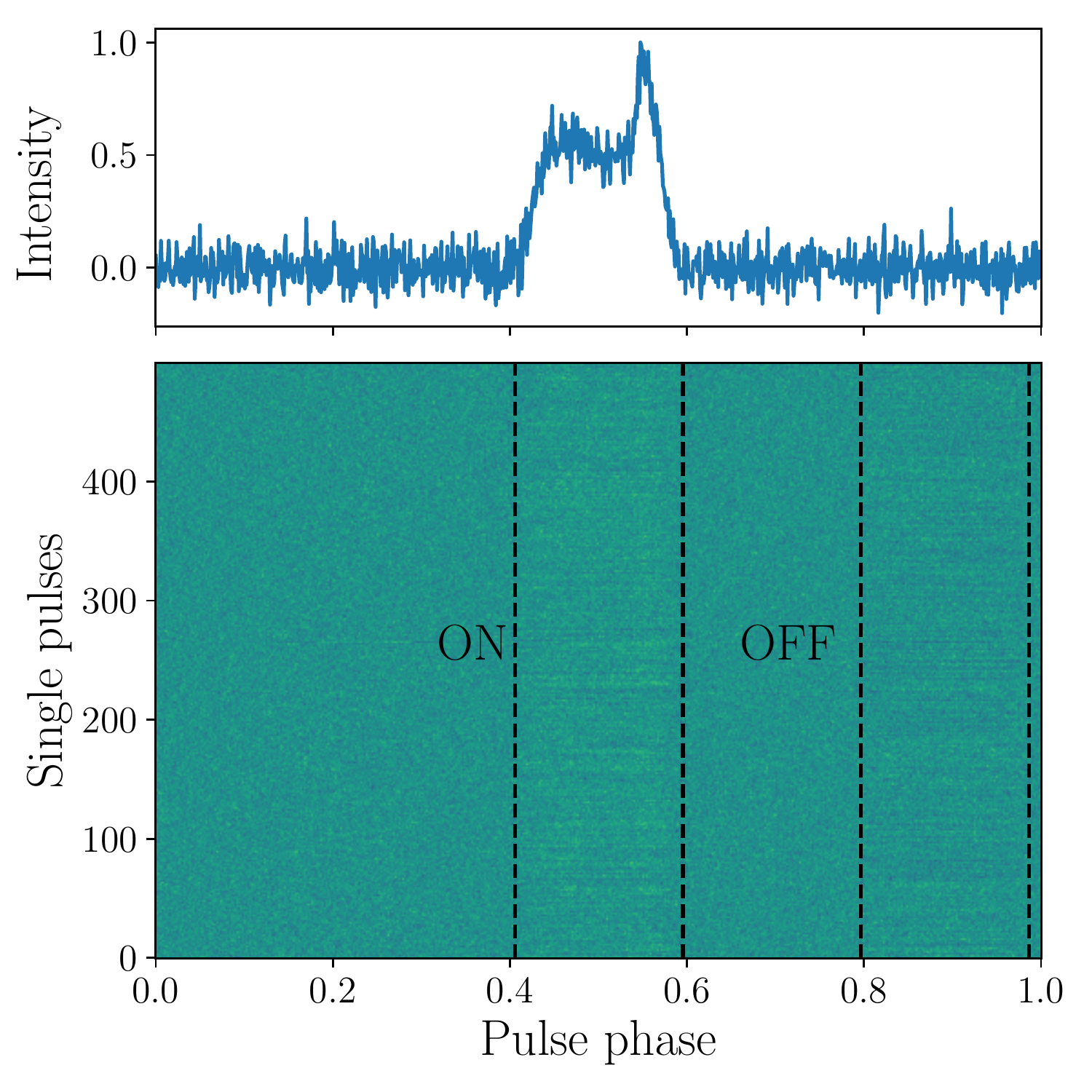}
      \includegraphics[width=0.5\textwidth]{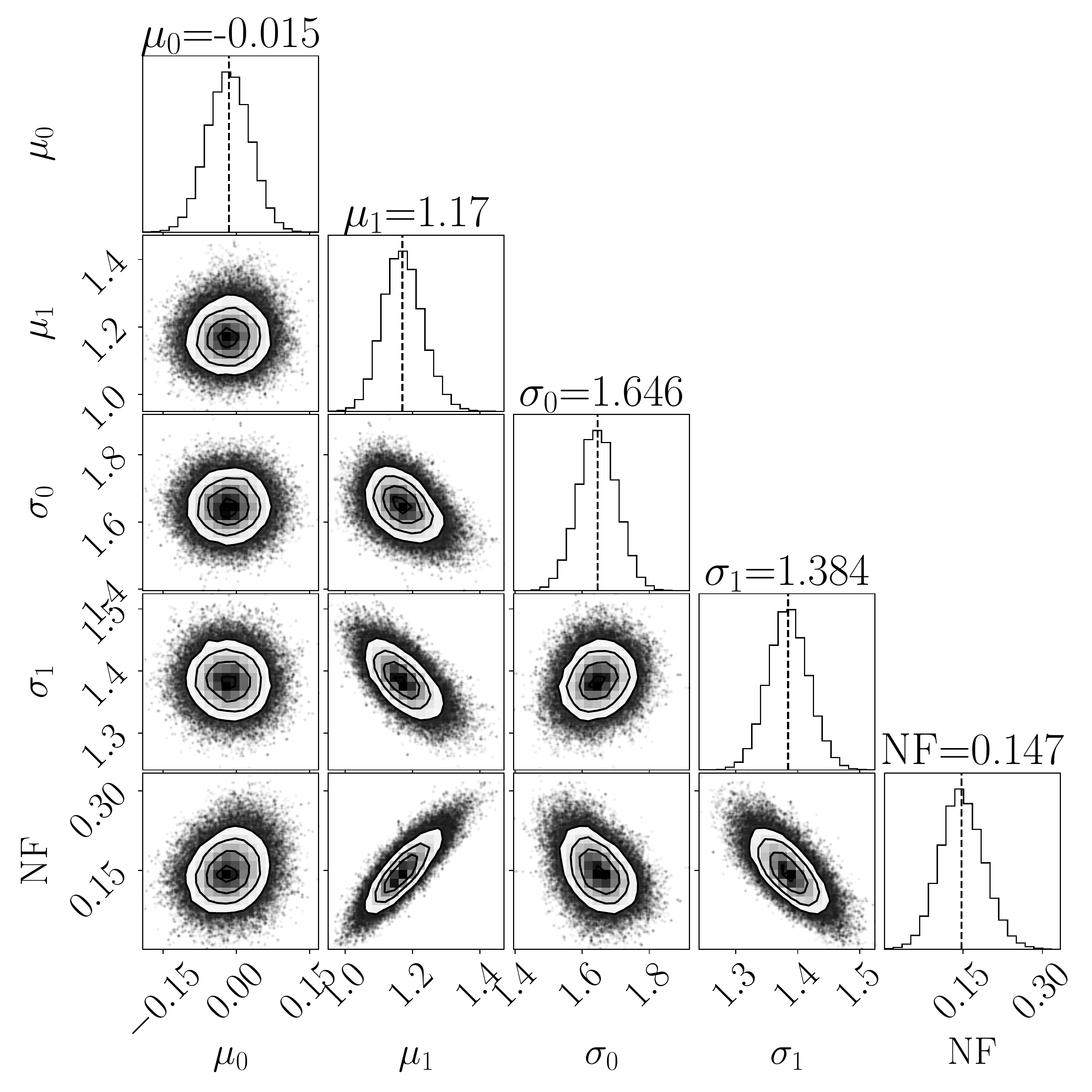}
      \centerline{\includegraphics[width=0.67\textwidth]{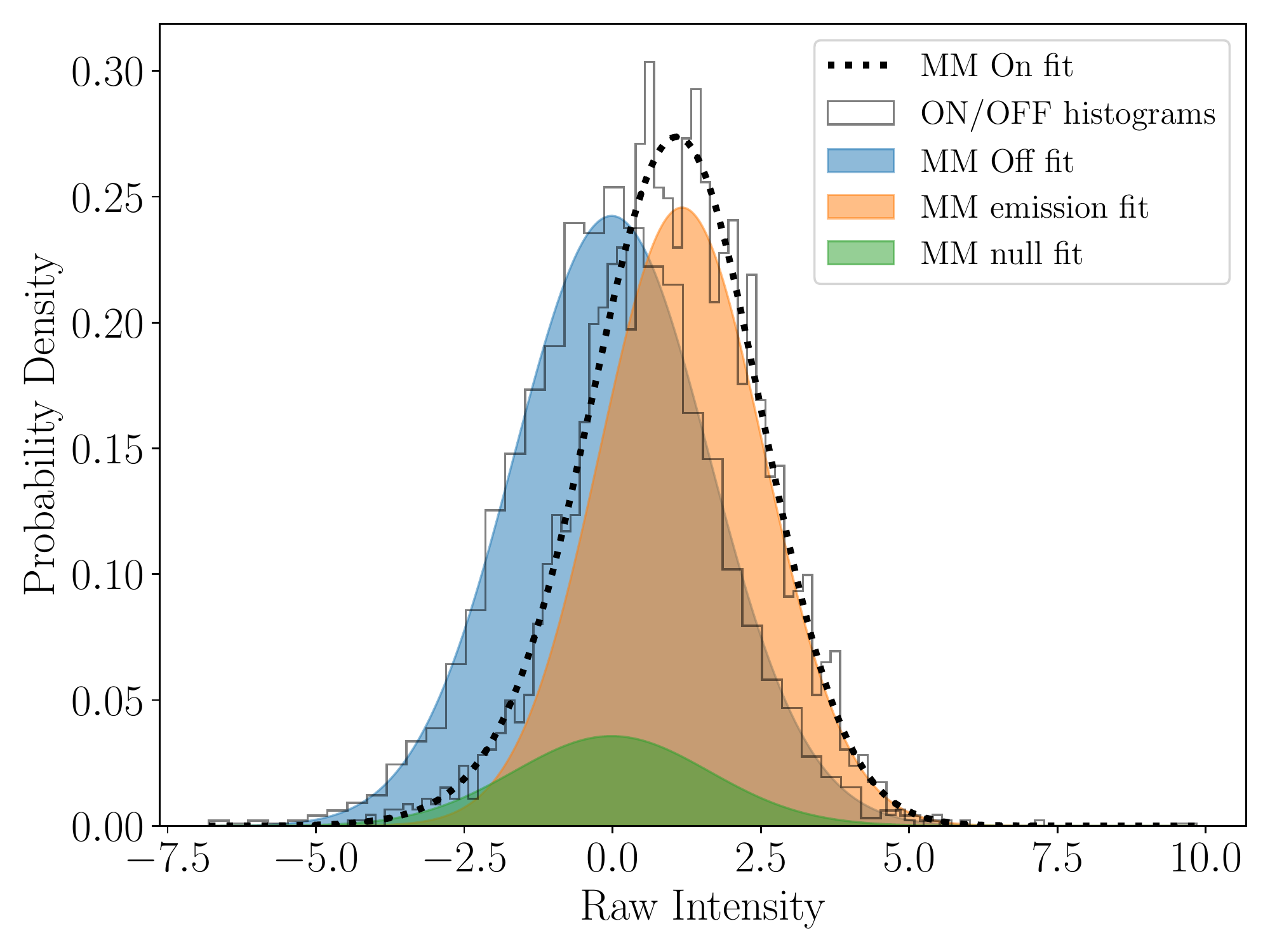}}          
      \caption{Single pulse stack (upper left), MCMC corner plot (bottom), and pulse intensity histogram (upper right) for PSR J1901-04. In this case the best fit model is a 2-component Gaussian mixture}
 \end{figure*}

\begin{figure*}
      \includegraphics[width=0.5\textwidth]{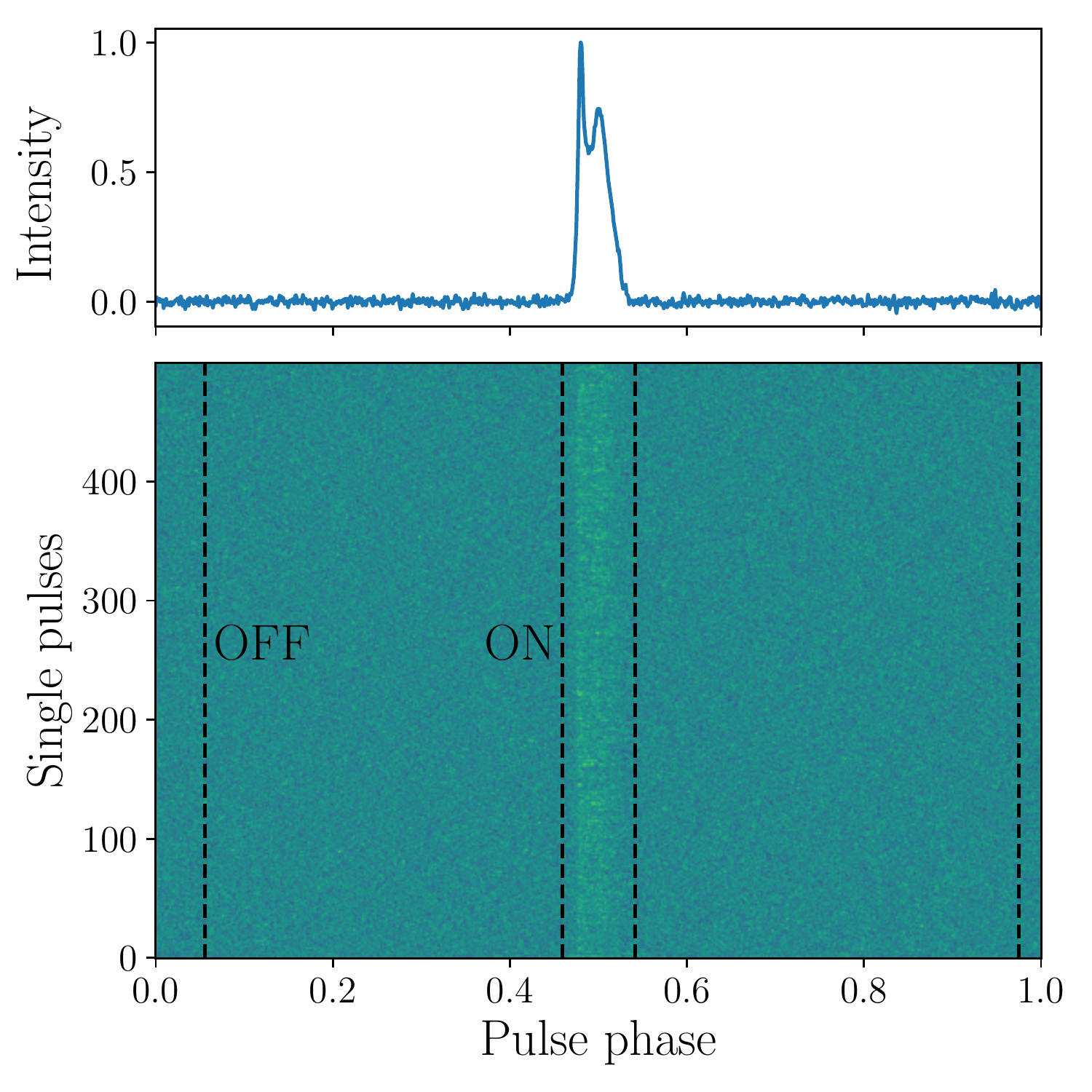}
      \includegraphics[width=0.5\textwidth]{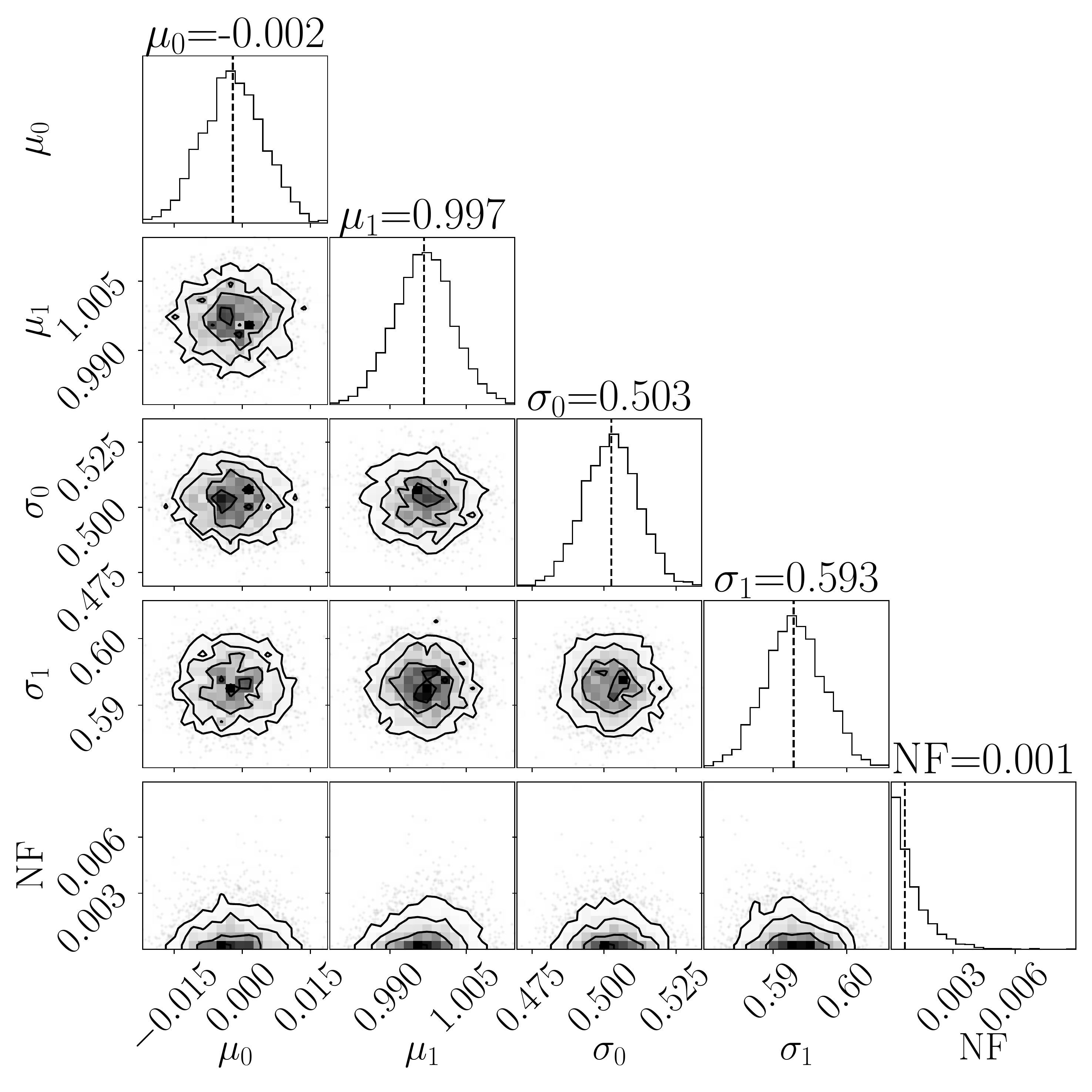}
      \centerline{\includegraphics[width=0.67\textwidth]{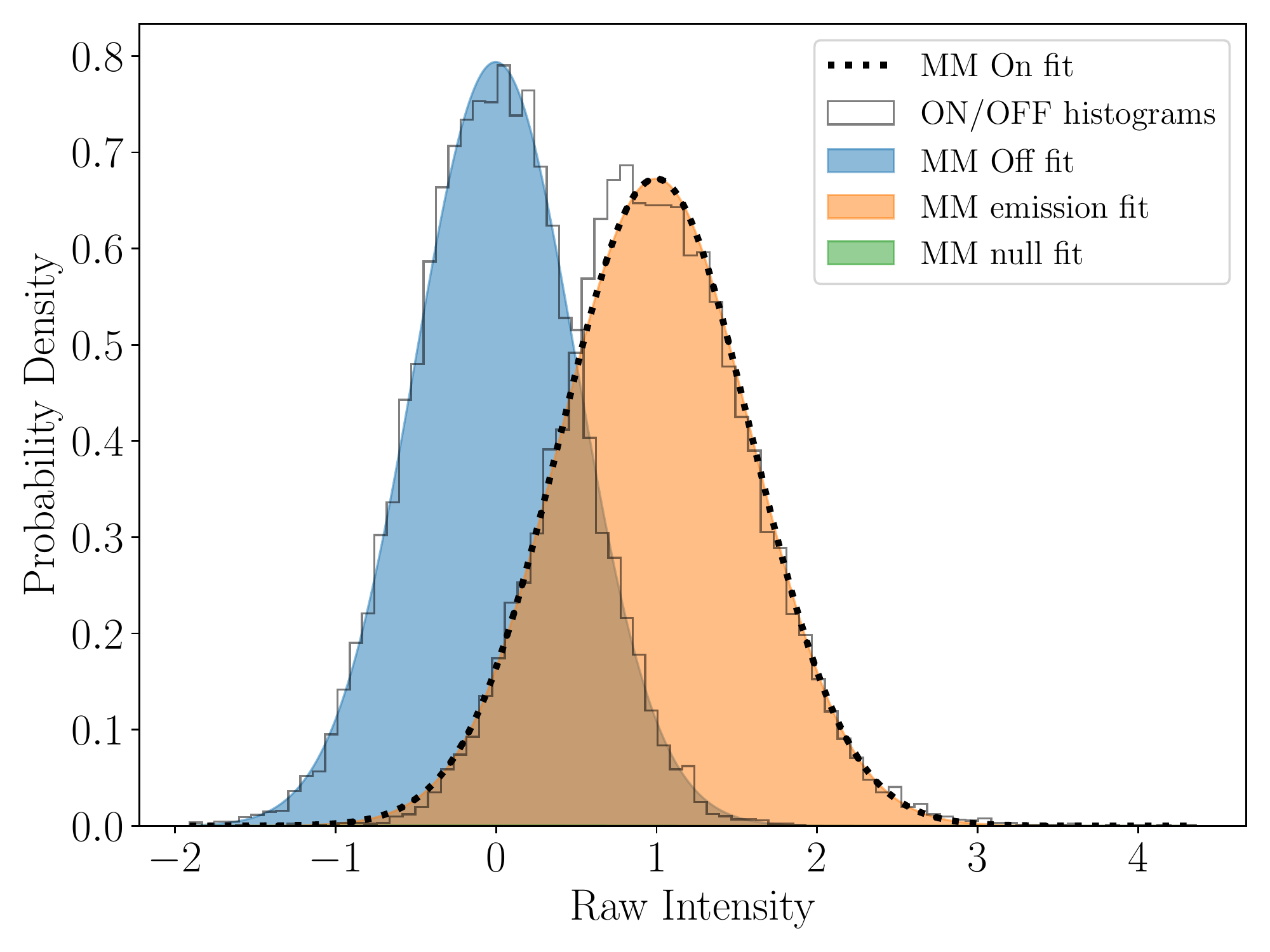}}          
      \caption{Single pulse stack (upper left), MCMC corner plot (bottom), and pulse intensity histogram (upper right) for PSR J1904+33. In this case the best fit model is a 2-component Gaussian mixture}
 \end{figure*}

\begin{figure*}
      \includegraphics[width=0.5\textwidth]{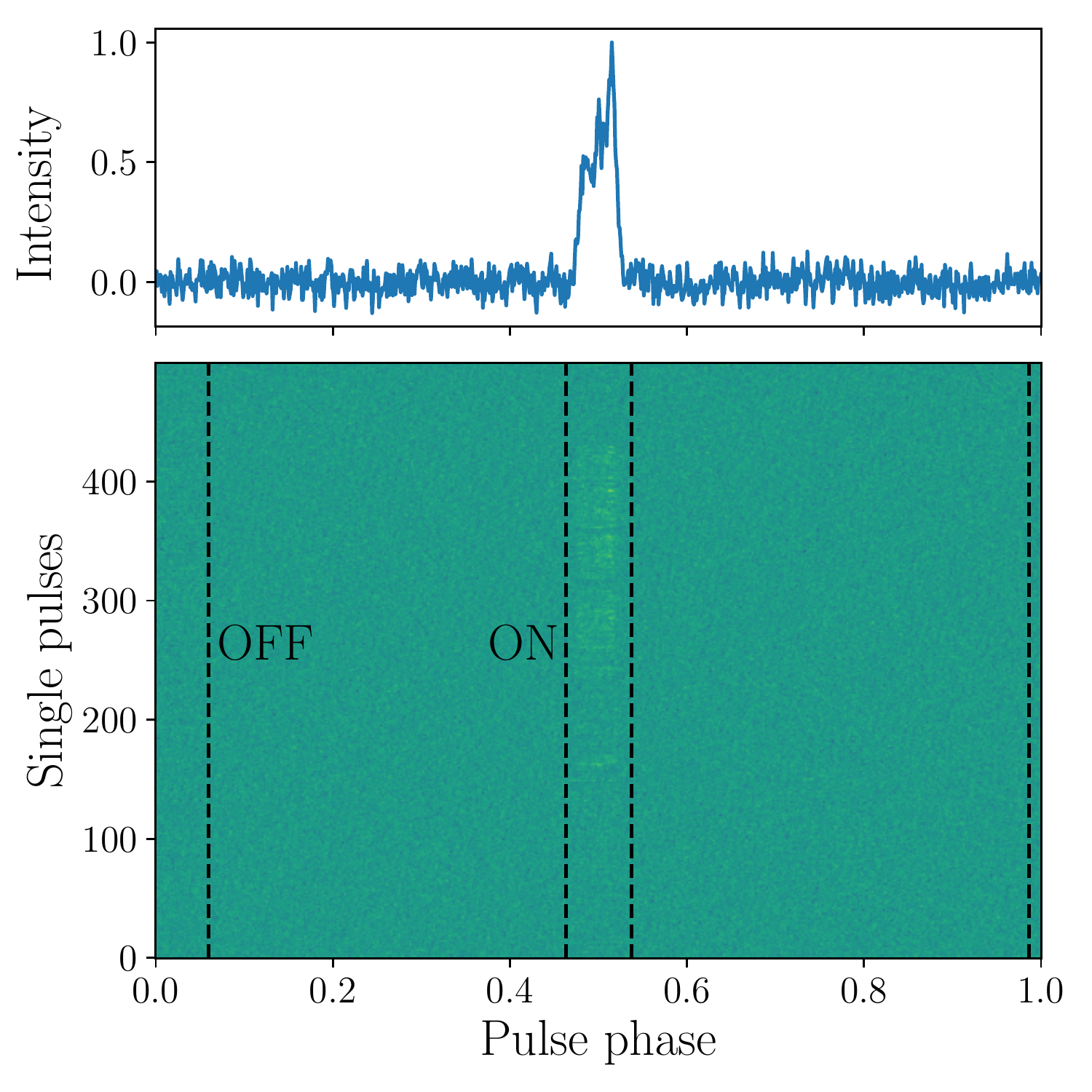}
      \includegraphics[width=0.5\textwidth]{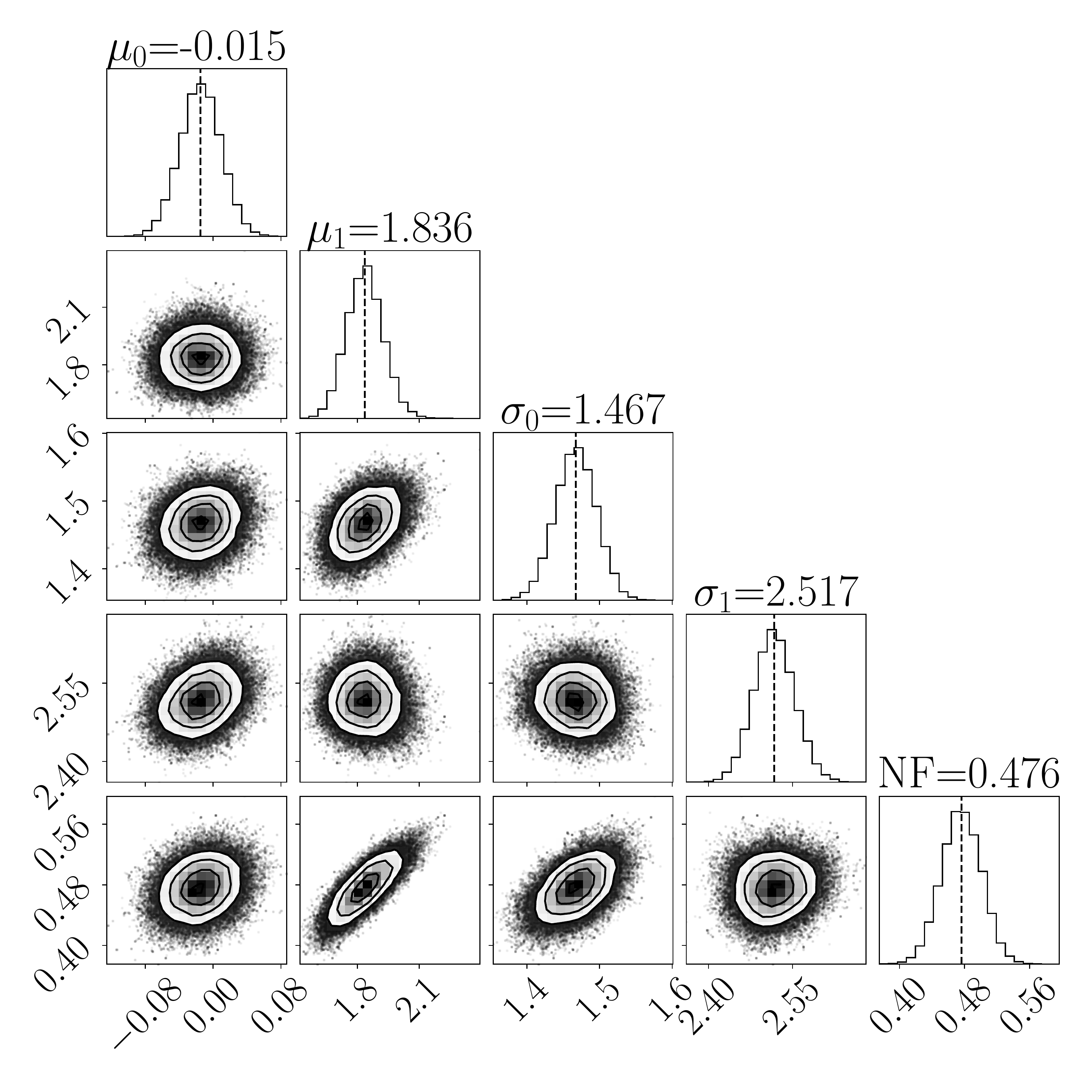}
      \centerline{\includegraphics[width=0.67\textwidth]{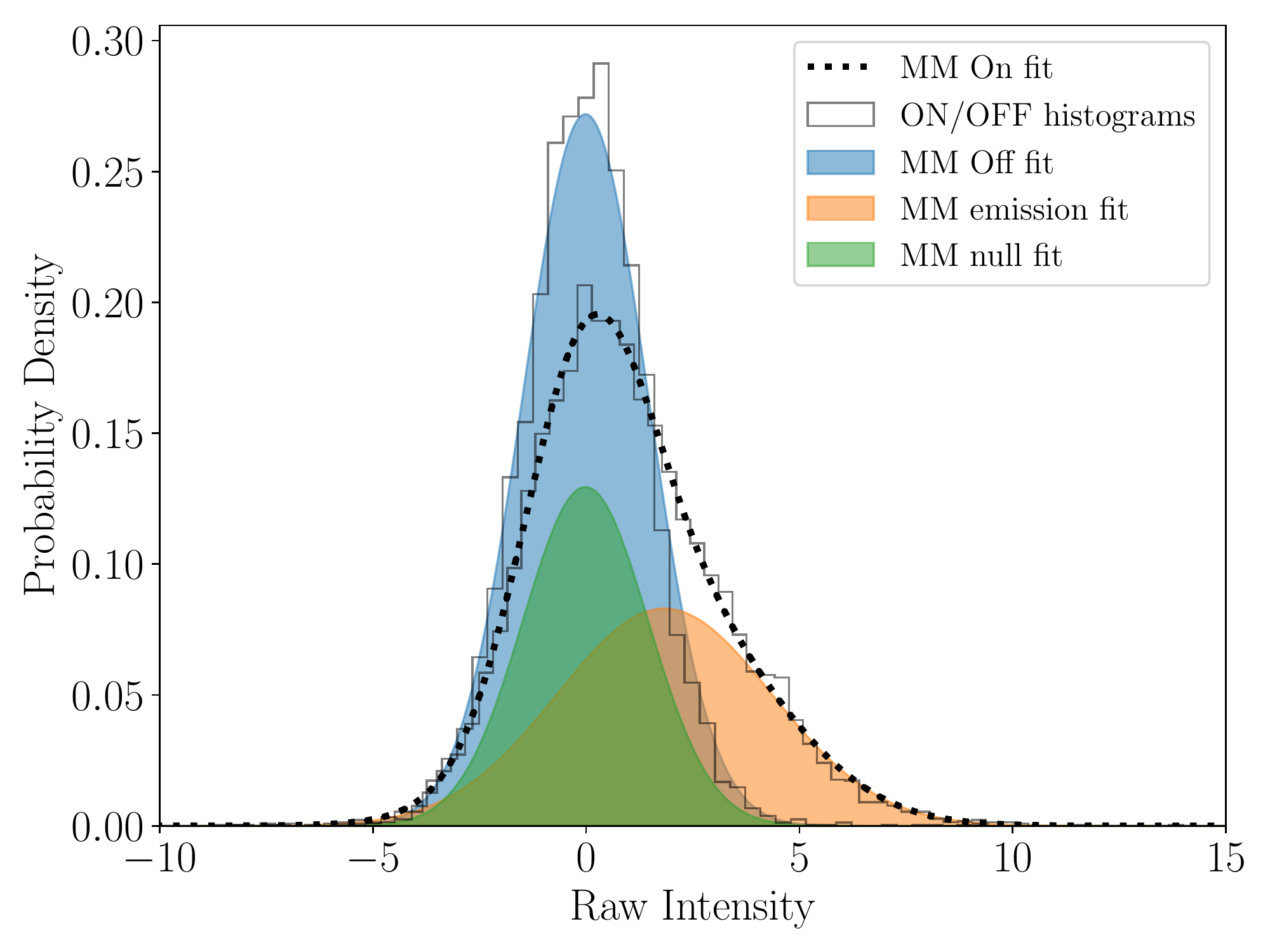}}          
      \caption{Single pulse stack (upper left), MCMC corner plot (bottom), and pulse intensity histogram (upper right) for PSR J1928+28. In this case the best fit model is a 2-component Gaussian mixture}
 \end{figure*}

\begin{figure*}
      \includegraphics[width=0.5\textwidth]{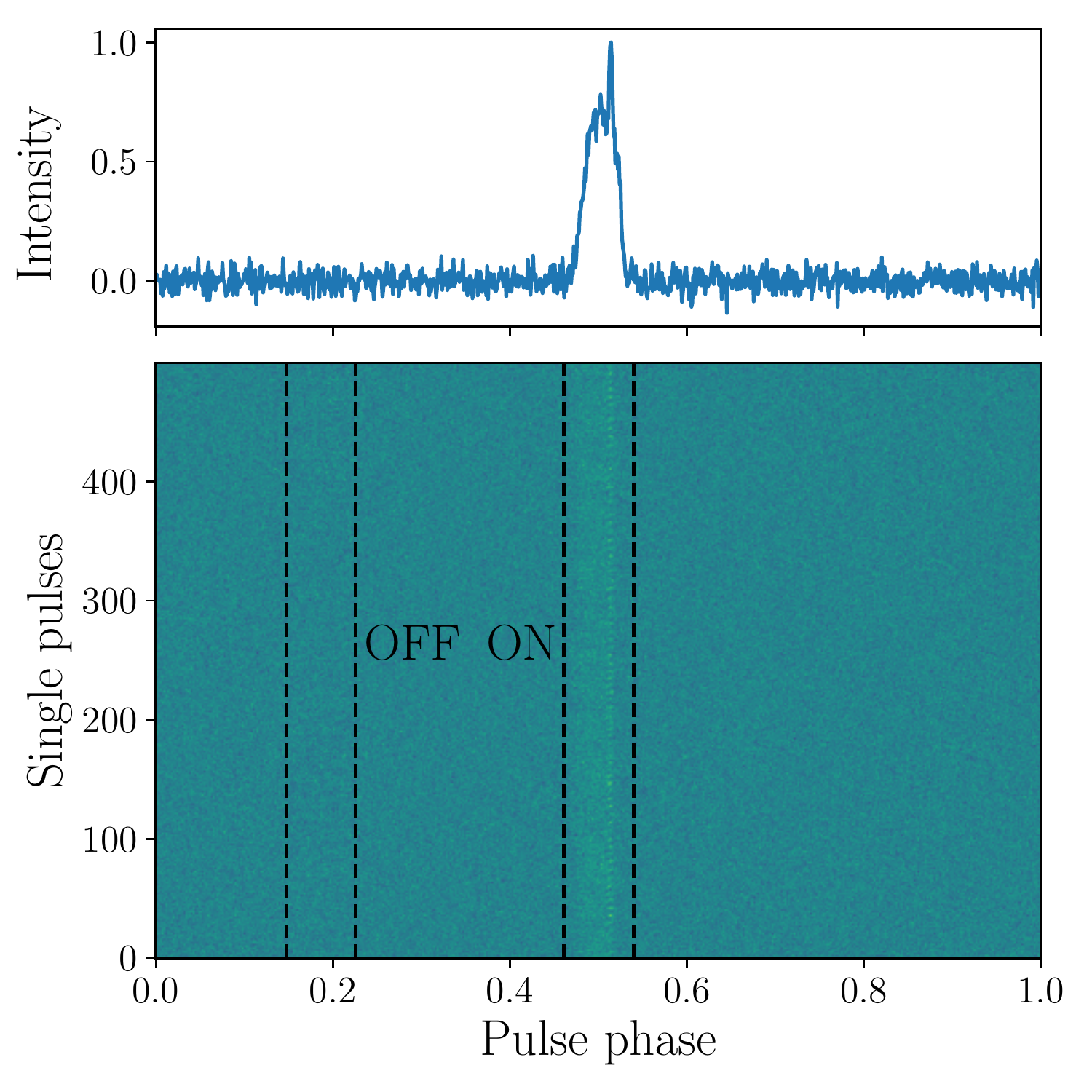}
      \includegraphics[width=0.5\textwidth]{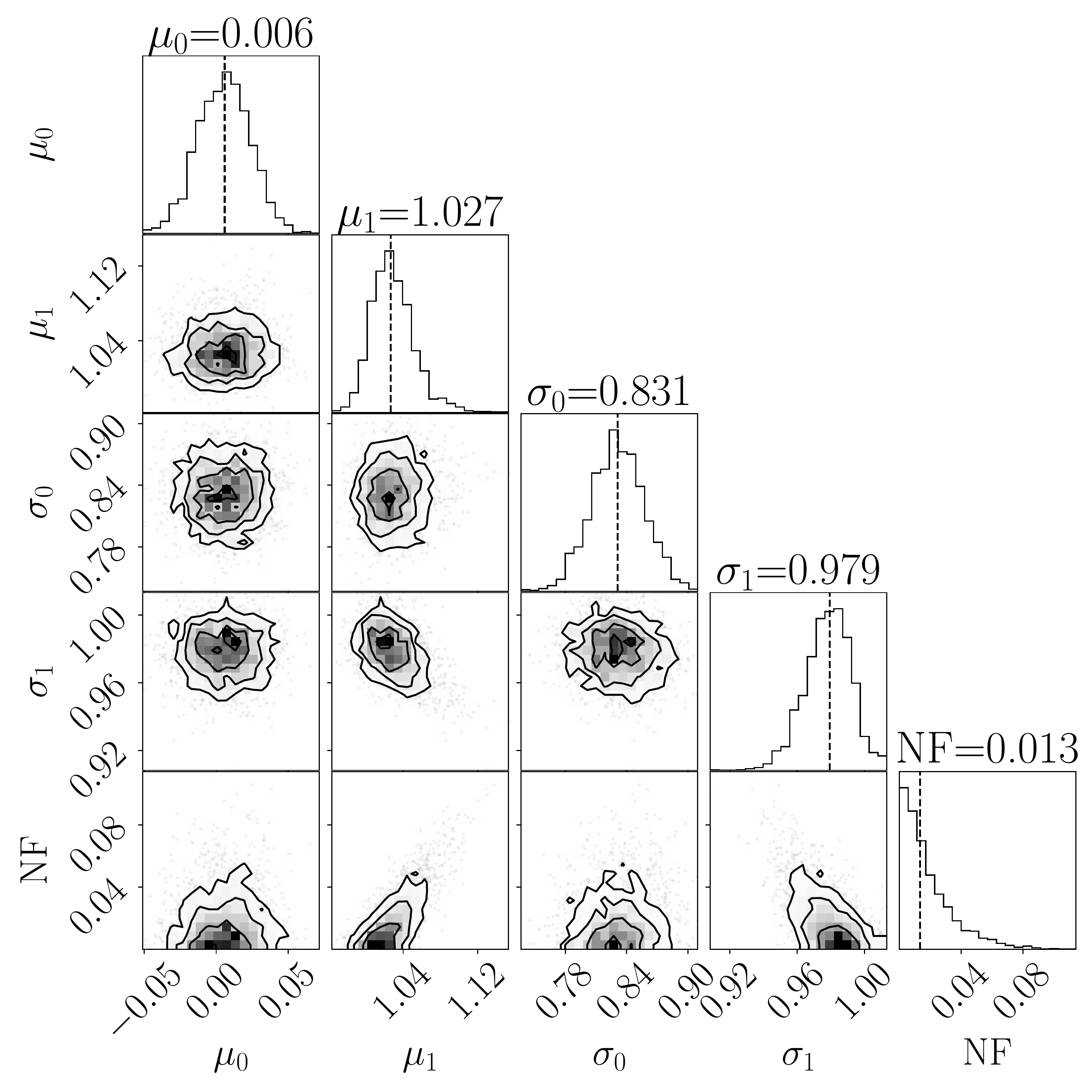}
      \centerline{\includegraphics[width=0.67\textwidth]{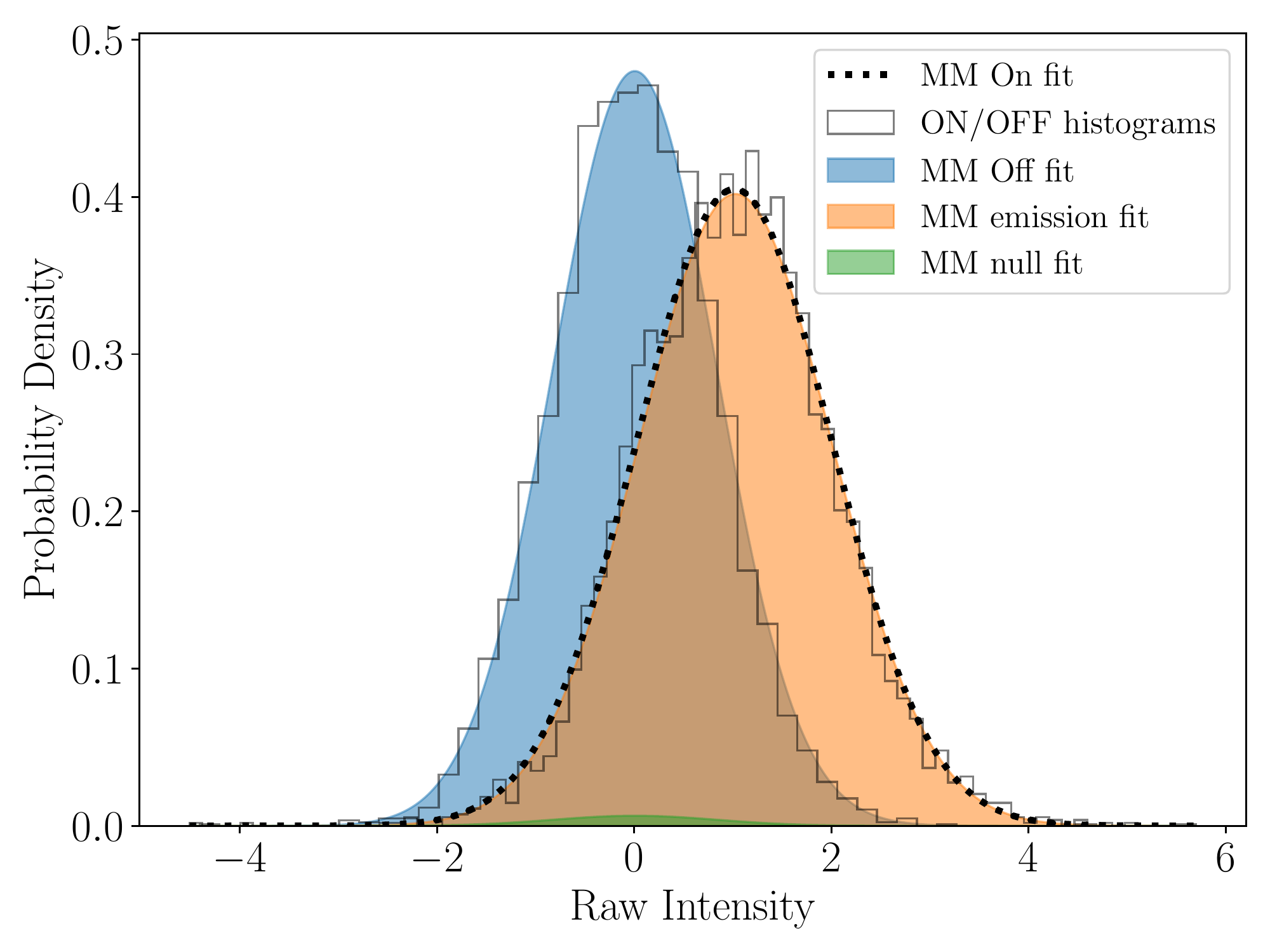}}          
      \caption{Single pulse stack (upper left), MCMC corner plot (bottom), and pulse intensity histogram (upper right) for PSR J1941+02. In this case the best fit model is a 2-component Gaussian mixture}
 \end{figure*}

\begin{figure*}
      \includegraphics[width=0.5\textwidth]{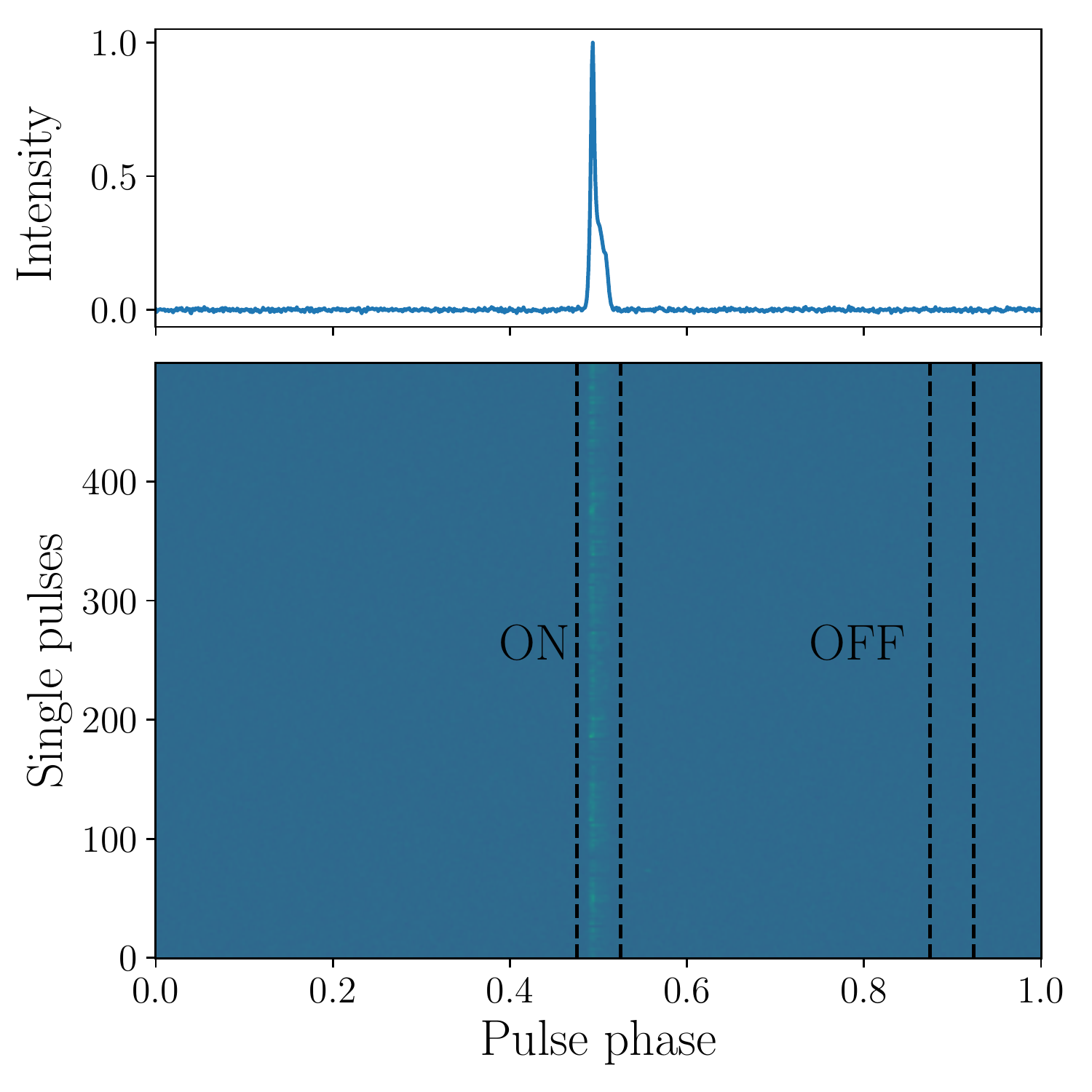}
      \includegraphics[width=0.5\textwidth]{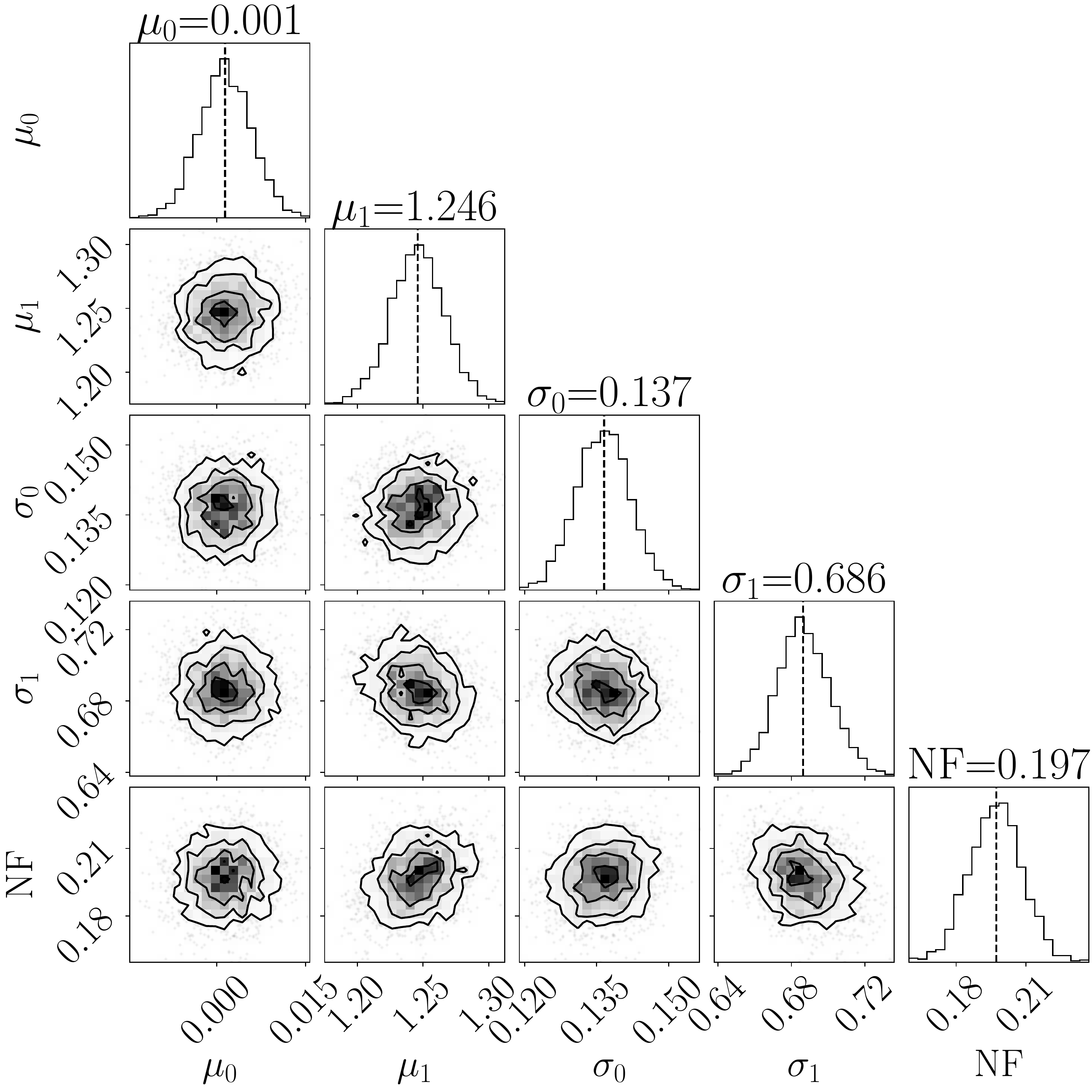}
      \centerline{\includegraphics[width=0.67\textwidth]{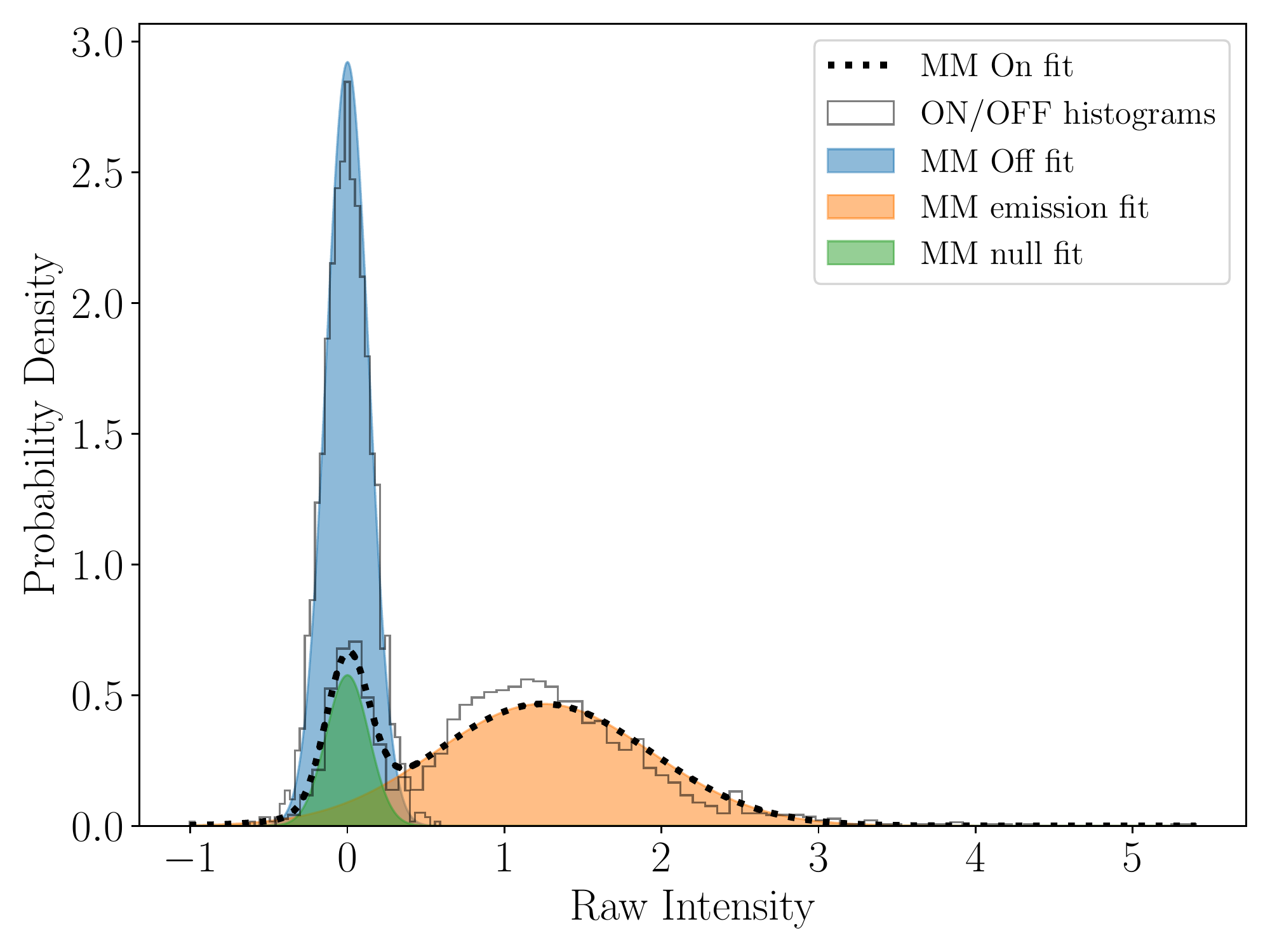}}          
      \caption{Single pulse stack (upper left), MCMC corner plot (bottom), and pulse intensity histogram (upper right) for PSR J2000+29. In this case the best fit model is a 2-component Gaussian mixture}
 \end{figure*}

\begin{figure*}
      \includegraphics[width=0.5\textwidth]{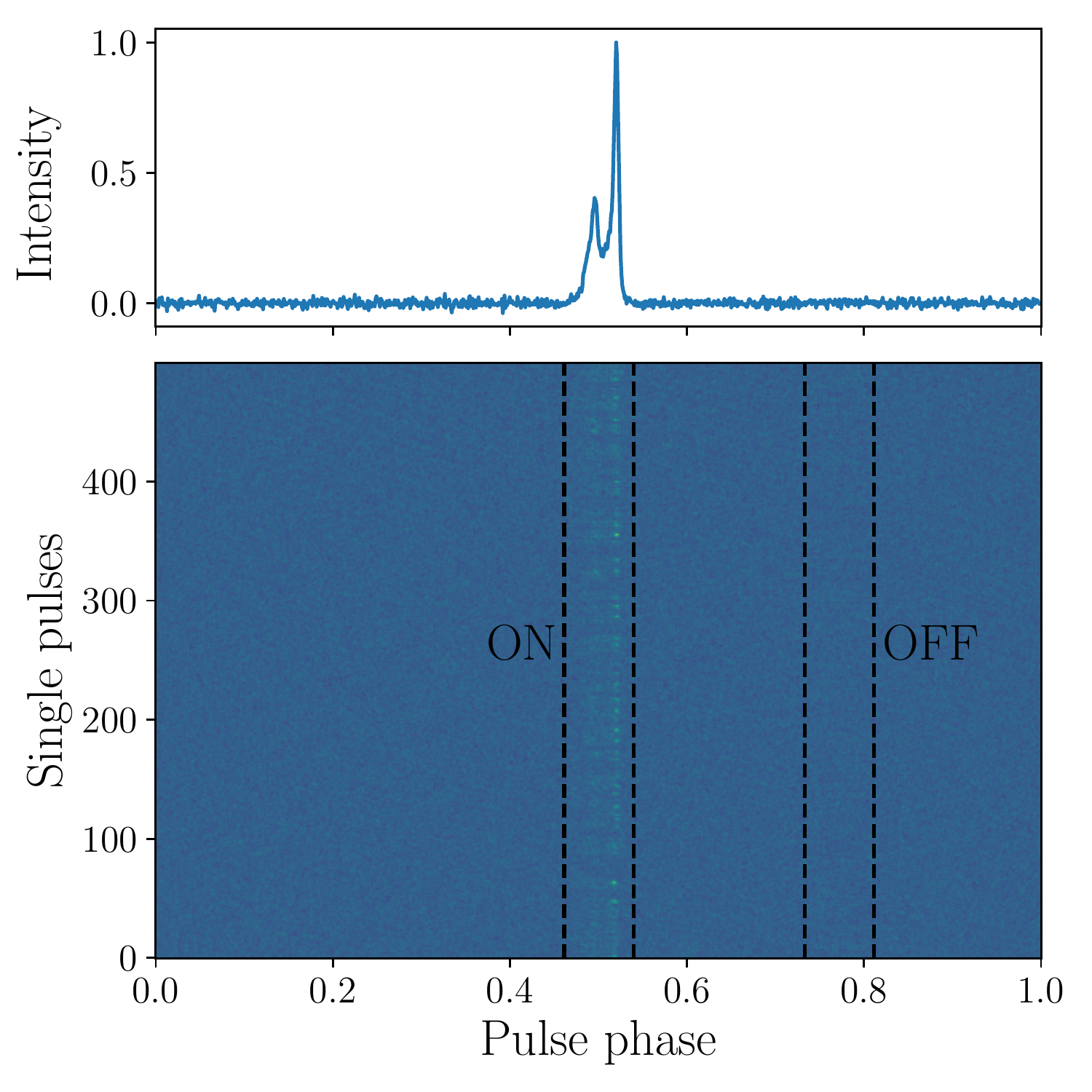}
      \includegraphics[width=0.5\textwidth]{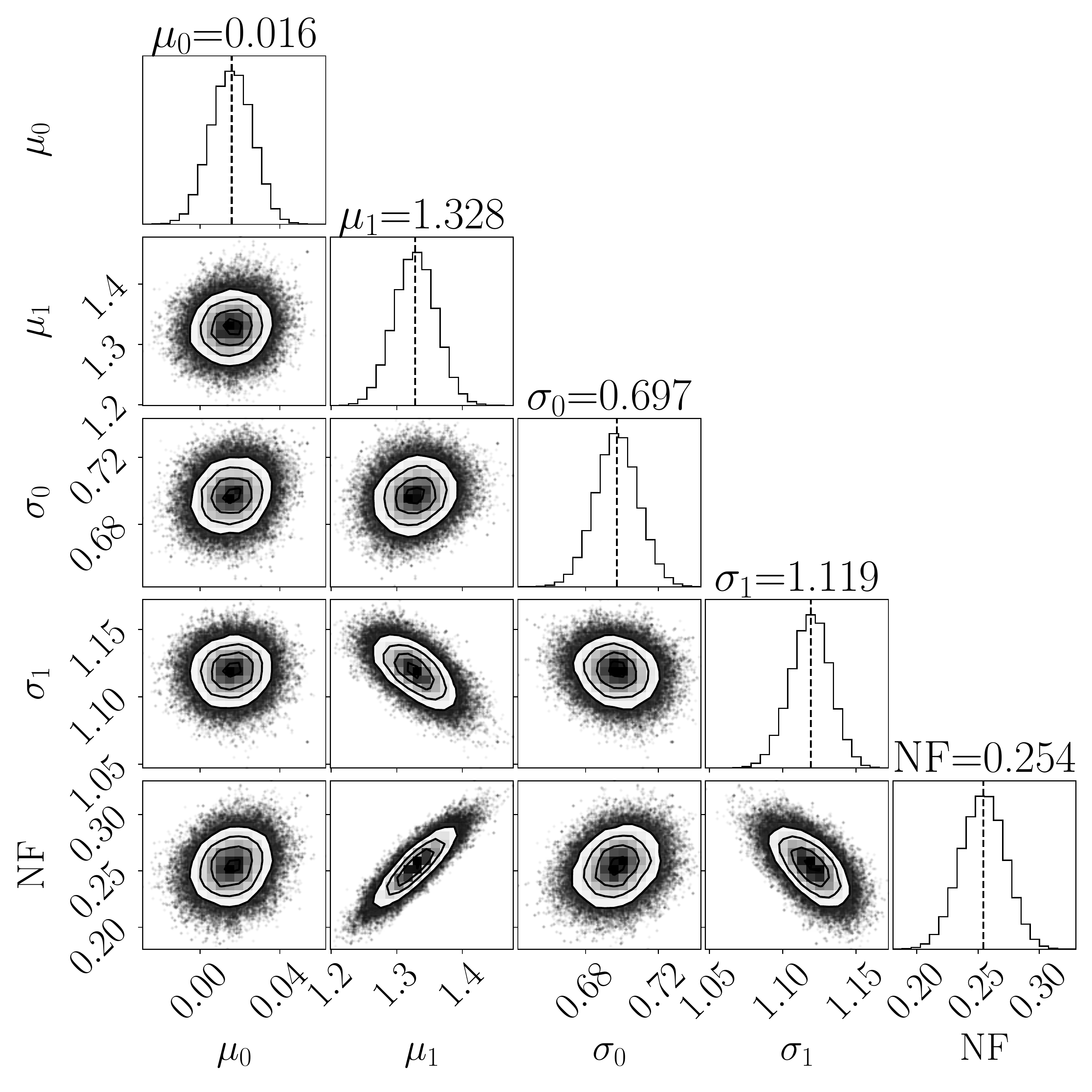}
      \centerline{\includegraphics[width=0.67\textwidth]{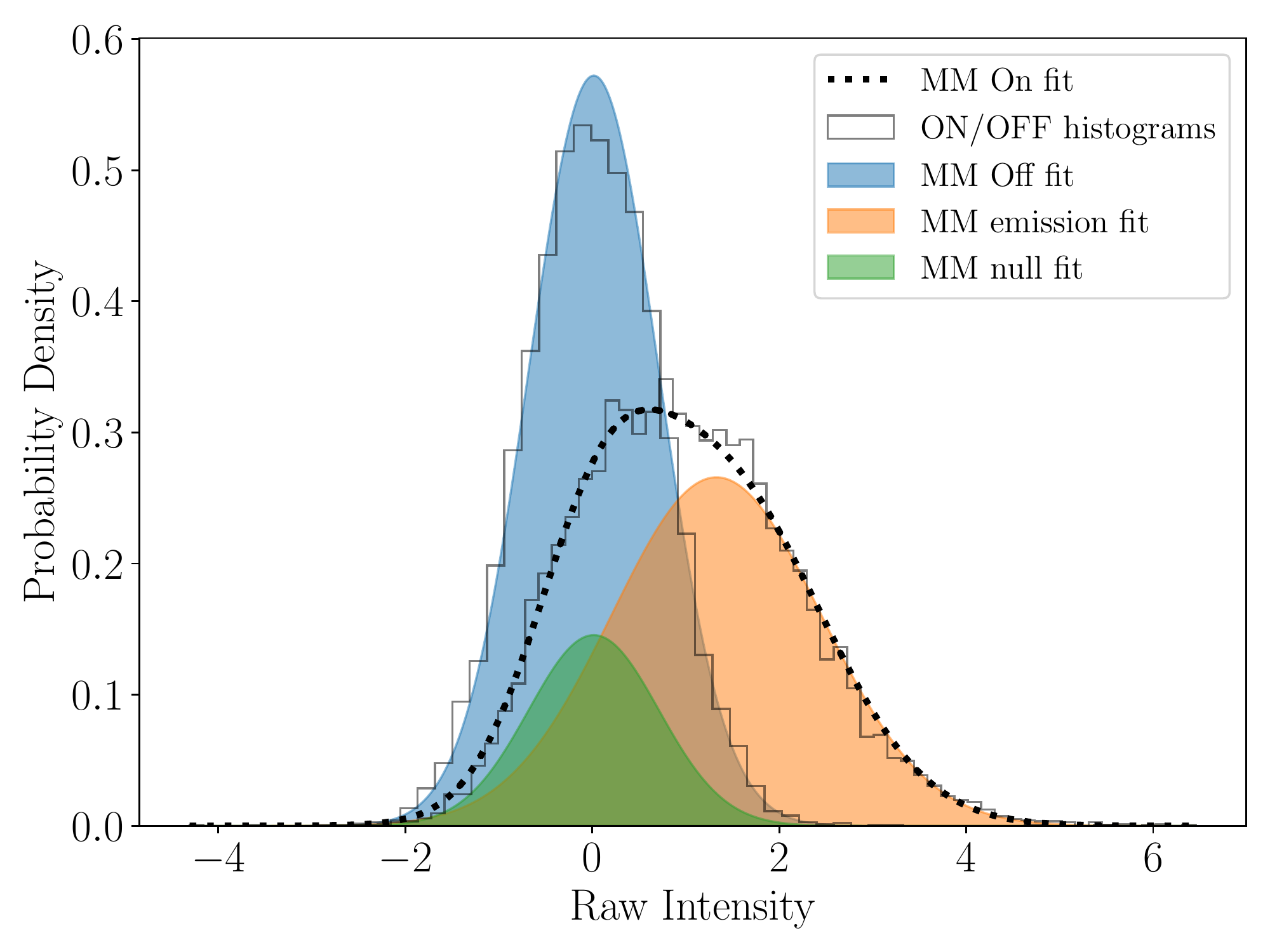}}          
      \caption{Single pulse stack (upper left), MCMC corner plot (bottom), and pulse intensity histogram (upper right) for PSR J2040-21. In this case the best fit model is a 2-component Gaussian mixture}
 \end{figure*}

\begin{figure*}
      \includegraphics[width=0.5\textwidth]{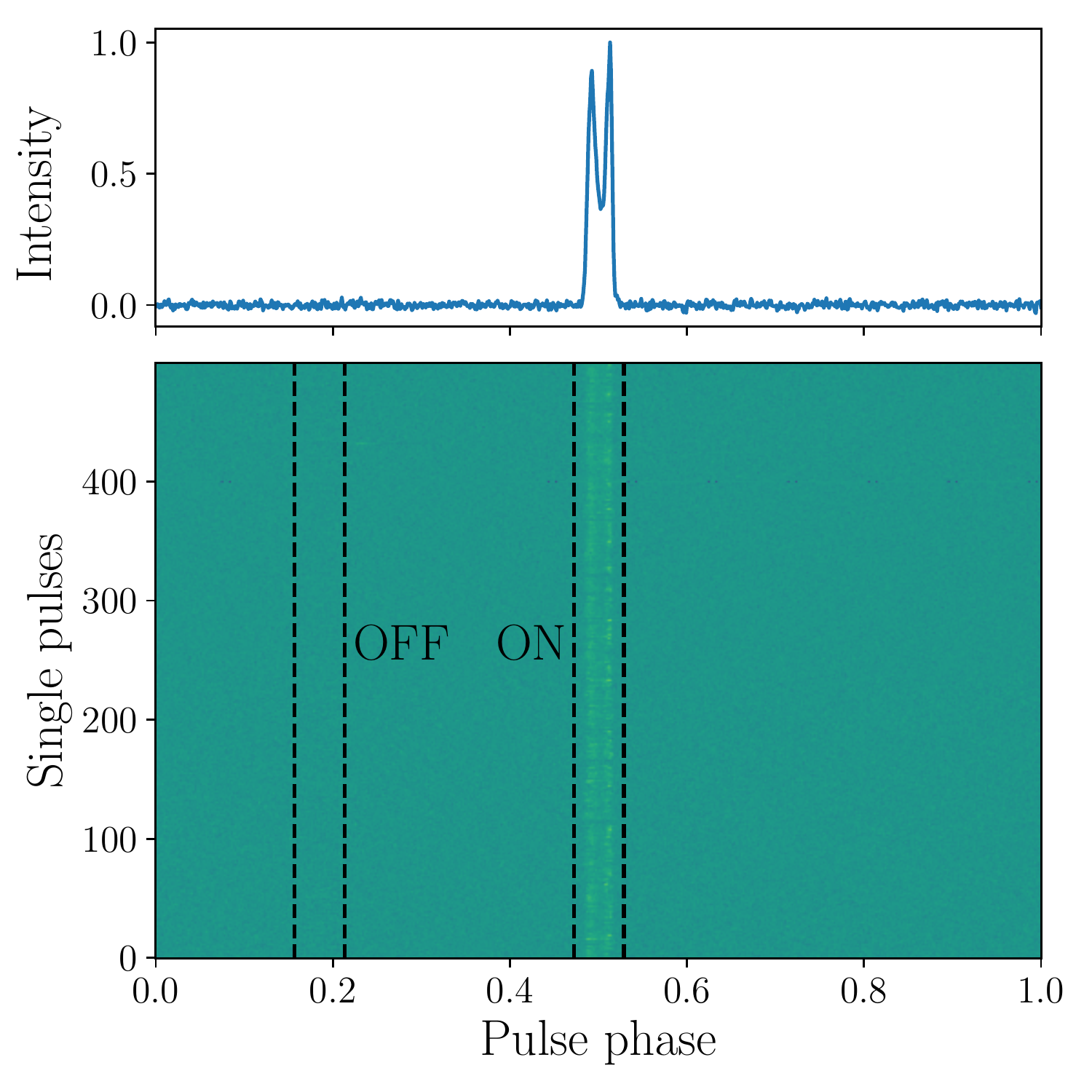}
      \includegraphics[width=0.5\textwidth]{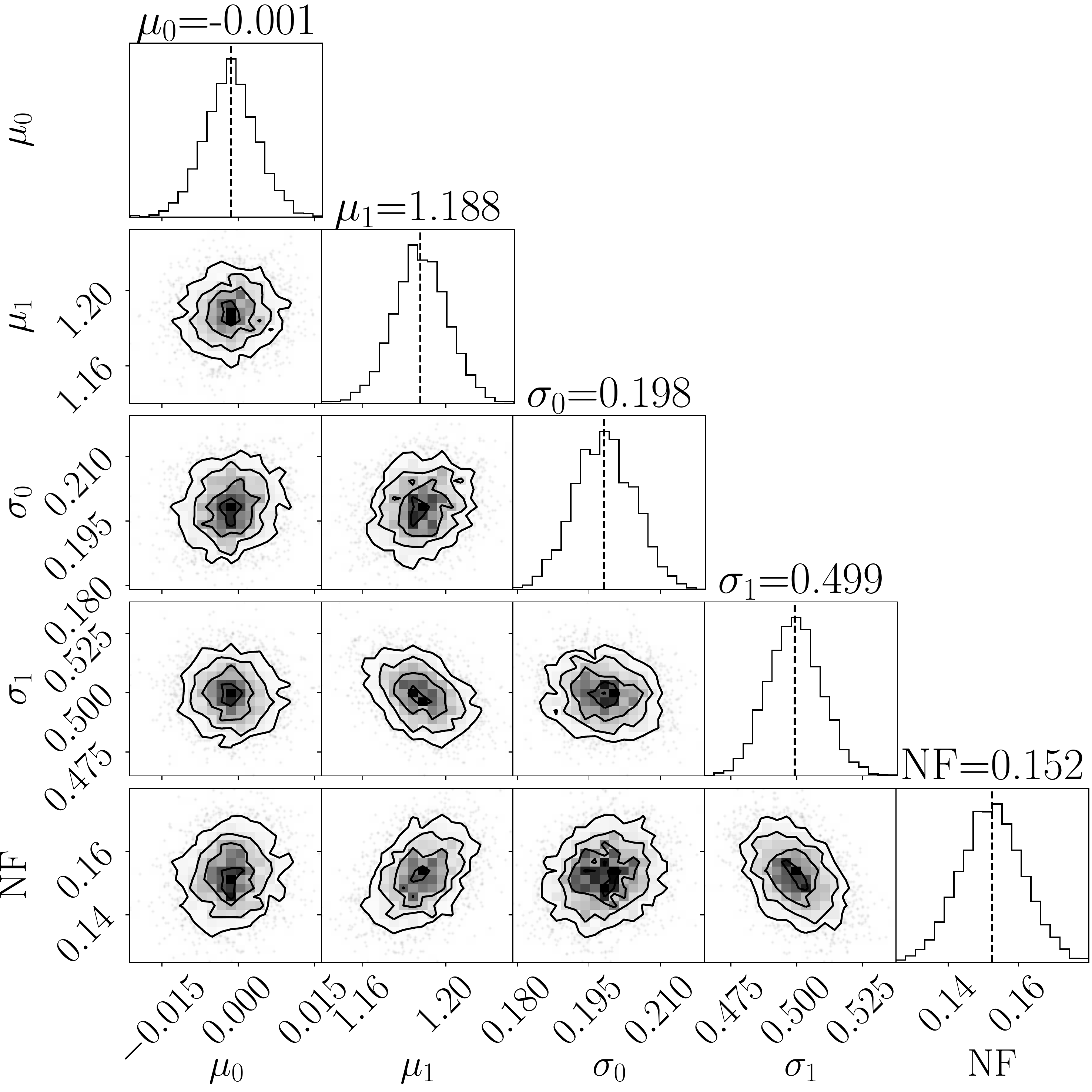}
      \centerline{\includegraphics[width=0.67\textwidth]{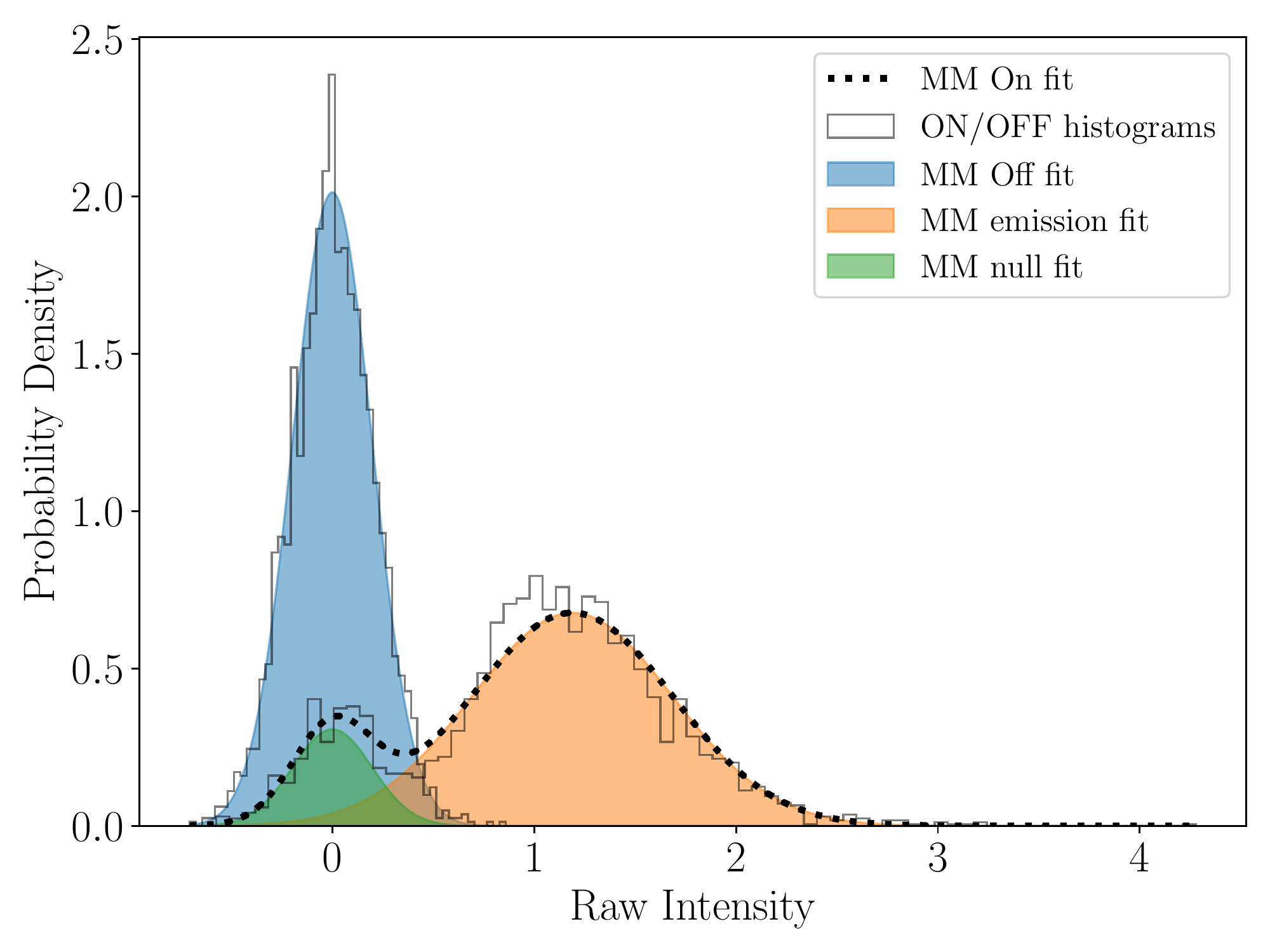}}          
      \caption{Single pulse stack (upper left), MCMC corner plot (bottom), and pulse intensity histogram (upper right) for PSR J2044+28. In this case the best fit model is a 2-component Gaussian mixture}
 \end{figure*}

\begin{figure*}
      \includegraphics[width=0.5\textwidth]{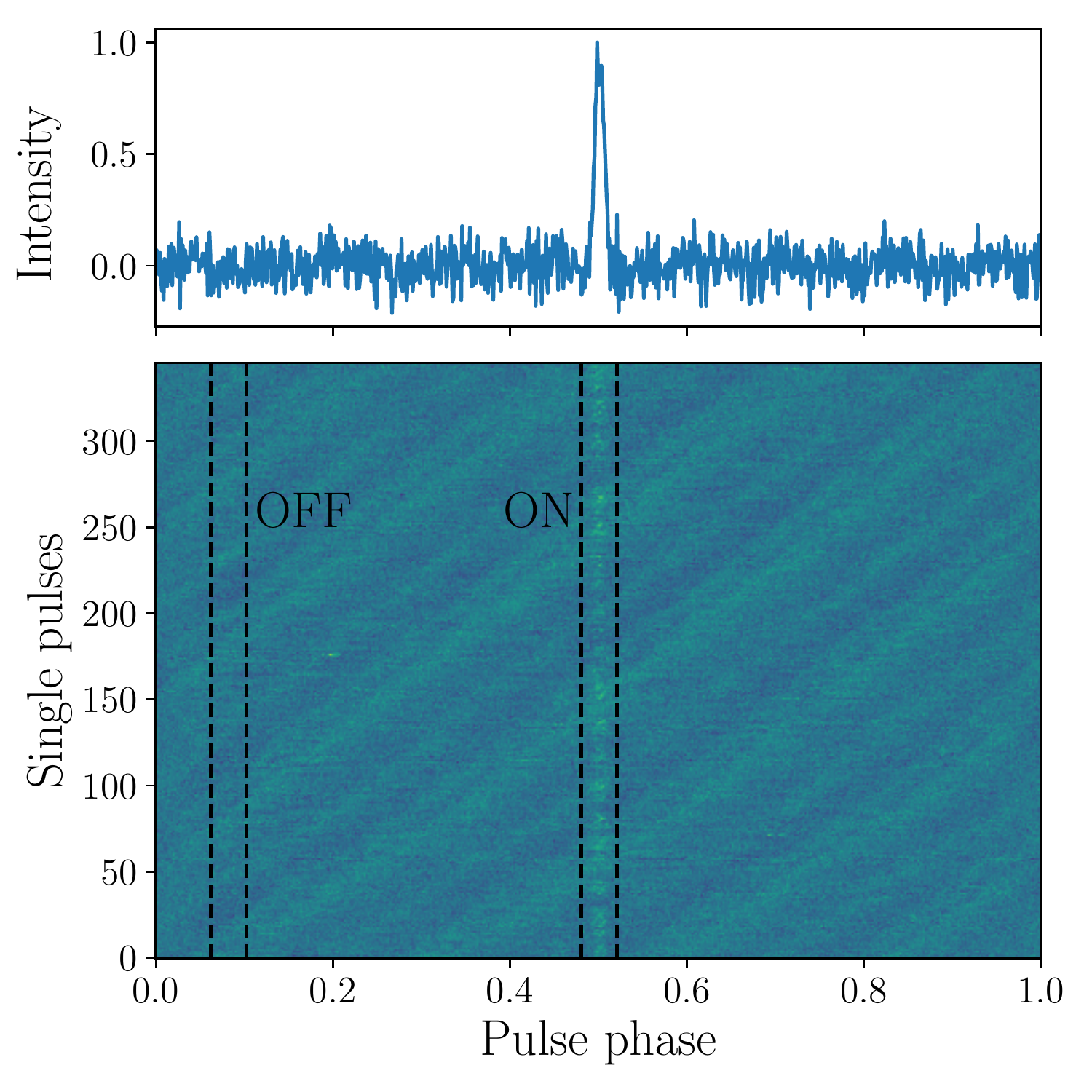}
      \includegraphics[width=0.5\textwidth]{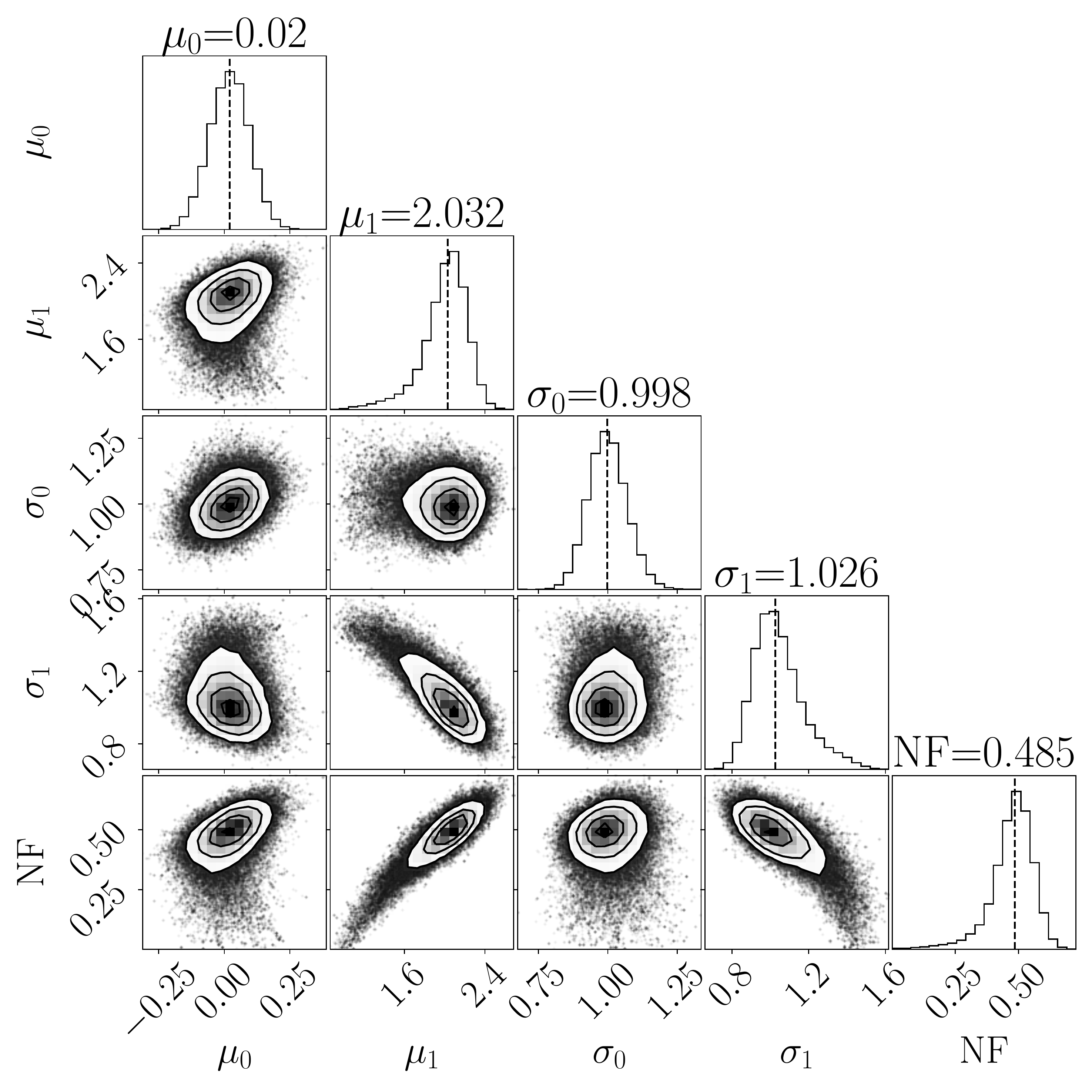}
      \centerline{\includegraphics[width=0.67\textwidth]{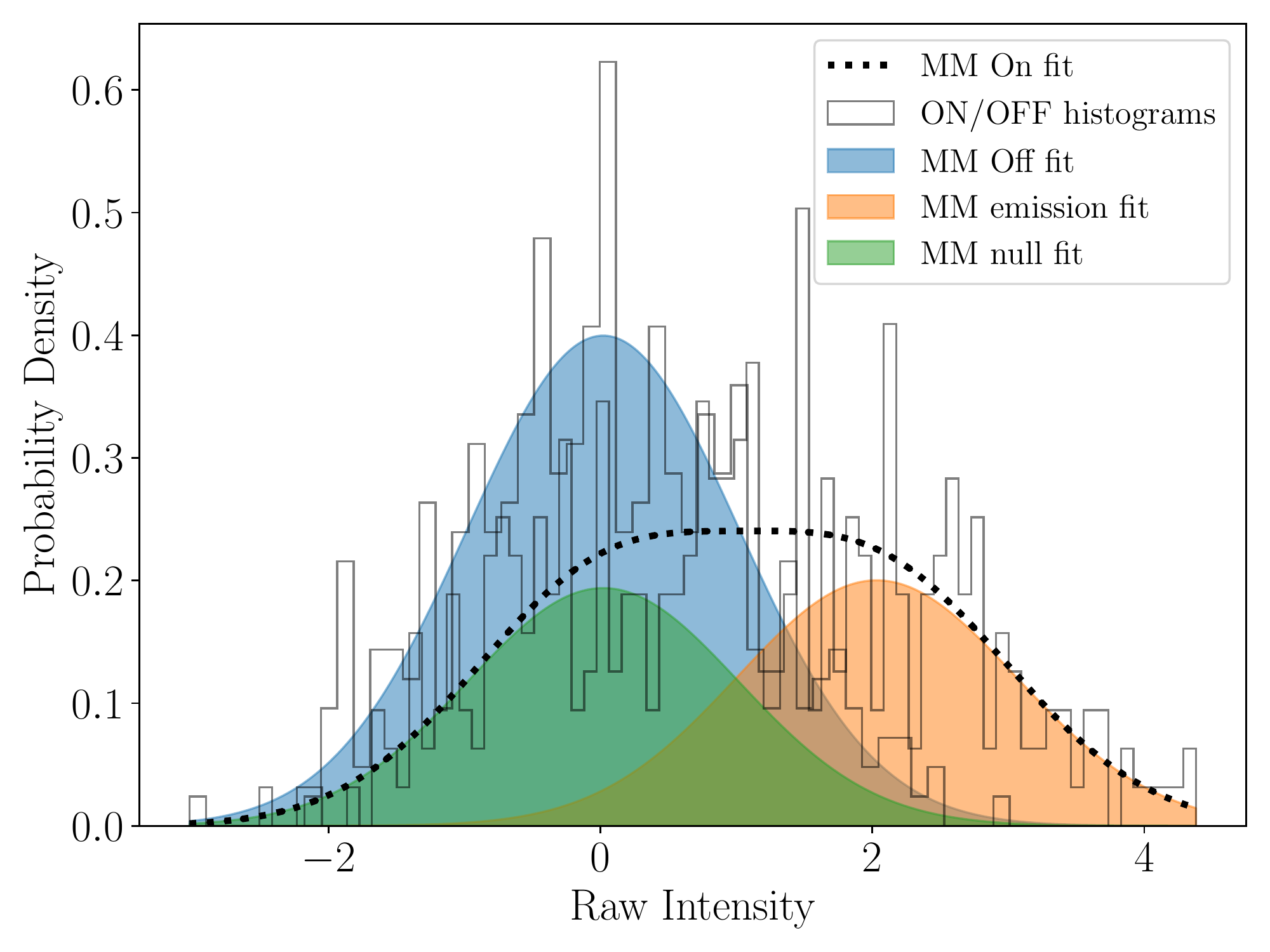}}          
      \caption{Single pulse stack (upper left), MCMC corner plot (bottom), and pulse intensity histogram (upper right) for PSR J2131-31. In this case the best fit model is a 2-component Gaussian mixture}
 \end{figure*}

\begin{figure*}
      \includegraphics[width=0.5\textwidth]{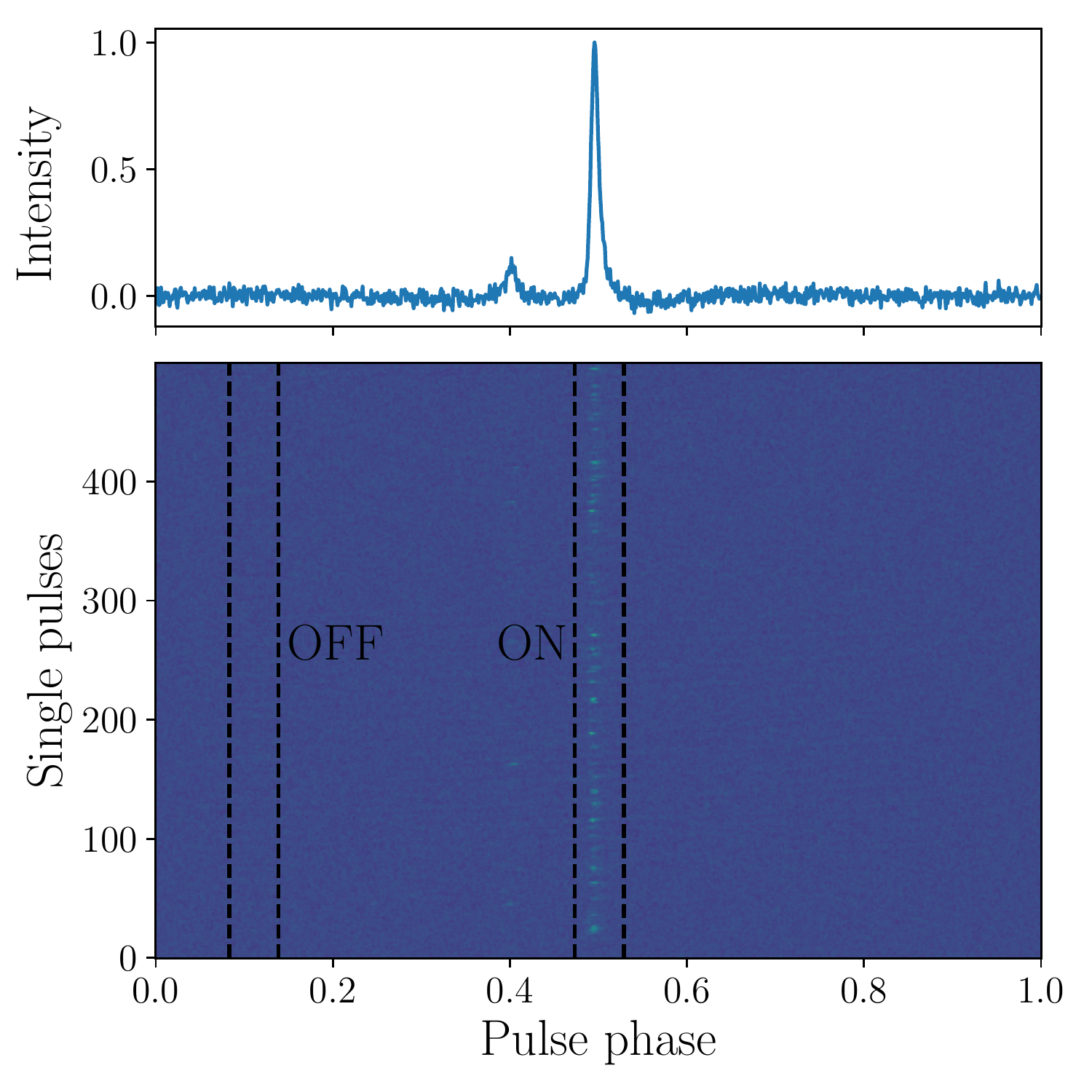}
      \includegraphics[width=0.5\textwidth]{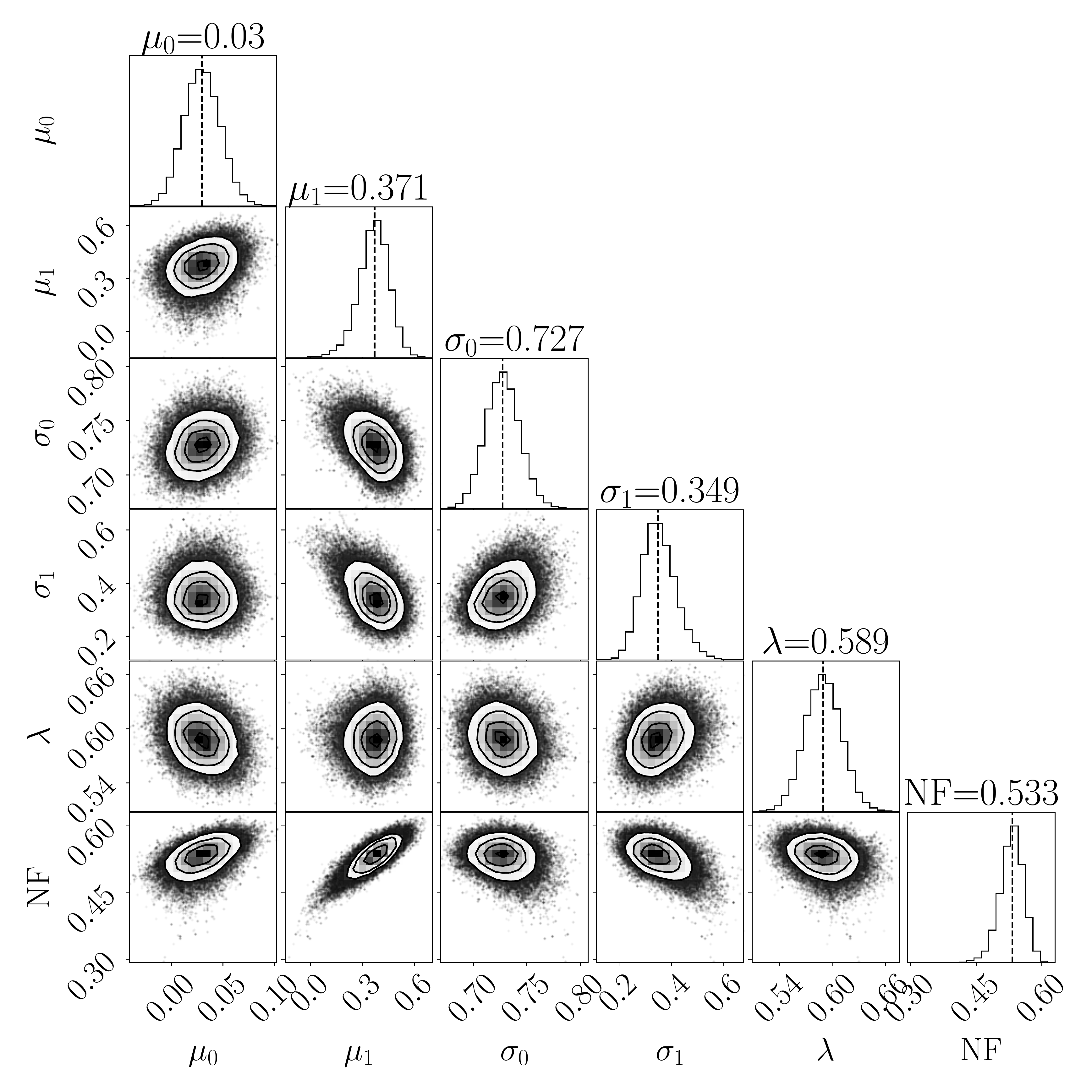}
      \centerline{\includegraphics[width=0.67\textwidth]{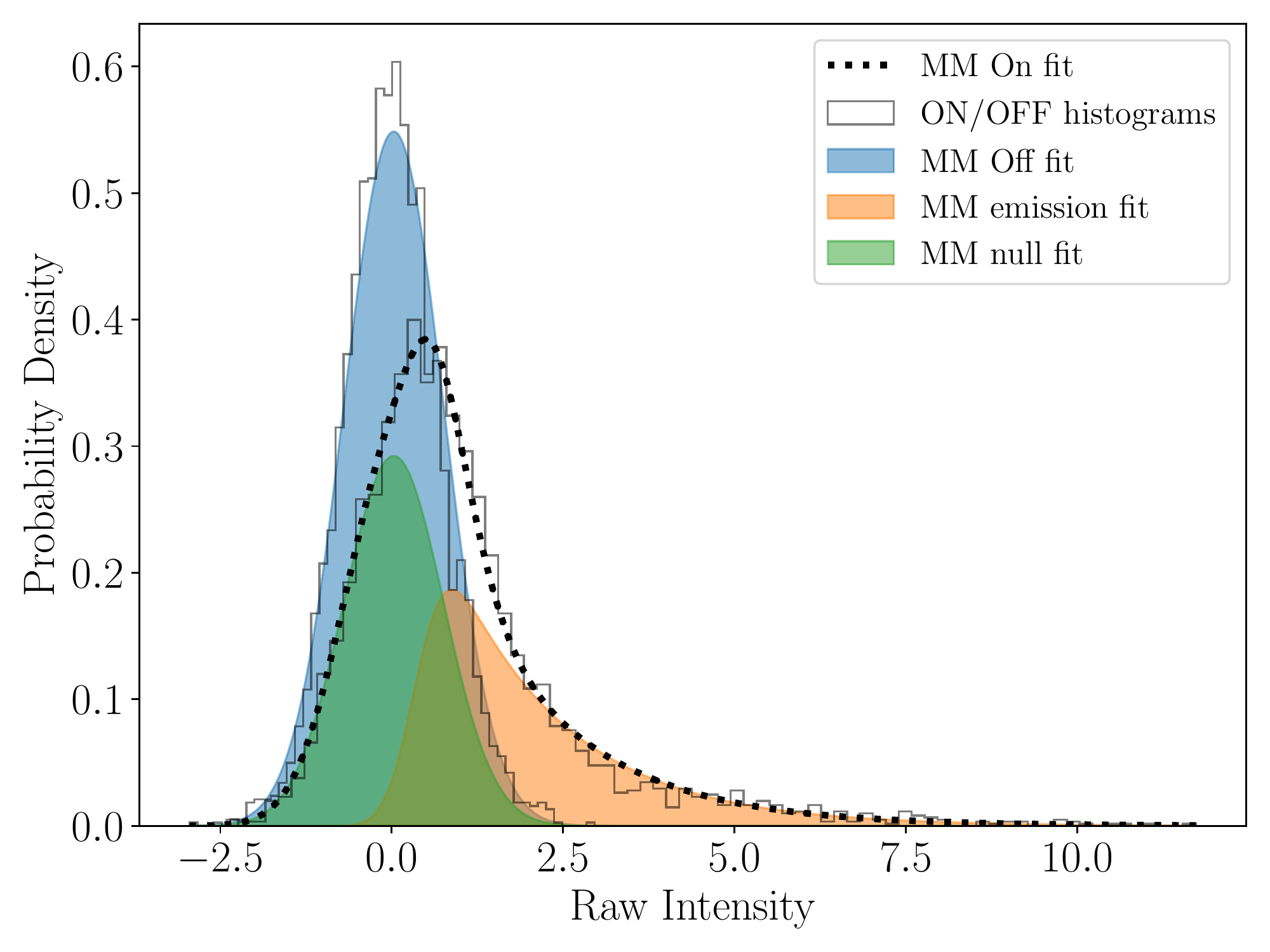}}          
      \caption{Single pulse stack (upper left), MCMC corner plot (bottom), and pulse intensity histogram (upper right) for PSR J2310+6706. In this case the best fit model is a 2-component Exponential convolved Gaussian mixture}
 \end{figure*}

%% file: nulling.bbl
\begin{thebibliography}{}
\expandafter\ifx\csname natexlab\endcsname\relax\def\natexlab#1{#1}\fi
\providecommand{\url}[1]{\href{#1}{#1}}
\providecommand{\dodoi}[1]{doi:~\href{http://doi.org/#1}{\nolinkurl{#1}}}
\providecommand{\doeprint}[1]{\href{http://ascl.net/#1}{\nolinkurl{http://ascl.net/#1}}}
\providecommand{\doarXiv}[1]{\href{https://arxiv.org/abs/#1}{\nolinkurl{https://arxiv.org/abs/#1}}}

\bibitem[{{Arjunwadkar} {et~al.}(2014){Arjunwadkar}, {Rajwade}, \&
  {Gupta}}]{Arjunwadkar2014}
{Arjunwadkar}, M., {Rajwade}, K., \& {Gupta}, Y. 2014, in Astronomical Society
  of India Conference Series, Vol.~13, Astronomical Society of India Conference
  Series, 79--81

\bibitem[{{Astropy Collaboration} {et~al.}(2013){Astropy Collaboration},
  {Robitaille}, {Tollerud}, {Greenfield}, {Droettboom}, {Bray}, {Aldcroft},
  {Davis}, {Ginsburg}, {Price-Whelan}, {Kerzendorf}, {Conley}, {Crighton},
  {Barbary}, {Muna}, {Ferguson}, {Grollier}, {Parikh}, {Nair}, {Unther},
  {Deil}, {Woillez}, {Conseil}, {Kramer}, {Turner}, {Singer}, {Fox}, {Weaver},
  {Zabalza}, {Edwards}, {Azalee Bostroem}, {Burke}, {Casey}, {Crawford},
  {Dencheva}, {Ely}, {Jenness}, {Labrie}, {Lim}, {Pierfederici}, {Pontzen},
  {Ptak}, {Refsdal}, {Servillat}, \& {Streicher}}]{astropy1}
{Astropy Collaboration}, {Robitaille}, T.~P., {Tollerud}, E.~J., {et~al.} 2013,
  \aap, 558, A33, \dodoi{10.1051/0004-6361/201322068}

\bibitem[{{Astropy Collaboration} {et~al.}(2018){Astropy Collaboration},
  {Price-Whelan}, {Sip{\H{o}}cz}, {G{\"u}nther}, {Lim}, {Crawford}, {Conseil},
  {Shupe}, {Craig}, {Dencheva}, {Ginsburg}, {VanderPlas}, {Bradley},
  {P{\'e}rez-Su{\'a}rez}, {de Val-Borro}, {Aldcroft}, {Cruz}, {Robitaille},
  {Tollerud}, {Ardelean}, {Babej}, {Bach}, {Bachetti}, {Bakanov}, {Bamford},
  {Barentsen}, {Barmby}, {Baumbach}, {Berry}, {Biscani}, {Boquien}, {Bostroem},
  {Bouma}, {Brammer}, {Bray}, {Breytenbach}, {Buddelmeijer}, {Burke},
  {Calderone}, {Cano Rodr{\'\i}guez}, {Cara}, {Cardoso}, {Cheedella}, {Copin},
  {Corrales}, {Crichton}, {D'Avella}, {Deil}, {Depagne}, {Dietrich}, {Donath},
  {Droettboom}, {Earl}, {Erben}, {Fabbro}, {Ferreira}, {Finethy}, {Fox},
  {Garrison}, {Gibbons}, {Goldstein}, {Gommers}, {Greco}, {Greenfield},
  {Groener}, {Grollier}, {Hagen}, {Hirst}, {Homeier}, {Horton}, {Hosseinzadeh},
  {Hu}, {Hunkeler}, {Ivezi{\'c}}, {Jain}, {Jenness}, {Kanarek}, {Kendrew},
  {Kern}, {Kerzendorf}, {Khvalko}, {King}, {Kirkby}, {Kulkarni}, {Kumar},
  {Lee}, {Lenz}, {Littlefair}, {Ma}, {Macleod}, {Mastropietro}, {McCully},
  {Montagnac}, {Morris}, {Mueller}, {Mumford}, {Muna}, {Murphy}, {Nelson},
  {Nguyen}, {Ninan}, {N{\"o}the}, {Ogaz}, {Oh}, {Parejko}, {Parley}, {Pascual},
  {Patil}, {Patil}, {Plunkett}, {Prochaska}, {Rastogi}, {Reddy Janga},
  {Sabater}, {Sakurikar}, {Seifert}, {Sherbert}, {Sherwood-Taylor}, {Shih},
  {Sick}, {Silbiger}, {Singanamalla}, {Singer}, {Sladen}, {Sooley},
  {Sornarajah}, {Streicher}, {Teuben}, {Thomas}, {Tremblay}, {Turner},
  {Terr{\'o}n}, {van Kerkwijk}, {de la Vega}, {Watkins}, {Weaver}, {Whitmore},
  {Woillez}, {Zabalza}, \& {Astropy Contributors}}]{astropy2}
{Astropy Collaboration}, {Price-Whelan}, A.~M., {Sip{\H{o}}cz}, B.~M., {et~al.}
  2018, \aj, 156, 123, \dodoi{10.3847/1538-3881/aabc4f}

\bibitem[{{Backer}(1970{\natexlab{a}})}]{backer}
{Backer}, D.~C. 1970{\natexlab{a}}, \nat, 228, 42, \dodoi{10.1038/228042a0}

\bibitem[{{Backer}(1970{\natexlab{b}})}]{modechanging}
---. 1970{\natexlab{b}}, \nat, 228, 1297, \dodoi{10.1038/2281297a0}

\bibitem[{{Backer}(1970{\natexlab{c}})}]{backer1970drift}
---. 1970{\natexlab{c}}, \nat, 228, 752, \dodoi{10.1038/228752a0}

\bibitem[{{Bates} {et~al.}(2014){Bates}, {Lorimer}, {Rane}, \&
  {Swiggum}}]{bates14}
{Bates}, S.~D., {Lorimer}, D.~R., {Rane}, A., \& {Swiggum}, J. 2014, \mnras,
  439, 2893, \dodoi{10.1093/mnras/stu157}

\bibitem[{{Bhat} {et~al.}(2003){Bhat}, {Cordes}, \& {Chatterjee}}]{bhat2003}
{Bhat}, N.~D.~R., {Cordes}, J.~M., \& {Chatterjee}, S. 2003, \apj, 584, 782,
  \dodoi{10.1086/345775}

\bibitem[{{Bhat} {et~al.}(2007){Bhat}, {Gupta}, {Kramer}, {Karastergiou},
  {Lyne}, \& {Johnston}}]{Bhat2007}
{Bhat}, N.~D.~R., {Gupta}, Y., {Kramer}, M., {et~al.} 2007, \aap, 462, 257,
  \dodoi{10.1051/0004-6361:20053157}

\bibitem[{{Davies} {et~al.}(1984){Davies}, {Lyne}, {Smith}, {Izvekova},
  {Kuzmin}, \& {Shitov}}]{Davies1984}
{Davies}, J.~G., {Lyne}, A.~G., {Smith}, F.~G., {et~al.} 1984, \mnras, 211, 57,
  \dodoi{10.1093/mnras/211.1.57}

\bibitem[{{Drake} \& {Craft}(1968)}]{subpulsedrift}
{Drake}, F.~D., \& {Craft}, H.~D. 1968, \nat, 220, 231,
  \dodoi{10.1038/220231a0}

\bibitem[{{Faucher-Gigu{\`e}re} \& {Kaspi}(2006)}]{faucher-giguere06}
{Faucher-Gigu{\`e}re}, C.-A., \& {Kaspi}, V.~M. 2006, \apj, 643, 332,
  \dodoi{10.1086/501516}

\bibitem[{{Filippenko} \& {Radhakrishnan}(1982)}]{Filip&Radha}
{Filippenko}, A.~V., \& {Radhakrishnan}, V. 1982, \apj, 263, 828,
  \dodoi{10.1086/160553}

\bibitem[{{Foreman-Mackey} {et~al.}(2013){Foreman-Mackey}, {Hogg}, {Lang}, \&
  {Goodman}}]{emcee}
{Foreman-Mackey}, D., {Hogg}, D.~W., {Lang}, D., \& {Goodman}, J. 2013, \pasp,
  125, 306, \dodoi{10.1086/670067}

\bibitem[{{Gajjar} {et~al.}(2012){Gajjar}, {Joshi}, \& {Kramer}}]{Vishal2012}
{Gajjar}, V., {Joshi}, B.~C., \& {Kramer}, M. 2012, \mnras, 424, 1197,
  \dodoi{10.1111/j.1365-2966.2012.21296.x}

\bibitem[{{Gajjar} {et~al.}(2014{\natexlab{a}}){Gajjar}, {Joshi}, {Kramer},
  {Karuppusamy}, \& {Smits}}]{Vishal2014}
{Gajjar}, V., {Joshi}, B.~C., {Kramer}, M., {Karuppusamy}, R., \& {Smits}, R.
  2014{\natexlab{a}}, \apj, 797, 18, \dodoi{10.1088/0004-637X/797/1/18}

\bibitem[{{Gajjar} {et~al.}(2014{\natexlab{b}}){Gajjar}, {Joshi}, \&
  {Wright}}]{Vishal2013}
{Gajjar}, V., {Joshi}, B.~C., \& {Wright}, G. 2014{\natexlab{b}}, \mnras, 439,
  221, \dodoi{10.1093/mnras/stt2389}

\bibitem[{Harris {et~al.}(2020)Harris, Millman, van~der Walt, Gommers,
  Virtanen, Cournapeau, Wieser, Taylor, Berg, Smith, Kern, Picus, Hoyer, van
  Kerkwijk, Brett, Haldane, del R{\'{i}}o, Wiebe, Peterson,
  G{\'{e}}rard-Marchant, Sheppard, Reddy, Weckesser, Abbasi, Gohlke, \&
  Oliphant}]{numpy}
Harris, C.~R., Millman, K.~J., van~der Walt, S.~J., {et~al.} 2020, Nature, 585,
  357, \dodoi{10.1038/s41586-020-2649-2}

\bibitem[{{Herfindal} \& {Rankin}(2007)}]{herfindal07}
{Herfindal}, J.~L., \& {Rankin}, J.~M. 2007, \mnras, 380, 430,
  \dodoi{10.1111/j.1365-2966.2007.12089.x}

\bibitem[{{Herfindal} \& {Rankin}(2009)}]{herfindal09}
---. 2009, \mnras, 393, 1391, \dodoi{10.1111/j.1365-2966.2008.14119.x}

\bibitem[{{Honnappa} {et~al.}(2012){Honnappa}, {Lewandowski}, {Kijak},
  {Deshpande}, {Gil}, {Maron}, \& {Jessner}}]{sneha2012}
{Honnappa}, S., {Lewandowski}, W., {Kijak}, J., {et~al.} 2012, \mnras, 421,
  1996, \dodoi{10.1111/j.1365-2966.2012.20424.x}

\bibitem[{Hunter(2007)}]{matplotlib}
Hunter, J.~D. 2007, Computing in Science \& Engineering, 9, 90,
  \dodoi{10.1109/MCSE.2007.55}

\bibitem[{{Ivezi{\'c}} {et~al.}(2020){Ivezi{\'c}}, {Connolly}, {VanderPlas}, \&
  {Gray}}]{ivezic_book}
{Ivezi{\'c}}, {\v{Z}}., {Connolly}, A.~J., {VanderPlas}, J.~T., \& {Gray}, A.
  2020, {Statistics, Data Mining, and Machine Learning in Astronomy. A
  Practical Python Guide for the Analysis of Survey Data, Updated Edition}

\bibitem[{{Kaplan} {et~al.}(2018){Kaplan}, {Swiggum}, {Fichtenbauer}, \&
  {Vallisneri}}]{Kaplan2018}
{Kaplan}, D.~L., {Swiggum}, J.~K., {Fichtenbauer}, T.~D.~J., \& {Vallisneri},
  M. 2018, \apj, 855, 14, \dodoi{10.3847/1538-4357/aaab62}

\bibitem[{{Konar} \& {Deka}(2019)}]{Konar2019}
{Konar}, S., \& {Deka}, U. 2019, Journal of Astrophysics and Astronomy, 40, 42,
  \dodoi{10.1007/s12036-019-9608-z}

\bibitem[{{Kramer} {et~al.}(2006){Kramer}, {Lyne}, {O'Brien}, {Jordan}, \&
  {Lorimer}}]{kramer2006}
{Kramer}, M., {Lyne}, A.~G., {O'Brien}, J.~T., {Jordan}, C.~A., \& {Lorimer},
  D.~R. 2006, Science, 312, 549, \dodoi{10.1126/science.1124060}

\bibitem[{Lomax(2007)}]{lomax2007statistical}
Lomax, R. 2007, Statistical Concepts: A Second Course (Lawrence Erlbaum
  Associates).
\newblock \url{https://books.google.com/books?id=p17rT373FNAC}

\bibitem[{{Lorimer} \& {Kramer}(2004)}]{handbook}
{Lorimer}, D.~R., \& {Kramer}, M. 2004, {Handbook of Pulsar Astronomy}, Vol.~4

\bibitem[{{Luo} {et~al.}(2019){Luo}, {Ransom}, {Demorest}, {van Haasteren},
  {Ray}, {Stovall}, {Bachetti}, {Archibald}, {Kerr}, {Colen}, \&
  {Jenet}}]{pint}
{Luo}, J., {Ransom}, S., {Demorest}, P., {et~al.} 2019, {PINT: High-precision
  pulsar timing analysis package}, Astrophysics Source Code Library, record
  ascl:1902.007.
\newblock \doeprint{1902.007}

\bibitem[{{Lynch} {et~al.}(2013){Lynch}, {Boyles}, {Ransom}, {Stairs},
  {Lorimer}, {McLaughlin}, {Hessels}, {Kaspi}, {Kondratiev}, {Archibald},
  {Berndsen}, {Cardoso}, {Cherry}, {Epstein}, {Karako-Argaman}, {McPhee},
  {Pennucci}, {Roberts}, {Stovall}, \& {van Leeuwen}}]{Lynch2013}
{Lynch}, R.~S., {Boyles}, J., {Ransom}, S.~M., {et~al.} 2013, \apj, 763, 81,
  \dodoi{10.1088/0004-637X/763/2/81}

\bibitem[{{Lynch} {et~al.}(2018){Lynch}, {Swiggum}, {Kondratiev}, {Kaplan},
  {Stovall}, {Fonseca}, {Roberts}, {Levin}, {DeCesar}, {Cui}, {Cenko},
  {Gatkine}, {Archibald}, {Banaszak}, {Biwer}, {Boyles}, {Chawla}, {Dartez},
  {Day}, {Ford}, {Flanigan}, {Hessels}, {Hinojosa}, {Jenet}, {Karako-Argaman},
  {Kaspi}, {Leake}, {Lunsford}, {Martinez}, {Mata}, {McLaughlin}, {Noori},
  {Ransom}, {Rohr}, {Siemens}, {Spiewak}, {Stairs}, {van Leeuwen}, {Walker}, \&
  {Wells}}]{Lynch2018}
{Lynch}, R.~S., {Swiggum}, J.~K., {Kondratiev}, V.~I., {et~al.} 2018, \apj,
  859, 93, \dodoi{10.3847/1538-4357/aabf8a}

\bibitem[{{Lyne}(2009)}]{Lyne2009}
{Lyne}, A.~G. 2009, in Astrophysics and Space Science Library, Vol. 357,
  Astrophysics and Space Science Library, ed. W.~{Becker}, 67,
  \dodoi{10.1007/978-3-540-76965-1_4}

\bibitem[{{Manchester} {et~al.}(2005){Manchester}, {Hobbs}, {Teoh}, \&
  {Hobbs}}]{atnfcat}
{Manchester}, R.~N., {Hobbs}, G.~B., {Teoh}, A., \& {Hobbs}, M. 2005, \aj, 129,
  1993, \dodoi{10.1086/428488}

\bibitem[{{McKinnon}(2014)}]{scatteringtail}
{McKinnon}, M.~M. 2014, \pasp, 126, 476, \dodoi{10.1086/676975}

\bibitem[{{McLaughlin} {et~al.}(2006){McLaughlin}, {Lyne}, {Lorimer}, {Kramer},
  {Faulkner}, {Manchester}, {Cordes}, {Camilo}, {Possenti}, {Stairs}, {Hobbs},
  {D'Amico}, {Burgay}, \& {O'Brien}}]{rrats}
{McLaughlin}, M.~A., {Lyne}, A.~G., {Lorimer}, D.~R., {et~al.} 2006, \nat, 439,
  817, \dodoi{10.1038/nature04440}

\bibitem[{Pedregosa {et~al.}(2011)Pedregosa, Varoquaux, Gramfort, Michel,
  Thirion, Grisel, Blondel, Prettenhofer, Weiss, Dubourg, Vanderplas, Passos,
  Cournapeau, Brucher, Perrot, \& {{\'E}}douard Duchesnay}]{scikit}
Pedregosa, F., Varoquaux, G., Gramfort, A., {et~al.} 2011, Journal of Machine
  Learning Research, 12, 2825.
\newblock \url{http://jmlr.org/papers/v12/pedregosa11a.html}

\bibitem[{{Rajwade} {et~al.}(2014){Rajwade}, {Gupta}, {Kumar}, \&
  {Arjunwadkar}}]{Kaustubh2014}
{Rajwade}, K., {Gupta}, Y., {Kumar}, U., \& {Arjunwadkar}, M. 2014, in
  Astronomical Society of India Conference Series, Vol.~13, Astronomical
  Society of India Conference Series, 73--77

\bibitem[{{Ransom} {et~al.}(2009){Ransom}, {Demorest}, {Ford}, {McCullough},
  {Ray}, {DuPlain}, \& {Brandt}}]{guppi}
{Ransom}, S.~M., {Demorest}, P., {Ford}, J., {et~al.} 2009, in American
  Astronomical Society Meeting Abstracts, Vol. 214, American Astronomical
  Society Meeting Abstracts \#214, 605.08

\bibitem[{{Ransom} {et~al.}(2002){Ransom}, {Eikenberry}, \&
  {Middleditch}}]{scott2002}
{Ransom}, S.~M., {Eikenberry}, S.~S., \& {Middleditch}, J. 2002, \aj, 124,
  1788, \dodoi{10.1086/342285}

\bibitem[{{Redman} \& {Rankin}(2009)}]{Redman2009}
{Redman}, S.~L., \& {Rankin}, J.~M. 2009, \mnras, 395, 1529,
  \dodoi{10.1111/j.1365-2966.2009.14632.x}

\bibitem[{{Ritchings}(1976)}]{Ritchings76}
{Ritchings}, R.~T. 1976, \mnras, 176, 249, \dodoi{10.1093/mnras/176.2.249}

\bibitem[{{Rosen} {et~al.}(2013){Rosen}, {Swiggum}, {McLaughlin}, {Lorimer},
  {Yun}, {Heatherly}, {Boyles}, {Lynch}, {Kondratiev}, {Scoles}, {Ransom},
  {Moniot}, {Cottrill}, {Weaver}, {Snider}, {Thompson}, {Raycraft},
  {Dudenhoefer}, {Allphin}, {Thorley}, {Meadows}, {Marchiny}, {Liska},
  {O'Dwyer}, {Butler}, {Bloxton}, {Mabry}, {Abate}, {Boothe}, {Pritt},
  {Alberth}, {Green}, {Crowley}, {Agee}, {Nagley}, {Sargent}, {Hinson},
  {Smith}, {McNeely}, {Quigley}, {Pennington}, {Chen}, {Maynard}, {Loope},
  {Bielski}, {McGough}, {Gural}, {Colvin}, {Tso}, {Ewen}, {Zhang},
  {Ciccarella}, {Bukowski}, {Novotny}, {Gore}, {Sarver}, {Johnson},
  {Cunningham}, {Collins}, {Gardner}, {Monteleone}, {Hall}, {Schweinhagen},
  {Ayers}, {Jay}, {Uosseph}, {Dunkum}, {Pal}, {Dydiw}, {Sterling}, \&
  {Phan}}]{Rosen2013}
{Rosen}, R., {Swiggum}, J., {McLaughlin}, M.~A., {et~al.} 2013, \apj, 768, 85,
  \dodoi{10.1088/0004-637X/768/1/85}

\bibitem[{{Ruderman} \& {Sutherland}(1975)}]{ruderman}
{Ruderman}, M.~A., \& {Sutherland}, P.~G. 1975, \apj, 196, 51,
  \dodoi{10.1086/153393}

\bibitem[{{Sheikh} \& {MacDonald}(2021)}]{Sofia2021}
{Sheikh}, S.~Z., \& {MacDonald}, M.~G. 2021, \mnras, 502, 4669,
  \dodoi{10.1093/mnras/stab282}

\bibitem[{{Smith}(1973)}]{exptails}
{Smith}, F.~G. 1973, \mnras, 161, 9P, \dodoi{10.1093/mnras/161.1.9P}

\bibitem[{{Stovall} {et~al.}(2014){Stovall}, {Lynch}, {Ransom}, {Archibald},
  {Banaszak}, {Biwer}, {Boyles}, {Dartez}, {Day}, {Ford}, {Flanigan}, {Garcia},
  {Hessels}, {Hinojosa}, {Jenet}, {Kaplan}, {Karako-Argaman}, {Kaspi},
  {Kondratiev}, {Leake}, {Lorimer}, {Lunsford}, {Martinez}, {Mata},
  {McLaughlin}, {Roberts}, {Rohr}, {Siemens}, {Stairs}, {van Leeuwen},
  {Walker}, \& {Wells}}]{gbncc}
{Stovall}, K., {Lynch}, R.~S., {Ransom}, S.~M., {et~al.} 2014, \apj, 791, 67,
  \dodoi{10.1088/0004-637X/791/1/67}

\bibitem[{{Taylor} {et~al.}(1975){Taylor}, {Manchester}, \&
  {Huguenin}}]{Taylor1975}
{Taylor}, J.~H., {Manchester}, R.~N., \& {Huguenin}, G.~R. 1975, \apj, 195,
  513, \dodoi{10.1086/153351}

\bibitem[{{van Straten} \& {Bailes}(2011)}]{dspsr}
{van Straten}, W., \& {Bailes}, M. 2011, \pasa, 28, 1, \dodoi{10.1071/AS10021}

\bibitem[{{van Straten} {et~al.}(2011){van Straten}, {Demorest}, {Khoo},
  {Keith}, {Hotan}, \& {et al.}}]{psrchive}
{van Straten}, W., {Demorest}, P., {Khoo}, J., {et~al.} 2011, {PSRCHIVE:
  Development Library for the Analysis of Pulsar Astronomical Data},
  Astrophysics Source Code Library, record ascl:1105.014.
\newblock \doeprint{1105.014}

\bibitem[{{Wang} {et~al.}(2007){Wang}, {Manchester}, \& {Johnston}}]{Wang2007}
{Wang}, N., {Manchester}, R.~N., \& {Johnston}, S. 2007, \mnras, 377, 1383,
  \dodoi{10.1111/j.1365-2966.2007.11703.x}

\bibitem[{Wilks(2008)}]{wilks2008mathematical}
Wilks, S. 2008, Mathematical Statistics (Read Books).
\newblock \url{https://books.google.com/books?id=iMDWgCcqswkC}

\end{thebibliography}
